\documentclass[11pt,a4paper,twoside]{article}

\usepackage[utf8]{inputenc}
\usepackage[T1]{fontenc}

\usepackage{amsmath,amssymb,amsthm,amsfonts}
\usepackage{bm}           
\usepackage{mathtools}    
\usepackage[makeroom]{cancel}
\everymath{\displaystyle}
\allowdisplaybreaks

\usepackage[dvipsnames]{xcolor}   
\usepackage{graphicx}             
\usepackage{rotating}             
\usepackage{float}                
\usepackage{subcaption}           
\usepackage[labelfont=bf,font=small]{caption}

\usepackage[dvipsnames]{xcolor}

\usepackage{tikz}
\usetikzlibrary{shapes.geometric, arrows}

\usepackage[margin=1in]{geometry}

\tikzstyle{process} = [rectangle, minimum width=3cm, minimum height=1cm, text centered, draw=blue!50, fill=blue!20, text width=3cm]

\usepackage{booktabs,threeparttable,longtable,siunitx}
\usepackage{tabularx}
\usepackage{multirow,multicol,blkarray,array}
\usepackage[section]{placeins}  
\usepackage{numprint}

\usepackage{epigraph}
\usepackage{marginnote}
\usepackage{soul}         
\usepackage{lscape, pdflscape}
\usepackage{indentfirst}
\usepackage{enumitem}
\usepackage{cases}
\usepackage{url}
\usepackage{setspace}
\usepackage{ulem}         
\usepackage{framed}
\usepackage{adjustbox}
\usepackage{here}
\usepackage{listings}
\usepackage{xstring}
\usepackage{flafter}
\usepackage{comment}

\usepackage[
  colorlinks   = true,
  linkcolor    = red,
  citecolor    = blue,
  urlcolor     = cyan,
  filecolor    = black,
  bookmarks    = false,
  pdfauthor    = {Simon Scheidegger},
  pdfsubject   = {},
  pdftitle     = {}
]{hyperref}

\usepackage{cleveref}
\crefname{equation}{Eq.}{Eqs.}
\crefname{figure}  {Figure}{Figures}
\crefname{table}   {Table}{Tables}
\crefname{line}    {Algorithm}{Algorithms}
\crefname{asmp}    {Assumption}{Assumptions}
\crefname{section} {Section}{Sections}
\crefname{chapter} {Chapter}{Chapters}
\crefname{appsec}  {Appendix}{Appendixes}

\renewcommand{\ref}{\cref}  

\usepackage[utf8]{inputenc}
\usepackage[authoryear,round]{natbib}
\usepackage{fancyhdr}
\fancypagestyle{firstpagestyle}{
  
  \fancyhf{}
  \fancyfoot[C]{\thepage}
}

\fancypagestyle{CVfirstpagestyle}{
  
  \fancyhf{}
  \fancyfoot[C]{\thepage}
}

\DeclareFontFamily{U}{mathx}{}
\DeclareFontShape{U}{mathx}{m}{n}{<-> mathx10}{}
\DeclareSymbolFont{mathx}{U}{mathx}{m}{n}
\DeclareMathAccent{\widehat}{0}{mathx}{"70}
\DeclareMathAccent{\widecheck}{0}{mathx}{"71}

\newcommand{\expnum}[2]{\ifnum#1=1 10^{#2} \else #1\!\cdot\!10^{#2}\fi}

\renewcommand{\emph}[1]{\textit{#1}}

\begin{document} 
\title{Using Machine Learning to Compute Constrained  Optimal Carbon Tax Rules\thanks{We thank Simon Dietz, Lars Hansen, Armon Rezai, Tom Sargent, Christian Träger, Rick van der Ploeg, Frank Venmans, as well as seminar participants at the University of Chicago, the Grantham Institute at LSE, the Toulouse School of Economics, the Mohammed VI Polytechnic University, the University of Zürich, Peking University, and the University of Lund for their valuable support and insights. This work was generously supported by the Swiss National Science Foundation (SNF), under project ID  \lq\lq Can Economic Policy Mitigate Climate-Change\rq\rq, and the Becker Friedman Institute for Economics at the University of Chicago.}
}
\author{
    Felix Kübler\thanks{Department for Banking and Finance, University of Zürich, Switzerland, and SFI; Email: \href{mailto:felix.kuebler@bf.uzh.ch}{felix.kuebler@bf.uzh.ch}} \and 
    Simon Scheidegger\thanks{Department of Economics, University of Lausanne; Grantham Research Institute, London School of Economics; Email: \href{mailto:simon.scheidegger@unil.ch}{simon.scheidegger@unil.ch}} \and
    Oliver Surbek\thanks{Department of Economics, HEC Lausanne; Email: \href{oliver.surbek@unil.ch}{oliver.surbek@unil.ch}}
}
\date{\today}

\maketitle

\begin{abstract}
We develop a computational framework for deriving Pareto-improving and constrained optimal carbon tax rules in a stochastic overlapping generations (OLG) model with climate change. By integrating Deep Equilibrium Networks for fast policy evaluation and Gaussian process surrogate modeling with Bayesian active learning, the framework systematically locates optimal carbon tax schedules for heterogeneous agents exposed to climate risk. We apply our method to a 12-period OLG model in which exogenous shocks affect the carbon intensity of energy production, as well as the damage function. Constrained optimal carbon taxes consist of tax rates that are simple functions of observables and revenue‑sharing rules that guarantee that the introduction of the taxes is Pareto improving. This reveals that a straightforward policy is highly effective: a Pareto-improving linear tax on cumulative emissions alone yields a 0.42\% aggregate welfare gain in consumption-equivalent terms while adding further complexity to the tax provides only a marginal increase to 0.45\%. The application demonstrates that the proposed approach produces scalable tools for macro-policy design in complex stochastic settings. Beyond climate economics, the framework offers a template for systematically analyzing welfare-improving policies in various heterogeneous‑agent problems.

\end{abstract}


\newpage

\section{Introduction}
\label{sec:intro}

\paragraph{Motivation and Challenges.}
Anthropogenic climate change presents grave uncertainties for current and future generations.
As with other externalities, climate change reflects a market failure that can be addressed with appropriate taxes, and carbon taxation is the canonical
candidate to correct for this externality.
However, it is unclear how to transfer carbon tax revenues across generations to make all generations better off and thus generate the necessary political support from the current old.
Even if one postulates a social welfare function (SWF) that aggregates the welfare of all current and future generations, determining the optimal (SWF maximizing) time–state path for carbon taxes in the presence of heterogeneous agents, stochastic shocks to technology and climate, and nonlinear damage functions remains computationally prohibitive. 
Most integrated assessment models (IAMs), therefore, either do not consider uncertainty (in which case the so-called social cost of carbon can often be used to determine optimal taxes) or invoke a representative agent,
linearize dynamics, or search heuristically over a handful of handpicked tax schedules (see, e.g., \citealp{DIETZ20241} and~\citealp{annurev:/content/journals/10.1146/annurev-economics-091124-045357} for recent reviews). As a consequence, the computational barrier has prevented a systematic search for tax-and-transfer schemes that are not only economically optimal but also politically viable, that is, policies guaranteed to make every generation, including the currently decisive elderly, better off.
\vspace{-1.25em}
\paragraph{Objectives and Model Overview.}
This paper confronts this computational barrier directly. We develop a novel, machine learning-based framework to systematically search for optimal and Pareto-improving carbon tax schedules for heterogeneous agents exposed to climate risk.\footnote{Throughout the paper, we assume that agents’ welfare is evaluated at an ex ante stage, that is, an agent is identified by the date of his birth and his welfare is evaluated as the time-zero expected utility of his cohort (see, e.g. \citealp{gottardi2011social}).} We postulate a standard SWF that consists of the weighted average of expected utilities of all current and future generations and search for carbon taxes and transfers of tax revenue between agents that are Pareto-improving and that maximize the SWF. 
We require carbon taxes to be simple (linear) functions of current observable aggregate variables. This restriction of taxes to simple functional forms enables us to clearly identify aggregate variables that serve as sufficient statistics for welfare-improving carbon taxes. In this setting, we examine how
(i) optimal taxes and transfers of tax revenues redistribute risk and welfare across overlapping generations (OLG), and (ii) 
how large the potential welfare gains from \emph{Pareto‑improving} policies could be.  
By construction, an allocation that maximizes our social welfare function is Pareto-efficient; in that sense, our method finds one specific constrained Pareto-efficient carbon tax schedule whose introduction is Pareto-improving.
\vspace{-1.25em}
\begingroup
\paragraph{Our Contribution: A Framework for Finding Pareto-improving Policies.}
The primary contribution of this paper is a generic, three-step, machine learning-based methodology for computing constrained optimal policies in complex stochastic models with heterogeneous agents. Our approach alleviates the curse of dimensionality~\citep{bellman1961adaptive} that renders such problems computationally prohibitive for traditional methods. The framework consists of three integrated steps:
\begin{enumerate}[label=\textbf{\arabic*.}, wide, labelwidth=!, labelindent=0pt]
\item \textbf{Global Solution as a Deep Surrogate:} We first compute a global solution for the stochastic OLG model across a continuum of possible tax-and-transfer rules. This is achieved using Deep Equilibrium Nets (DEQNs; see~\citealp{azinovicDEEPEQUILIBRIUMNETS2022}), which treat policy parameters as ``pseudo-states'' to generate a comprehensive ``deep surrogate'' for the model's policy functions.\footnote{A ``surrogate model'', also known as an emulator or response surface, is a statistical or machine-learning approximation that replicates the input–output behaviour of an expensive simulation at a fraction of the computational cost (see, e.g.,~\citealp{chen2025Deep}, and references therein).} 
    \item \textbf{Efficient Welfare Evaluation:} We then train a Gaussian Process (GP) surrogate model (see, e.g.,~\citealp{Rasmussen:2005:GPM:1162254} for a textbook treatment) to create a computationally cheap and accurate approximation of the SWF. Due to the data efficiency of GPs, this step requires remarkably few simulations of the full, expensive model.
    \item \textbf{Rapid Policy Optimization:} Finally, we leverage the speed of the GP surrogate to solve the planner’s complex, non-convex optimization problem. This allows us to rapidly find the specific tax-and-transfer parameters that maximize social welfare, including those subject to the crucial constraint that the policy must be Pareto-improving.\footnote{A key advantage of our framework is its computational efficiency. The entire three-step procedure completes in a matter of hours on a standard laptop. This stands in stark contrast to traditional methods for such complex models, which can require tens of thousands of core-hours on high-performance computing clusters and would render the systematic search for Pareto-improving policies computationally infeasible (see, e.g.,~\citealp{BKS1}, and references therein).}
\end{enumerate}
This integrated framework produces scalable tools for macropolicy design. To demonstrate its power, we apply it to a 12-period OLG model where agents are exposed to stochastic shocks to carbon intensity and the climate damage function.\footnote{In layman’s terms, carbon intensity is how much CO$_2$ pollution comes out for every dollar’s worth of goods and services we produce: the higher the number, the dirtier the production; the lower the number, the cleaner.}
\endgroup

\vspace{-1.25em}
\paragraph{Preview of Results.}
We apply our three-step computational framework to a series of increasingly complex policy design problems, demonstrating its ability to systematically derive and evaluate optimal carbon tax rules. First, we establish a BAU baseline, which, in the absence of policy, projects a mean global warming of approximately 3°C over a 150-year horizon, with worst-case economic damages potentially exceeding 15\% of GDP.

Our first policy experiment considers a simple, welfare-improving tax rule that is linear in cumulative emissions. The optimal policy under this scheme significantly mitigates climate risk, stabilizing the average Global warming at around 2.7°C and substantially reducing the probability of extreme damage scenarios. This policy, however, induces a trade-off across generations: while it increases overall social welfare by 1.6\% in consumption-equivalent terms, the gains are concentrated among future cohorts, whereas the initial generations bear welfare losses of up to 5\%.

To address this distributional conflict, our second experiment finds a Pareto-improving policy by simultaneously optimizing the linear tax on cumulative emissions and an accompanying intergenerational transfer scheme, subject to the additional constraint that all generations must be weakly better off. Our framework successfully identifies a tax-and-transfer rule that ensures no generation is made worse off than under the BAU scenario. The optimal transfers effectively shield the initial cohorts from welfare losses, enabling a modest but strictly positive climate mitigation policy. This policy yields an overall social welfare gain of 0.42\% in consumption-equivalent terms, without harming any generation, demonstrating that politically feasible climate action is achievable within the model. However, this political viability necessitates a significantly less ambitious policy path, with much lower taxes, more modest climate mitigation, and substantially smaller welfare gains for future generations.

Finally, we explore whether more complex tax rules provide significant additional benefits. We expand the tax base to be a linear function of not only cumulative emissions but also carbon intensity and proximity to a climate tipping point. While our framework again finds a Pareto-improving policy, the additional welfare gain is minimal. The overall social welfare increases to 0.45\% in consumption-equivalent terms, a marginal improvement of only 7\% over the simpler tax-and-transfer scheme. This key result suggests that, in our calibration, the bulk of the welfare gains can be captured by a simple, well-designed tax on cumulative emissions, and that adding further complexity to the tax base yields diminishing returns.

\paragraph{Related Literature.}
This paper is related to three strands of literature: (i) studies analyzing global warming through the lens of OLG models, (ii) research on optimal taxation with heterogeneous agents, and (iii) the growing literature on solving dynamic economic models using machine learning techniques.  
  
(i) Examples of OLG models with environmental concerns include \cite{Burton1993}, \cite{Heijdra2006}, \cite{Karp2014}, \cite{kotlikoff2021making}, \cite{Karp_Peri_Rezai_2024}, and \cite{kotlikoffCanTodaysTomorrows2024}, who incorporate clean and dirty energy sectors, but abstract from climate uncertainty. Multiple studies, including \cite{weitzman2012ghg}, \cite{golosov2014optimal}, \cite{Barnett2020}, and \cite{kotlikoff2020pareto}, emphasize the importance of accounting for climate risks in optimal policy design.
Our paper is most closely related to \cite{kotlikoff2020pareto}, who study optimal Pareto-improving taxes in an OLG model with climate uncertainty. While we abstract from modelling clean and dirty energy sectors separately, we expand on their work by proposing a global solution method to compute the Pareto-optimal state-dependent carbon taxes in an OLG model with climate uncertainty.

(ii) We contribute to the literature on optimal Ramsey problems in quantitative, heterogeneous-agent models. \cite{dyrda2023optimal}, \cite{bhandari2013}, and \cite{feng2024optimal} analyze optimal fiscal policy in incomplete-markets models with heterogeneous agents. \cite{itskhoki2019Optimal} analyze the Ramsey optimal development policy. \cite{nuno2020Optimal} find the Ramsey optimal monetary policy. \cite{douenne_hummel_pedroni_2024} are probably closest to our work in that they consider optimal policy in a model with climate change. The main difference between these papers and our work is that we focus on simple tax rules and only search for tax rules that are optimal within this small set. This simplifies the computational analysis, allows us to solve for optimal policy in much more complex environments, and makes an interpretation of our results much more straightforward.

(iii) Finally, we also contribute to the rapidly growing literature on using machine learning techniques to solve dynamic stochastic models.\footnote{See \citet{fernandez2024taming} for a recent review.} We combine methods developed in \cite{SCHEIDEGGER201968}, \cite{azinovicDEEPEQUILIBRIUMNETS2022}, and \cite{chen2025Deep}. Conceptually, our workflow resembles that of \citet{friedlDeep2023}: they (i) use DEQNs to solve a stochastic planner’s problem and (ii) fit a Gaussian-process surrogate to the social cost of carbon, which they then exploit for parametric uncertainty quantification~\citep{Harenberg17}. Our framework departs from theirs in three key ways.  First, we work with a stochastic overlapping-generations (SOLG) model instead of a representative-agent setting. Second, rather than treating structural parameters such as risk aversion as pseudo-states, we embed the coefficients of the carbon-tax and transfer rules directly in the state vector and solve for the resulting policy functions with a single DEQN run. Third, we construct GP surrogates for the SWF, not for the social cost of carbon, and embed this surrogate in an optimizer to recover welfare-optimal tax and transfer schedules.
\vspace{-1.25em}
\paragraph{Organization of the Article.}
Section~\ref{sec:model} introduces a SOLG framework with an explicit climate externality. 
Section~\ref{sec:calibration} explains our choice of key parameters and functional forms of the model.
Section~\ref{sec:methodology} develops a generic, machine-learning-based procedure for computing Pareto-improving policies in heterogeneous-agent economies with various sources of uncertainty. Section~\ref{sec:results} illustrates the method’s performance through a series of numerical experiments, and Section ~\ref{sec:conclusion} concludes.

\section{A Stochastic OLG Model with Climate Change}
\label{sec:model}
We consider a stochastic OLG economy with climate change caused by industrial emissions, which leads to a decrease in future total factor productivity.
There is an infinite time-horizon, $ t=0,\ldots $, and exogenous Markovian shocks realize every period.
The economic block consists of 12 overlapping generations of selfish agents, in which each period spans five calendar years; households enter the economy at age~20 and exit at age~80~\citep{kotlikoff2020pareto}. 
A representative firm produces a single consumption good using capital and labor as inputs and emitting CO$_2$ during the production process. We assume exogenously declining stochastic carbon intensity dynamics.
The climate block is represented by a reduced-form, cumulative-emissions emulator~\citep{dietz2019cumulative}.\footnote{Climate emulators are simplified models that link CO$_2$ emissions to global mean temperature at negligible computational cost and are therefore well suited for policy analysis. For more details, see, e.g.,~\cite{doi:10.1086/713977,Folini_2021}, and~\cite{eftekhari2024building}.} 
Climate damages follow the convex specification of~\citet{weitzman2012ghg}, augmented with stochastic tipping points in the spirit of~\citet{kotlikoff2020pareto}.
We outline the model in the following steps: Section~\ref{sec:firms} presents the firm block, Section~\ref{sec:households} explains household behaviour, Section~\ref{sec:government} details the role of the government, Section~\ref{sec:climate_module} describes the climate module, and Section~\ref{sec:equilbrium} defines the competitive equilibrium.

\subsection{Firms}
\label{sec:firms}
The representative firm produces consumption good $Y$ and emissions $e_t$ at each time $t$:
\begin{equation}
(Y_t,e_t)=f_t(K_t,L_t,\mu_t)=\left( \Omega(T^{AT}_t)\Phi(1-\mu_t) K_t^{\alpha} L_t^{1-\alpha} + (1 - \delta) K_t, (1-\mu_t) \kappa_t K_t^{\alpha} L_t^{1-\alpha}\right).
\end{equation}
Production is given by a Cobb-Douglas production function, where $\alpha$ represents the elasticity of substitution of capital, $K$, $L$, and $\delta$ are capital, labour, and depreciation, respectively. $ \mu $ denotes abatement, and $ \Phi:[0,1] \rightarrow [0,1] $ is a mitigation cost function. $\kappa$ is the exogenous carbon intensity of production and $\Omega(T^{AT})$ are damages arising from climate change, where $ T^{AT} $ denotes current average temperature. Firms take as given a carbon tax rate $\tau_t$ and maximize profits:
\begin{equation}
    \max_{\mu_t, K_t, L_t} \Pi  =\Omega_t(T_{t}) (1-\theta_{1}\mu_t^{\theta_2}) K_t^{\alpha} L_t^{1-\alpha} + (1- \delta) K_t -w_tL_t-(1+r_t)K_t - \tau_t (1-\mu_t) \kappa_t K_t^{\alpha} L_t^{1-\alpha}
    \label{eq:profit}
\end{equation}
$$\text{s.t.} \quad 0 \leq \mu_t \leq 1.$$

The carbon intensity ($\kappa_t$) follows an exogenous time-varying stochastic process. In the calibration, our objective will be to capture the uncertainty in CO$_2$ emissions and account for the possibility that emissions remain elevated over the next 100 years. 

We assume an abatement function of the form $ \Phi (1- \mu) = 1 - \theta_{1} \mu_t^{\theta_2}$ in the spirit of \cite{nordhausRevisitingSocialCost2017}. Abatement is restricted to be $0 \leq \mu \leq 1$. Unabated emissions lead to an increase in atmospheric temperature, which triggers damages that reduce output. We employ a damage function as in \cite{weitzman2012ghg} and follow  \cite{kotlikoff2020pareto} by enhancing damages with stochastic tipping:
\begin{align}
    \label{eq:damage_func_tipping_new}
    \Omega_t \left( T_t^{AT} \right) =  \frac{1}{1 + \left(\frac{1}{\psi_1} T_t^{AT} \right)^{2} + \left( \frac{1}{2 \cdot TP_{t}} T_t^{AT} \right)^{6.754}},
\end{align}
where $T_t^{AT}$ is atmospheric temperature and $TP_t$ is the tipping point which follows a random walk.\footnote{A tipping point is a critical threshold in the climate system beyond which a small additional change triggers large, abrupt, and often irreversible shifts in the state of the system (examples include a collapse of the Atlantic Meridional Overturning Circulation, Amazon rainforest dieback, or disintegration of the West Antarctic Ice Sheet; see, e.g., \citealp{lenton2008tipping,doi:10.1086/701890}). 
Note that, whereas we refer to a tipping point in Equation~\eqref{eq:damage_func_tipping_new}, \citet{DIETZ2012321} consider such a damage specification, with a changing exponent, as an instance of epistemic uncertainty.}

\subsection{Households}
\label{sec:households}

In each period, a new household enters the economy with zero assets. Households live for A =12 periods, where each period corresponds to 5 years. There is no uncertainty regarding lifetime. An agent born at time $t$ maximizes expected lifetime utility given by:
\begin{equation}
     {\mathbb E}_t \sum\limits_{j = 1}^{A} \beta^j \frac{C_{t+j-1, j}^{1-\sigma}}{1-\sigma}, 
    \label{eq:agentproblem}
\end{equation} 
subject to: 
\begin{equation}
C_{t,j}+a_{t+1, j+1}=(1+r_t)a_{t,j} + w_{t} l_{j} + \mathbb{T}_{t,j},
\label{eq:BC_main}
\end{equation} 
where $\beta$ is the discount factor, $C_{t,j}$, $a_{t, j}$, $l_{j}$ are consumption, assets, and labour endowment of generation $j$ at time $t$. $r_t$ and $w_t$ are the factor prices of production (interest rate and wages), and $\mathbb{T}_{t,j}$ is the transfer from the government.

In each period, households sell their assets to the firm and supply labor inelastically. Households choose consumption $C_{j,t}$ and savings $a_{t+1, j+1}$ in period $t$. We follow \citet{benzell2015simulating} and assume that agents enter the workforce with $0$ assets, work for $8$ periods ($40$ years), and then retire. After retirement, agents' labor endowment drops to $40\%$ of the endowment in period 8, which serves as a proxy for a pension system. 
Further details on the recursive formulation of the household problem are given in Appendix~\ref{sec:appendix_recursive}.

We define the time-zero expected utilities of a future generation, born at time $t$ as 
\begin{equation}
\tilde{U}_t = \mathbb{E}_0 \sum\limits_{j = 1}^{A} \beta^j \frac{C_{t+j-1, j}^{1-\sigma}}{1-\sigma}, 
\label{eq:utility} 
\end{equation}
where $ C_{t+j-1,j} $ denotes the agent's optimal consumption under a policy scenario with taxes. This distinguishes it from the utility in the BAU case, denoted by $U_t$, which assumes zero taxes and is defined analogously.

\subsection{Government}
\label{sec:government}
The government levies a carbon tax on firms and redistributes the tax income among concomitant households. It sets the path for the tax rate $\tau_t$ and redistribution schedule $\mathbb{T}_{t}$ in period 0 to maximize ex-ante welfare given Welfare weights $\gamma$. Specifically, the government maximizes the ex-ante sum of weighted utilities:
\begin{equation}
    \max_{\tau_t} \sum_{t=-10}^{29} \gamma_t \tilde{U_t},
    \label{eq:planner_objective_general}
\end{equation}
subject to the resource constraint:
\begin{equation}
    \sum_{j=1}^{A} \mathbb{T}_{t,j} = \tau_{t} e_t \quad  \forall t .
\label{eq:resource_constraint_planner}
\end{equation}
In the numerical experiments performed in Section~\ref{sec:results}, we consider linear, state-dependent taxes, $\tau_t(\cdot)$, on cumulative emissions, carbon intensity, and the climate-tipping variable.
We evaluate welfare gains by computing the consumption equivalence factor:
  \begin{equation}
       \left(\frac{\tilde{U}_t}{U_t}\right)^{\frac{1}{1-\sigma}}-1,
       \label{eq:cons_equiv_variation}
  \end{equation}   
In addition to considering the maximization problem (\ref{eq:planner_objective_general}), we also consider the additional constraint 
\begin{equation}
\label{eq:paretoim}
  \left(\frac{\tilde{U}_t}{U_t}\right)^{\frac{1}{1-\sigma}}-1 \ge 0, \mbox{ for all } t=-10,\ldots,29 .
       \end{equation}
%

\subsection{Climate Module}
\label{sec:climate_module}
Climate science has established a near-linear relationship between cumulative CO$_2$ emissions and the resulting global warming, which we adopt: 
\begin{equation}
    T_t^{AT} \approx \sigma_{CCR} \sum_{s=0}^{t-1} e_s ,
\end{equation}
where $T_t^{AT}$ denotes the global mean surface temperature, $e_s$ represents global emissions, and $ \sigma_{CCR}$ is the constant of proportionality, referred to as the carbon-climate response (CCR) or the Transient Climate Response to Cumulative Carbon Emissions (TCRE; \citealp{matthews2009proportionality, dietz2019cumulative}). We report emissions in terms CO$_2$. For notational convenience, we also define the stock of carbon in the atmosphere as
\begin{equation}
    E_t = \frac{1}{3.666}\sum_{s=0}^{t-1} e_s, 
\end{equation}
where $3.666$ is the conversion factor from Carbon C to CO$_2$.
Additional details can be found in Appendix~\ref{sec:Climate_module_appendix}.\footnote{We report cumulative emissions in thousand gigatonnes of carbon (th Gt C) and, for brevity, use the terms ``stock of carbon'' and ``cumulative emissions'' interchangeably.}

\subsection{Equilibrium}
\label{sec:equilbrium}

A competitive equilibrium, given initial conditions ($E_0,T_0^{AT},TP_0,\kappa_0,\{a_{0,j} \}_1^A$), a tax function (e.g., $\tau(E,\kappa,TP)$) and transfer scheme ($\mathbb{T}$), is a collection of choices for households ($\{c_{0,j} \}_{j=1}^A$,$\{a_{0,j+1} \}_{j=1}^{A}$), the representative firm $(K_t,L_t)_{t=0}^\infty$, and prices $(r_t,w_t)_{t=0}^\infty$ such that:
\begin{enumerate}
    \item Given $(r_t, w_t, \mathbb{T})_{t=0}^\infty$, households maximize \eqref{eq:utility} subject to \eqref{eq:BC_main}.
    \item Given $(r_t, w_t, \tau_t)_{t=0}^\infty$, firms maximize \eqref{eq:profit}.
    \item All markets clear, that is,
    \begin{equation}
\label{eq:capmarket}
    \sum_{j = 1}^{A} a_{t,j} = K_t \quad \forall t,
\end{equation}
\begin{equation}
\label{eq:labmarket}
    \sum_{j = 1}^{A} l_j = L_t \quad \forall t.
\end{equation}
\end{enumerate}
The first-order conditions of the firm imply that
\begin{equation}
    \label{eq:FOC_mu}
\Omega_t(T_{t}^{AT}) (\theta_1\theta_2\mu_t^{\theta_2-1})  = \tau_t \kappa_t,
\end{equation}
\begin{equation}
 r_t = \alpha K_t^{\alpha-1} L_t^{1-\alpha}\left(\Omega_t(T_{t}^{AT}) (1-\theta_1\mu_t^{\theta_2}) - \tau_t \kappa_t(1-\mu_t)  \right)  - \delta,
\end{equation}
\begin{equation}
    w_t = (1-\alpha) K_t^{\alpha} L_t^{-\alpha} \left(\Omega_t(T_{t}^{AT}) (1-\theta_1\mu_t^{\theta_2}) -\tau_t \kappa_t (1-\mu_t) \right) .
\end{equation}
The Euler equations for the households $j \in 1,...,A-1$ are given by:
\begin{equation}
\label{eq:euler}
  C_{t,j}^{-\sigma} = \beta \mathbb{E}_t \left[ (1+r_{t+1}) C_{t+1,j+1}^{-\sigma}  \right]. 
\end{equation}
Note that there is no Euler equation for $j=A$ because agents consume their entire endowment in the final period.

\section{Calibration}
\label{sec:calibration}

This section presents the model calibration, and the corresponding parameter values are summarized in Table~\ref{tab:Parameters}.

\paragraph{Households.} Households live for 12 periods, where each period corresponds to 5 years.  
Labour endowments follow the functional form proposed by \citet{benzell2015simulating}: 
\begin{equation}
\label{eq:labendow}
    l_{j} =  \left\{ \begin{array}{ll} l_0 e^{4.47+0.033 j - 0.00067 j^2} & j \leq 40,\\
    l_j=1/3\cdot l_{40} & j > 40. \end{array} \right.
\end{equation}
The aggregate labor endowment is normalized to a value of 1. The preference parameters are standard and can be found in Table \ref{tab:Parameters}. To obtain the initial asset distribution, we simulate a deterministic model with constant temperature and damages but no climate change, using its equilibrium as the initial distribution.

\begin{table}[t!]
    \centering
    \begin{tabular}{l|lll}
    \hline \hline
 & \textbf{Symbol} & \textbf{Parameter} & \textbf{Value} \\
    \hline
    Households & $\beta$  & Discount factor& 0.9 \\
    & $\sigma$  & Relative risk aversion & 3 \\
    &$L$ & Aggregate Labor (to match initial emissions) & 600 \\
    \hline
   Firms &$\alpha$  & Capital share & 0.3 \\
    &$\delta$  &Depreciation & 0.2 \\
    &$\kappa_{0}$  & Initial carbon intensity & 0.35032 \\
    &$\rho_0$  & Initial AR(1) param for carbon intensity & 1.08 \\
    &$\rho_{\infty}$  & Final AR(1) param for carbon intensity & 0.91 \\
    &$\delta^{\rho}$  & AR(1) param depreciation for carbon intensity  & 0.04 \\
        &$\theta_1$  & Intercept of abatement cost function & 0.7 \\
    &$\theta_2$  & Exponent of abatement cost function & 2.6 \\
    \hline
   Climate &$E_0$ & initial stock of carbon thGtC& 0.851 \\
    &$\sigma_{CCR}$  & Climate sensitivity & 1.7 \\
    &$\psi_1$  & parameter damage function & 13.16 \\
    \bottomrule
\end{tabular}
    \caption{Summary of Model Parameters.}
    \label{tab:Parameters}
\end{table}

\vspace{-1.25em}
 \paragraph{Production.} 
The production function is Cobb-Douglas with standard parameters (see Table~\ref{tab:Parameters}).  
We consider uncertainty in the carbon intensity of production $\kappa_t$. 
We model carbon‐intensity uncertainty as a non-negative autoregressive process: 
\begin{equation}
  \kappa_t =
  \begin{cases}
    \rho_t\,\kappa_{t-1} + \epsilon_{\kappa,t}, & \text{if } \rho_t\,\kappa_{t-1} + \epsilon_{\kappa,t} > 0,\\[4pt]
    0, & \text{otherwise},
  \end{cases}
\end{equation}
where the innovation \(\epsilon_{\kappa,t}\) is i.i.d.\ and takes the following discrete values:
\begin{equation}   
  \epsilon_{\kappa,t} =
  \begin{cases}
    0.03, & \text{with probability } \tfrac{1}{3},\\
    0.00, & \text{with probability } \tfrac{1}{3},\\
   -0.03, & \text{with probability } \tfrac{1}{3}.
  \end{cases}
\end{equation}
Thus, \(\kappa_t\) follows a bounded random walk that cannot fall below zero: once it reaches zero, it remains at that absorbing state.

Carbon intensity declines over time, with $\kappa = 0$ serving as an absorbing state.  
The persistence parameter \(\rho_t\) evolves deterministically according to  
\begin{equation}
  \rho_t \;=\; \rho_0 + \bigl(\rho_\infty - \rho_0\bigr)\!\left(1 - e^{-\delta^\rho \,\Delta_t\, t}\right),
\end{equation}
where \(\rho_0\) denotes the initial level, \(\rho_\infty\) its long-run limit, and $\delta^\rho$ governs the speed of convergence to that limit (see Table~\ref{tab:Parameters} for their numerical values).

Following \citet{nordhausRevisitingSocialCost2017}, we initialise carbon intensity at $\kappa_0 = 0.35032$ and set the stock of cumulative emissions to 851 GtC.  
Total factor productivity is calibrated to $600^{1-\alpha}$; this parameter acts solely as a scaling factor that matches base-year CO$_2$ emissions to observed data.  
Figure~\ref{fig:results_BAU} below shows the resulting BAU emission path together with its uncertainty band.

We set the mitigation‐cost parameters to $\theta_1 = 0.7$ and $\theta_2 = 2.6$. The deliberately high value of $\theta_1$ prevents full abatement from ever becoming optimal, while $\theta_2$ follows the estimate in \citet{nordhausRevisitingSocialCost2017}.

\vspace{-1.25em}
\paragraph{Climate Module.}
To obtain roughly $3^{\circ}\text{C}$ of average warming under BAU while approximating the RCP 4.5 emissions path, we calibrate the carbon-intensity parameters to $\rho_0 = 1.08$, $\delta^{\rho} = 0.04$, and $\rho_{\infty} = 0.91$.

\vspace{-1.25em}
\paragraph{Damages.}
Following \citet{kotlikoff2020pareto}, we model climate damages with a stochastic tipping point.  
The threshold $TP_t$ evolves as a bounded random walk provided that the contemporaneous atmospheric temperature $T_t^{\mathrm{AT}}$ remains below the threshold,
\begin{equation}
      TP_t = TP_{t-1} + \epsilon_{TP,t},
\end{equation}
where the innovation $\epsilon_{TP,t}$ is i.i.d.\ and takes the following discrete values:
\begin{equation}
  \epsilon_{TP,t} =
  \begin{cases}
    0.1, & \text{with probability } \tfrac{1}{3},\\
    0,   & \text{with probability } \tfrac{1}{3},\\
   -0.1, & \text{with probability } \tfrac{1}{3}.
  \end{cases}
\end{equation}
Reflective barriers at $\underline{TP}=2.5$ and $\overline{TP}=3.5$ prevent implausible extremes.  
Once the realized temperature reaches the threshold $\bigl(T_t^{\mathrm{AT}} \ge TP_t\bigr)$, the tipping point becomes absorbing: $TP_{t+s}=TP_t$ for all $s \ge 0$.  
The curvature parameter of the damage function is set to $\psi_1 = 13.17$.

\section{Machine Learning for Computing Pareto
Optimal Policies}
\label{sec:methodology}

This section presents our generic, three-step machine learning-based methodology for computing welfare-maximizing or (constrained) Pareto-optimal policies in stochastic models with heterogeneous agents, using DEQNs and Gaussian Process Regression (GPR) to maximize the planner’s objective.
\vspace{-1.25em}
\paragraph{Step 1: Deep Equilibrium Nets Surrogate.}
In Step 1 (cf. Section~\ref{sec:DEQN} for details), we employ DEQNs~\citep{azinovicDEEPEQUILIBRIUMNETS2022} to solve the SOLG IAM over a continuum of parameterized tax (and transfer) schemes by treating their parameters as pseudostates in the model’s state vector~\citep{chen2025Deep}. This formulation enables the simultaneous computation of global solutions for all feasible schemes in a single model run, yielding policy functions as explicit mappings:
\begin{equation}
P\colon (x_{\texttt{state,t}},\vartheta)\,\mapsto\,\mathbb{R}^{M},
\end{equation}
where $x_{\texttt{state,t}}\in\mathbb{R}^{d_x}$ is the economic-climate state vector, and $\vartheta\in[a,b]\subset\mathbb{R}^{d_\vartheta}$ are the model parameters. We define the augmented state:
\begin{equation}
    x_t = \begin{bmatrix}
x_{\texttt{state,t}} \\[0.5em]
\vartheta
\end{bmatrix}
\in \mathbb{R}^{d_x + d_\vartheta},
\end{equation}
with overall problem dimension $d = d_x + d_\vartheta$. In our specific implementation of the SOLG model (see Section~\ref{sec:results}), the augmented state vector reads as:
\begin{equation}
  x_{t}
  =\bigl[
      t,\;
      TP_t,\;
      TP_{\text{reached},t},\;
      \kappa_t,\;
      \{a_{t,j}\}_{j=1}^{A},\;
      E_t,\;
      \vartheta
    \bigr]^{\top}
  \in \mathbb{R}^{A+5+d_\vartheta}.
\end{equation}
The block $\vartheta$ collects the pseudo–state variables, whose dimensions and interpretations depend on the tax and transfer scheme being analyzed.  
For example, in Section~\ref{sec:cum_emissions} we study a linear cumulative-emissions tax with parameters $\vartheta=(\vartheta_0,\vartheta_E)\in\mathbb{R}^2$, for which 
\begin{equation}
  \tau_t = \vartheta_0 + \vartheta_E\,E_t .
\end{equation}
While the preceding example is illustrative, we solve much richer model specifications below. The SOLG model discussed in Section~\ref{sec:emissions_carbon_intensity_damages_optimal_transfer}, for instance, assumes a tax policy based on cumulative emissions, carbon intensity, climate tipping, and optimal transfers, thereby introducing 16 additional pseudo-state variables, resulting in a state space of dimension $d=33$. We approximate the policy by a neural network, that is, $P(x_t)\approx\mathcal{N}(x_t)$, trained to generate the 22-dimensional output of the SOLG (cf. Section~\ref{sec:model}): 
\begin{equation}
  \mathcal{N}(x_t)
  \;=\;
  \bigl[
      \{a_{t+1,j}\}_{j=1}^{A-1},
      \{v_{t,j}\}_{j=1}^{A-1}
    \bigr]^{\top}
  \in \mathbb{R}^{22},
\end{equation}
which concatenates the next-period asset choices and contemporaneous, normalized value functions for all cohorts (see Appendices~\ref{sec:appendix_recursive} and~\ref{sec:DEQN_for_SOLG_IAM} for more details). 
Training the DEQN is computationally demanding; it typically requires several hours on a standard laptop, but once fitted, policy evaluations (i.e., interpolating the trained neural network for a single state) and forward simulations (i.e., chaining these queries to generate a full time path) are virtually cost-free, completing in a few seconds.
This illustrates the key advantage of surrogate models: the computational burden is front-loaded, and the one-off (seemingly high) cost is quickly amortised over subsequent analyses.\footnote{For perspective, state-of-the-art solvers for stochastic OLG models, which do not include a climate block, such as the time-iteration method with adaptive sparse grids (see~\citealp{IPDPS2018}, and references therein) require of the order of tens of thousands of core-hours to solve a single calibration (i.e., a fixed set of parameters) that features roughly the same number of state variables and policy functions as the largest model we consider (cf. Section~\ref{sec:emissions_carbon_intensity_damages_optimal_transfer}). In our framework, Step 3 evaluates thousands of tax–transfer parameter vectors; consequently, employing conventional techniques would render a systematic search for welfare- and Pareto-improving tax–transfer schemes computationally infeasible.}


\vspace{-1.25em}
\paragraph{Step 2: Gaussian-process Social Welfare Function Surrogate.}

In Step 2 (see Section~\ref{sec:welfare_surrogate} for details) we fit a GPR surrogate that approximates the expected SWF, our quantity of interest, $\widehat{\mathrm{QoI}}$, as a function of the policy parameter vector $\vartheta$:
\begin{equation}
  \widehat{\mathrm{QoI}}\colon \vartheta \;\longmapsto\; \mathbb{R}.
\end{equation}
For example, in the welfare-maximization exercise studied in Section~\ref{sec:cum_emissions}, we consider
$\widehat{\mathrm{QoI}}(\theta):= \mathbb{E}[W(\theta)]$, the expected value of the social-welfare function under the tax scheme parametrized with~$\theta$. The training data for these GPR surrogates are generated through computationally relatively expensive simulations that leverage the policy functions $P(x_{t}) \approx \mathcal{N}(x_{t})$, computed in Step 1 using DEQNs. To mitigate the relatively high computational cost of generating training data, we employ GPR due to its ability to accurately interpolate the SWF using only a limited number of efficiently selected simulation points via Bayesian active learning, a form of reinforcement learning (see~\citealp{rennerscheidegger_2018}, and references therein).

\vspace{-1.25em}
\paragraph{Step 3: Optimization of Welfare Parameters.}
In Step 3 (see Section~\ref{sec:welfare_surrogate} for details), the GPR surrogate’s low evaluation cost enables rapid solution of the planner’s optimization problem:
\begin{equation}
\begin{aligned}
\vartheta^{*} &= \arg\max_{\vartheta\in[a,b]} \widehat{\mathrm{QoI}}(\vartheta), \\
&\quad \text{s.t. constraints}.
\end{aligned}
\label{eq:Planner_optimization}
\end{equation}
Given the constraints and the GP surrogate objective obtained in Step 2, we can identify welfare-improving (Pareto-optimal) policy parameters $\vartheta^{\ast}$ for a given tax–and-transfer scheme at virtually no additional computational cost: the optimization is performed directly on the GP surrogate, without the need for further model simulations, and thus completes in seconds.\footnote{Appendix~\ref{appendix:sampling} provides further details on the model-dependent constraints for Equation~\eqref{eq:Planner_optimization} and the permissible ranges for the pseudo-states $\vartheta$.}

\begin{figure}[th!]
  \centering
  \begin{tikzpicture}[node distance=1.5cm]
    \node (deqn) at (-1,0) {Step 1: DEQN};
    \node (gp) at (-1,-1.5) {Step 2: GP($\theta$)};
    \node (optimizer) at (-0.7,-3) {Step 3: Optimizer};
    \node (pxtheta) [process] at (4,0) {${P}(x_{\text{state},t},\vartheta)$};
    \node (qoi) [process] at (4,-1.5) {${\widehat{\mathrm{QoI}}}(\vartheta)$};
    \node (argmax) [process] at (4,-3) {$\arg\max {\widehat{\mathrm{QoI}}}(\vartheta)$};
    \draw [->, gray] (pxtheta) -- (qoi);
    \draw [->, gray] (qoi) -- (argmax);
  \end{tikzpicture}
\caption{Schematic of the solution scheme: first, DEQNs are used to solve the SOLG to obtain the optimal policy, $P(x_{\text{state},t},\vartheta)\approx \mathcal{N}(x_{\text{state},t},\vartheta)$, as a function of the economic and climate state variables, as well as the parameters (which have been added to the state space as so-called pseudo-state variables); second, a GP surrogate is trained on the SWF, simulated from $\mathcal{N}(x_{\text{state},t},\vartheta)$; third, an optimizer finds the policy parameters $\vartheta^*$ that maximize $\widehat{\mathrm{QoI}(\vartheta)}$.}
  \label{fig:solution_scheme}
\end{figure}
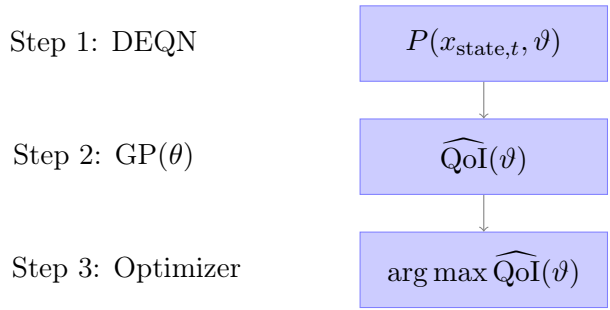

Figure~\ref{fig:solution_scheme} provides a schematic overview of the three-step procedure outlined above.
%

\subsection{Deep Equilibrium Nets} 
\label{sec:DEQN}

In this section, we provide a concise summary of the general concept of DEQNs, following the notation used by \cite{azinovicDEEPEQUILIBRIUMNETS2022} and~\cite{nuno2024monetary}. DEQNs are a simulation-based approach that employs deep neural networks to approximate the optimal policy function $P: X \to Y \subset \mathbb{R}^M$ for a dynamic model. This method assumes that the underlying economy can be described by discrete-time first-order equilibrium conditions, expressed as:
\begin{equation}
 G(x_t, P) = 0, \, \forall x_t \in X \subset \mathbb{R}^d.  \label{eq:Policy_Function_DEQN} 
\end{equation}

\vspace{-1.25em}
\paragraph{Neural-network approximation.} In essence, DEQNs operate by approximating an unknown policy function using a neural network. Specifically, the policy function \(P(x)\) is approximated as $P(x_t) \approx \mathcal{N}(x_t)$, where \(\mathcal{N}(\cdot)\) is a neural network with trainable parameters \(\nu\). These parameters are initially unknown and are determined by optimizing a suitable loss function that evaluates the accuracy of the approximation at a given economic state.

In our SOLG application, we use densely connected feed-forward neural networks (FNN). In accordance with the existing literature, we define an \( L \)-layer FNN as a function \(\mathcal{N}^L(x_t) : \mathbb{R}^{d_{\text{input}}} \to \mathbb{R}^{d_{\text{output}}}\), where the network consists of \(L-1\) hidden layers. The \(\ell\)-th layer contains \(N_\ell\) neurons, with \(N_0 = d_{\text{input}}\) and \(N_L = d_{\text{output}}\). For each layer \(1 \leq \ell \leq L\), we introduce a weight matrix \(\mathbf{W}^\ell \in \mathbb{R}^{N_\ell \times N_{\ell-1}}\) and a bias vector \(b^\ell \in \mathbb{R}^{N_\ell}\). The affine transformation in the \(\ell\)-th layer is defined as \(A^\ell(x_t) = \mathbf{W}^\ell x_t + b^\ell\). Using a nonlinear activation function \(\sigma(\cdot)\), such as ReLU, Swish, or GeLu, the FNN can be expressed as:
\begin{equation}
    P(x_t) \approx \mathcal{N}(x_t) = N^L(x_t) = A^L \circ \sigma_{L-1} \circ A^{L-1} \circ \ldots \circ \sigma_1 \circ A^1(x_t).
    \label{eq:NN_approx}
\end{equation}
The selection of hyper-parameters \(\{L, \{N_l\}_{l=1}^L, \{\sigma_l (\cdot)\}_{l=1}^L\}\) is known as the architecture design. Approaches to tuning these hyperparameters range from prior‐experience heuristics and manual, random, or grid search to more sophisticated techniques such as Bayesian optimization (see \citet{elsken2019neural} for a comprehensive review).

\vspace{-1.25em}
\paragraph{Training via Simulated Paths.} In the DEQN algorithm, an initial guess, by randomly initializing the FNN, is made for the unknown policy function. Following this, a sequence of \( N_{\text{path length}} \) states is simulated. Starting from a given state \( x_t \), the subsequent state \( x_{t+1} \) is determined by the policies encoded by the neural network, \( \mathcal{N}(x_t) \), and the remaining model-implied dynamics.

If the approximate policy function satisfying the equilibrium conditions were known, Equation~\eqref{eq:Policy_Function_DEQN} would hold along a simulated path. However, since the neural network is initialized with random coefficients, \( \mathbf{G}(x_t, \mathcal{N}(x_t)) \neq 0 \) along the simulated path of length \( N_{\text{path length}} \). This observation is used to refine the guessed policy function.

Specifically, DEQNs employ a loss function that quantifies the error in the equilibrium conditions, defined as:
\begin{equation}
\ell_v = \frac{1}{N_{\text{path length}}} \sum_{x_t \text{on sim. path}} \sum_{m=1}^{N_{\text{eq}}} \left( \mathbf{G}_m(x_t, \mathcal{N}(x_t)) \right)^2, \label{eq:LOSS_general}
\end{equation}
where \( \mathbf{G}_m(x_t, \mathcal{N}(x_t)) \) represents all the \( N_{\text{eq}} \) first-order conditions of a given model (as outlined in Section ~\ref{sec:model}), expressed as 
\begin{equation}
\mathbf{G}(x_t, \mathcal{N}(x_t)) = \sum_{m=1}^{N_{\text{eq}}} \mathbf{G}_m(x_t, \mathcal{N}(x_t)).
\end{equation}
Equation~\eqref{eq:LOSS_general} is utilized to update the weights of the network using any variant of (stochastic) gradient descent:
\begin{equation}
v_k' = v_k - \alpha \frac{\partial \ell (v)}{\partial v_k},
\label{sec:Adam}
\end{equation}
where \( v_k \) denotes the network weights, $v_k'$ are the updated weights, \( \alpha \) is the learning rate, and \( \frac{\partial \ell (v)}{\partial v_k} \) is the gradient of the loss function with respect to the weights.\footnote{In our numerical experiments, we employ the Adam optimizer~\citep{kingma2014adam}, an variant of stochastic gradient descent.} The updated neural network-based approximation of the policy functions is subsequently used to simulate the model again and evaluate the loss functions. This iterative procedure is pursued until $\ell_v < \epsilon \in \mathbb{R}$, that is, an approximate equilibrium policy, has been found. In our numerical experiments (see Section~\ref{sec:results}), we set $\epsilon = 10^{-6}$.

\vspace{-1.25em}
\paragraph{Four Key Ingredients.}
To summarize, the DEQN algorithm comprises four components:  
(i) a neural network approximator such as an FNN (cf. Equation~\eqref{eq:NN_approx}),  
(ii) the equilibrium-error loss (cf. Equation~\eqref{eq:LOSS_general}),  
(iii) a gradient-based update rule (cf. Equation~\ref{sec:Adam}), and (iv) simulated paths that supply training states.  
\vspace{-1.25em}
\paragraph{SOLG-specific Modifications to DEQN.} 

To adapt the baseline DEQN framework~\citep{azinovicDEEPEQUILIBRIUMNETS2022} for solving non-stationary SOLG IAMs across a continuum of tax and transfer schemes, we introduce several key modifications to the baseline DEQN algorithm. At the beginning of each simulation, we initialize the climate-economy state $x_{\texttt{state,t}}$ to a fixed starting point $x_{\texttt{state,0}}$, representing an initial condition such as the year 2015, and simulate forward over a horizon of 350 years. For each simulation, we sample a new set of pseudostates $\vartheta\in[a,b]\subset\mathbb{R}^{d_\vartheta}$ from the parameter space, which remain constant throughout this particular simulation (cf. Appendix~\ref{appendix:sampling} for more details). This approach enables the neural network to consistently learn policy functions $P(x_{\texttt{state,t}},\vartheta)\approx \mathcal{N}(x_{\texttt{state,t}},\vartheta)$, including household savings decisions and value functions. Implementation details, such as the explicit loss function \(\ell_v\) and the state-transition rules that map $x_t$ to $x_{t+1}$, are provided in Appendix~\ref{sec:DEQN_for_SOLG_IAM}.

\subsection{Welfare Surface Approximation via Gaussian Processes}
\label{sec:welfare_surrogate}

In this section, we explain how, in Step 2 of our method, we construct computationally efficient GP surrogates for approximating SWFs and, in Step 3, how we determine the tax-and-transfer parameters that maximize the planner’s objective (Equation~\eqref{eq:Planner_optimization}). Given a parametrized tax (and transfer) scheme fixed ex-ante, we consider two distinct QoIs: (i) a welfare-improving criterion and (ii) a Pareto-improving criterion.

\subsubsection{Social Welfare Objective and Pareto-optimal Parameters}
\label{sec:SWF_objective_sim}

When the goal is to maximize expected welfare, the planner solves
\begin{equation}
\label{eq:y_welfare}
\begin{aligned}
  \vartheta^{\ast}
  \;=\;
  & \arg\max_{\vartheta\in[a,b]}\,
  \widehat{\mathrm{QoI}}(\vartheta)
  \;=\;
  \arg\max_{\vartheta\in[a,b]}
  \mathbb{E}\!\bigl[W(\vartheta)\bigr], \\
& \quad \text{s.t. constraints,}
\end{aligned}
\end{equation}
where
\begin{equation}
\label{eq:surrogate_E}
  W(\vartheta)
  \;=\;
  \sum_{t=-10}^{29}
  \gamma_t\,\tilde{U}_t(\vartheta),
\end{equation}
$\gamma_t$ are again the welfare weights, and \(\tilde{U}_t(\vartheta)\) is a short-hand notation for the expected lifetime utility of generation \(t\) as of time~0 (see Equations~\eqref{eq:utility} and~\eqref{eq:planner_objective_general}), at a particular realization of the pseudo-state variables $\vartheta$.\footnote{We assume that the planner’s objective is to maximize the lifetime welfare of all cohorts that are already alive at $t=0$ as well as those that will be born over the next 150 years; formally, cohorts with birth dates $t\in[-10,29]$, where $t=-10$ denotes the oldest surviving cohort.}

Recall that the goal of our method's second step is to find the parameters $\vartheta$ that maximize the expected SWF (see, for instance, Equation~\eqref{eq:y_welfare}). A direct approach would involve using a black-box optimizer to evaluate the SWF at numerous candidate parameter vectors, $\{\vartheta_i\}_{i=1}^{n}$, until a desired numerical accuracy is reached. However, this is computationally prohibitive. Each evaluation of the SWF for a single $\vartheta_i$ is expensive, as it requires an estimate derived from 10,000 independent forward simulations over an approximately 150-year horizon. Repeating this costly process for the potentially thousands, if not tens of thousands, of queries required by a constrained optimization routine such as IPOPT~\citep{wachter2006implementation} would take hours of computation time.

To overcome this computational barrier, we ``front-load'' the intensive calculations by employing a surrogate modeling approach again. 
We build a GP surrogate (cf. Section~\ref{sec:GPR_intro} for an introduction) that approximates the objective function, using only a small number of true SWF evaluations as training points.\footnote{GPs offer three main advantages in our setting. First, they are highly data-efficient: by encoding smoothness and other prior beliefs through the kernel (for example, a Matérn kernel), a GP can learn nonlinear response surfaces from very few training points. In our own experiments, testing multiple deep neural network architectures, the GP achieved the desired out-of-sample accuracy with approximately ten times fewer model evaluations than the best deep neural net, which still produced slightly less accurate predictions (cf. Appendix~\ref{sec:HP_selection_GPs}). Second, GPs provide closed-form posterior variances, yielding well-calibrated uncertainty measures that are invaluable for acquisition rules in Bayesian active learning, enabling the efficient population of the GP's training set. Third, the GP's training, while involving a non-convex optimization to tune kernel hyperparameters, is often more stable and requires tuning far fewer parameters than the stochastic-gradient-based training of a deep neural network. Although deep networks can also be used as surrogates (see, e.g., \citealp{chen2025Deep} and~\citealp{kase2022estimating} for economic applications), the GP surrogate is markedly more effective in the high-cost, small-data regime of our welfare evaluations.}
The predictive mean of this surrogate, $\mu_*(\vartheta)$, then serves as a computationally cheap proxy for the true objective function:
\begin{equation}
    \mu_*(\vartheta) \approx \mathbb{E}[W(\vartheta)].
\end{equation}
This GP surrogate is then provided to the optimizer. The key advantage lies in the dramatic reduction of computational cost: evaluating the surrogate at a particular $\vartheta_i$ has a negligible interpolation cost of $\mathcal{O}(n)$~\citep{Rasmussen:2005:GPM:1162254}, where $n$ is the number of training points the GP was fitted to (which in our applications is typically of order 450-500 points (cf. Section~\ref{sec:results})). 
This is orders of magnitude cheaper than the $\mathcal{O}(\text{paths} \times \text{periods})$ cost of a full Monte Carlo evaluation, which, in our concrete applications, is $10,000\times 29$ per function evaluation.
This efficiency gain of about three orders of magnitude enables the rapid optimization in Step 3 (Section~\ref{sec:welfare_surrogate}) to recover welfare-improving parameters $\vartheta^*$ without further costly simulations.

The need for a surrogate becomes even more pronounced when the planner seeks Pareto improvements. In this case, the optimization problem reads as:
\begin{equation}
\label{eq:opt_pareto_y}
  \vartheta^{\ast}
  \;=\;
  \arg\max_{\vartheta\in[a,b]}
  \sum_{t=-10}^{29}\gamma_t\,\mathbb{E}_0\!\bigl[\tilde{U}_t(\vartheta)\bigr]
  \quad\text{s.t.}\quad
  \mathbb{E}_0\!\bigl[\tilde{U}_t(\vartheta)\bigr]
  \;\ge\;
  \mathbb{E}_0\!\bigl[U_t\bigr]
  \;\;\forall\,t,
\end{equation}
where $\mathbb{E}_0\!\bigl[U_t\bigr]$ denotes the lifetime utility of generation $t$ under the BAU scenario, which we compute by solving and simulating a separate version of the model without taxes and transfers. Evaluating the $40$ cohort-specific constraints by Monte-Carlo simulation at every candidate $\vartheta$ would be computationally prohibitive.
To render the problem tractable, we construct a vector-valued
surrogate (based on GPs) that approximates the mapping
\begin{equation}
\label{eq:pareto_map}
      \vartheta
  \;\longmapsto\;
  \bigl(
    \tilde{U}_{-10}(\vartheta),
    \ldots,
    \tilde{U}_{29}(\vartheta)
  \bigr)^{\top},
\end{equation}
that is, one GP per generation.  
These surrogates deliver instantaneous predictions of both the objective and the constraints, allowing the
constrained optimization in Equation~\eqref{eq:opt_pareto_y} to be solved at negligible marginal cost.
We next outline how GPR is used to approximate the various SWFs.

\subsubsection{Gaussian Process Regression.}
\label{sec:GPR_intro}

GPR models the target function as a realization of a GP with a mean function $m(\vartheta)$ and a covariance kernel $k(\vartheta,\vartheta')$. Given a training set
\begin{equation}    
\mathcal{D}=\bigl\{(\vartheta_i,y_i)\bigr\}_{i=1}^{n}=[\Theta,y],\quad \theta_i\in[a,b]\subset\mathbb{R}^{d_\vartheta},\; y_i\in\mathbb{R},
\end{equation}
we aim to learn a mapping $f(\vartheta)$ such that $y_i = f(\vartheta_i) + \epsilon_i$, where $\epsilon_i \sim \mathcal{N}(0,\sigma_\epsilon^{2})$ represents independent noise for numerical stability \citep{Rasmussen:2005:GPM:1162254}. Here, $\vartheta_i$ are the pseudo-states that parametrize tax (and transfer) schemes, and $y_i=\widehat{\mathrm{QoI}}(\vartheta_i)$ are the welfare outcomes computed via simulations using the neural network surrogate $\mathcal{N}(\cdot)$ for $P(\cdot)$.

In our numerical examples below, we set $m(\vartheta) \equiv 0$ and use the Matérn 5/2 kernel to ensure moderate smoothness and finite differentiability:
\begin{align}
\label{eq:def_matern52}
k(\theta,\theta') = \sigma_f^{2}\left(1+\frac{\sqrt{5}\,r}{\rho}+\frac{5r^{2}}{3\rho^{2}}\right) \exp\!\Bigl(-\frac{\sqrt{5}\,r}{\rho}\Bigr),
\end{align}
where $r = \lVert \theta - \theta' \rVert$, $\sigma_f^{2}$ is the signal variance, and $\rho$ is the length-scale.\footnote{For details on hyperparameter selection, see Appendix~\ref{sec:Appendix_Soc_welfare}.} Kernel hyperparameters are optimized by maximizing the marginal likelihood \citep[Ch.~18.2]{pml2Book}.\footnote{We maximize the marginal log-likelihood to tune the GP's kernel hyperparameters. Since this is a non-convex optimization problem, we use the Adam optimizer~\citep{kingma2014adam} with a multi-start strategy to mitigate the risk of converging to a poor local optimum; 
a common practice in the literature~\citep{pml2Book}.
Specifically, we perform the optimization from 10 randomly chosen starting points. For our applications, we found the resulting hyperparameters to be robust to this procedure, showing little sensitivity to the initial starting point.}
The posterior distribution at a test input $\theta_{\ast}$ is Gaussian:
\begin{align}
\label{eq:def_Bayes_GPR}
p\!\bigl(f_{\ast}\mid \theta_{\ast},\mathcal{D}\bigr) = \mathcal{N}\!\bigl(\mu_{\ast},\sigma_{\ast}^{2}\bigr),
\end{align}
with predictive mean and variance:
\begin{align}
\label{eq:predictive_mean}
\mu_{\ast}(\vartheta_{\ast}) &= k(\vartheta_{\ast},\varTheta) \bigl[K(\varTheta,\varTheta) + \sigma_{\epsilon}^{2}I\bigr]^{-1}y,\\
\label{eq:predictive_variance}
\sigma_{\ast}^{2}(\vartheta_{\ast}) &= k(\vartheta_{\ast},\vartheta_{\ast}) - k(\vartheta_{\ast},\varTheta)\bigl[K(\varTheta,\varTheta) + \sigma_{\epsilon}^{2}I\bigr]^{-1}k(\varTheta,\vartheta_{\ast}),
\end{align}
where $K(\varTheta,\varTheta')$ is the covariance matrix over training inputs. The predictive mean $\mu_{\ast}(\theta)$ serves as the surrogate $\widehat{\mathrm{QoI}}(\vartheta)$, scaling linearly with $n$ \citep{Rasmussen:2005:GPM:1162254}.

\vspace{-1.25em}
\paragraph{Generating Training Data for the GP.}
Constructing an accurate GP surrogate requires a training set that covers the admissible parameter space $[a,b]\subset\mathbb{R}^{d_\vartheta}$ using as few expensive simulations as possible. In other words, we seek the smallest feasible sample size $n$.  
We first generate an initial design of about $n=450$ points with the same uniform sampling scheme employed for DEQN training (see Appendix~\ref{appendix:sampling}).  
This training is then refined by adding roughly $50$ information-rich points selected through Bayesian active learning (see \citealp{rennerscheidegger_2018} and Appendix~\ref{sec:BAL_text}).

In our numerical experiments (cf.\ Section~\ref{sec:results}), generating a single observation  
$y_i=\widehat{\mathrm{QoI}}(\vartheta_i)$ on one CPU takes about \(2\;\text{seconds}\); refitting the GP to the current training set adds a further $20-30$ seconds.  
With a final training set of roughly \(500\) points, the complete surrogate–construction loop therefore finishes in $30-60$ minutes, the upper bound arising when several GP surrogates must be fitted (for example, when we search for Pareto-improving policy vectors).  
Because each \(y_i\) can be produced independently, these wall-clock times fall by one to two orders of magnitude on a modest multi-core or cluster setup.

\subsubsection{Surrogate-based Optimization of Welfare Parameters}
\label{sec:Max}

With the computationally intensive construction of the SWF surrogate, $\widehat{\mathrm{QoI}}(\vartheta)$, completed, the final step is to solve the constrained optimization problem~\eqref{eq:Planner_optimization}.
This problem is generally non-convex, and thus standard optimization algorithms are not guaranteed to converge to the global optimum. A key advantage of our approach is that, once the GP surrogate model is fitted, any suitable optimization algorithm can be applied to solve Equation~\eqref{eq:Planner_optimization}. This optimization is computationally negligible, typically finishing in under a second. For the specific problem addressed in this paper, the objective function appears to be concave. We confirm this property computationally for the simplest two-dimensional case. For higher-dimensional cases, we launch the optimizer from multiple starting points (see Appendix \ref{sec:HP_selection_GPs} for more details); all runs converge to the same solution, indicating that the results are robust.

\section{Results}
\label{sec:results}

In this section, we illustrate how our novel three-step machine learning methodology can systematically derive welfare- and Pareto-improving policies within a SOLG IAM. In Section~\ref{sec:solion_BAU}, we report results for the 12-dimensional SOLG IAM under the BAU scenario, which serves as a baseline for comparison with models incorporating climate policies and transfers. Furthermore, the BAU results provide the necessary benchmark for systematically deriving Pareto-improving tax and transfer schemes, as formulated in Equation~\eqref{eq:opt_pareto_y}. In Section~\ref{sec:cum_emissions}, we introduce a simple linear tax on cumulative emissions, accompanied by an exogenously prescribed transfer scheme, which adds two parameters to the state vector for the graphical illustration of our method. In Section~\ref{sec:cum_emissions_optimal_transfer}, we extend this to include optimal transfers as additional state variables, incorporating 12 transfer parameters alongside the two tax parameters, yielding a state space of \( d_x = 12 + 5 = 17 \) and parameter space of \( d_\vartheta = 14 \), for a total of 31 dimensions. In Section~\ref{sec:emissions_carbon_intensity_damages_optimal_transfer} we examine a richer tax specification that levies a linear tax on cumulative emissions, carbon intensity, and climate-tipping damages, complemented by
optimal transfers. This setting introduces $d_{\vartheta}=16$ pseudo-state variables.

\subsection{Solving the Business-As-Usual Model}
\label{sec:solion_BAU}

We compute global solutions to the SOLG IAM under a BAU scenario (see Appendix~\ref{sec:DEQN_for_SOLG_IAM} for implementation details). The solution algorithm terminates when the loss function falls below a preset threshold, $\ell_{\nu}^{BAU} < \epsilon = 10^{-6}$. We then evaluate the model's out-of-sample accuracy using two standard metrics: the relative Euler equation errors~\eqref{eq:EE_error_BAU_ERROR} and the relative errors in the value functions~\eqref{eq:value_error_BAU_ERROR}.

These errors are evaluated over 10,000 individual sample paths, each with a simulation length of 150 years. As reported in Table~\ref{tab:EE_errors_bau}, the mean out-of-sample relative errors of the first-order equations (defined as the 99.9th percentile) are in the order of $10^{-3}$ or smaller. This level of accuracy is highly competitive with other global solution methods for stochastic OLG models (see, e.g., \citealp{krueger2006pareto} and~\citealp{BKS1}).
\begin{table}[htbp]
\centering
\begin{adjustbox}{width=\textwidth}
\begin{tabular}{@{}lccccccccccc@{}}
\toprule
& Gen 1 & Gen 2 & Gen 3 & Gen 4 & Gen 5 & Gen 6 & Gen 7 & Gen 8 & Gen 9 & Gen 10 & Gen 11 \\
\midrule
Rel EE Mean          & 0.0003 & 0.0010 & 0.0005 & 0.0004 & 0.0004 & 0.0002 & 0.0013 & 0.0015 & 0.0003 & 0.0004 & 0.0008 \\
Rel EE 99.9     & 0.0022 & 0.0019 & 0.0024 & 0.0017 & 0.0018 & 0.0013 & 0.0019 & 0.0029 & 0.0016 & 0.0019 & 0.0025 \\
Rel Value Function Mean       & 0.0003 & 0.0002 & 0.0003 & 0.0002 & 0.0002 & 0.0002 & 0.0004 & 0.0006 & 0.0002 & 0.0004 & 0.0007 \\
Rel Value Function 99.9  & 0.0025 & 0.0027 & 0.0023 & 0.0025 & 0.0020 & 0.0021 & 0.0022 & 0.0019 & 0.0015 & 0.0020 & 0.0025 \\
\bottomrule
\end{tabular}
\end{adjustbox}
\caption{
    \textbf{Accuracy Metrics for the DEQN Model Solution under the BAU Scenario.}
    The table presents the mean and maximum (defined as the 99.9th percentile) for two key error metrics: the Relative Euler Equation Error (Rel EE) and the Relative Value Function Error across the various overlapping generations (Gen $i$). Statistics are computed over 10,000 simulated sample paths. 
}
\label{tab:EE_errors_bau}
\end{table}

\begin{figure}[th!]
    \centering
    \begin{subfigure}{0.45\textwidth}
        \includegraphics[width=0.95\textwidth]{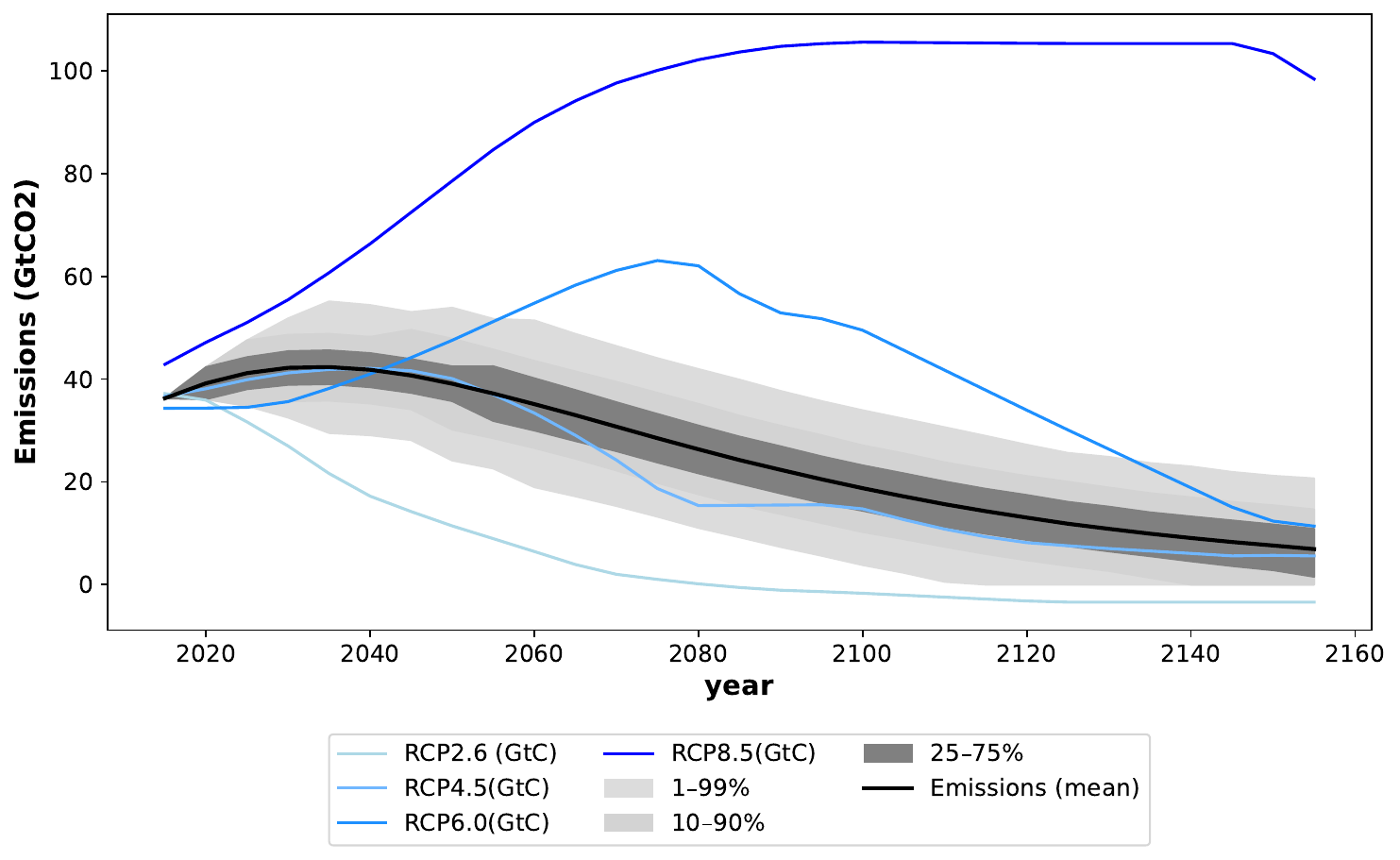} 
        \caption{Emissions}
    \end{subfigure}
    \hfill
    \begin{subfigure}{0.45\textwidth}
        \includegraphics[width=0.95\textwidth]{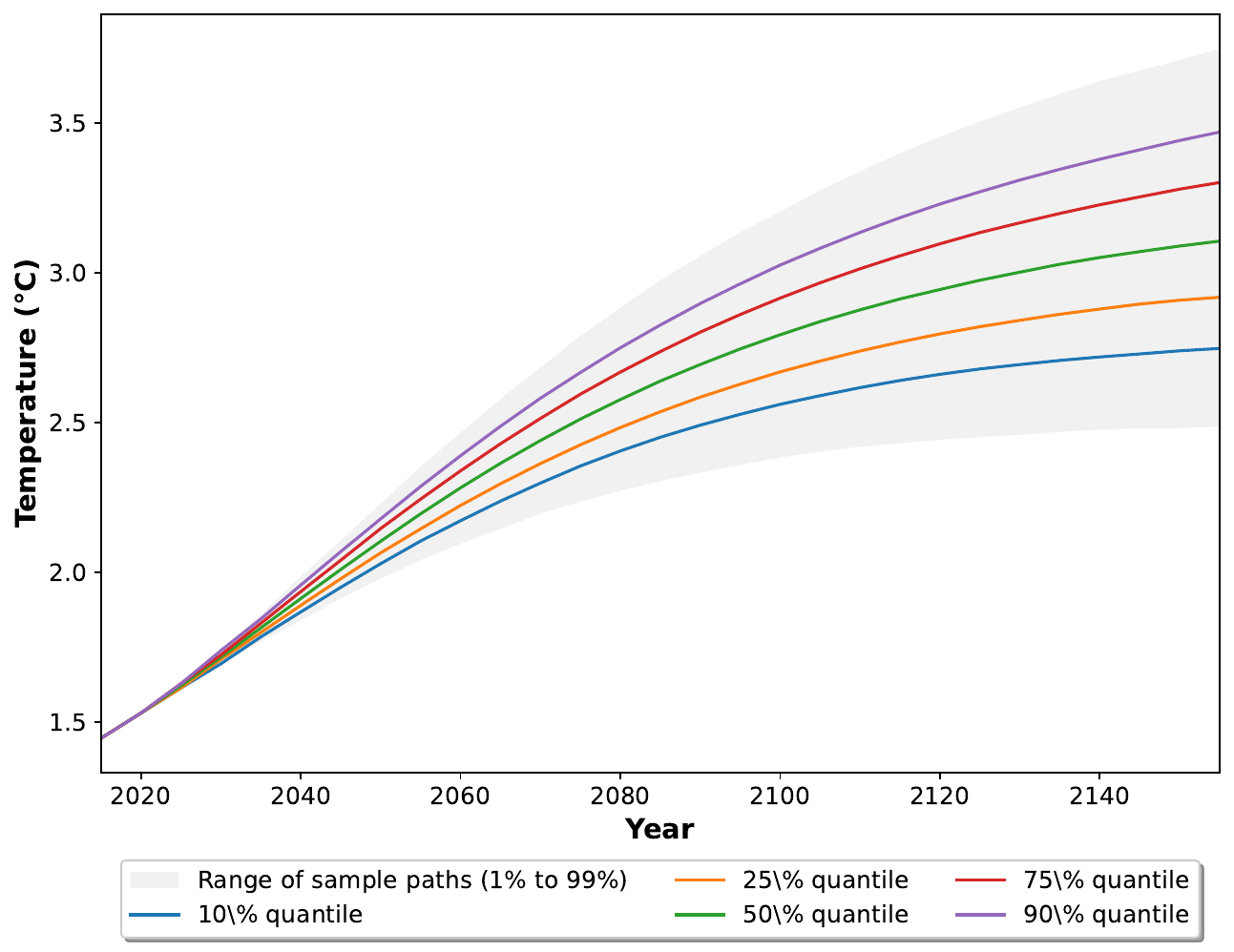} 
        \caption{Temperature}
    \end{subfigure}

    \begin{subfigure}{0.45\textwidth}
        \includegraphics[width=0.95\textwidth]{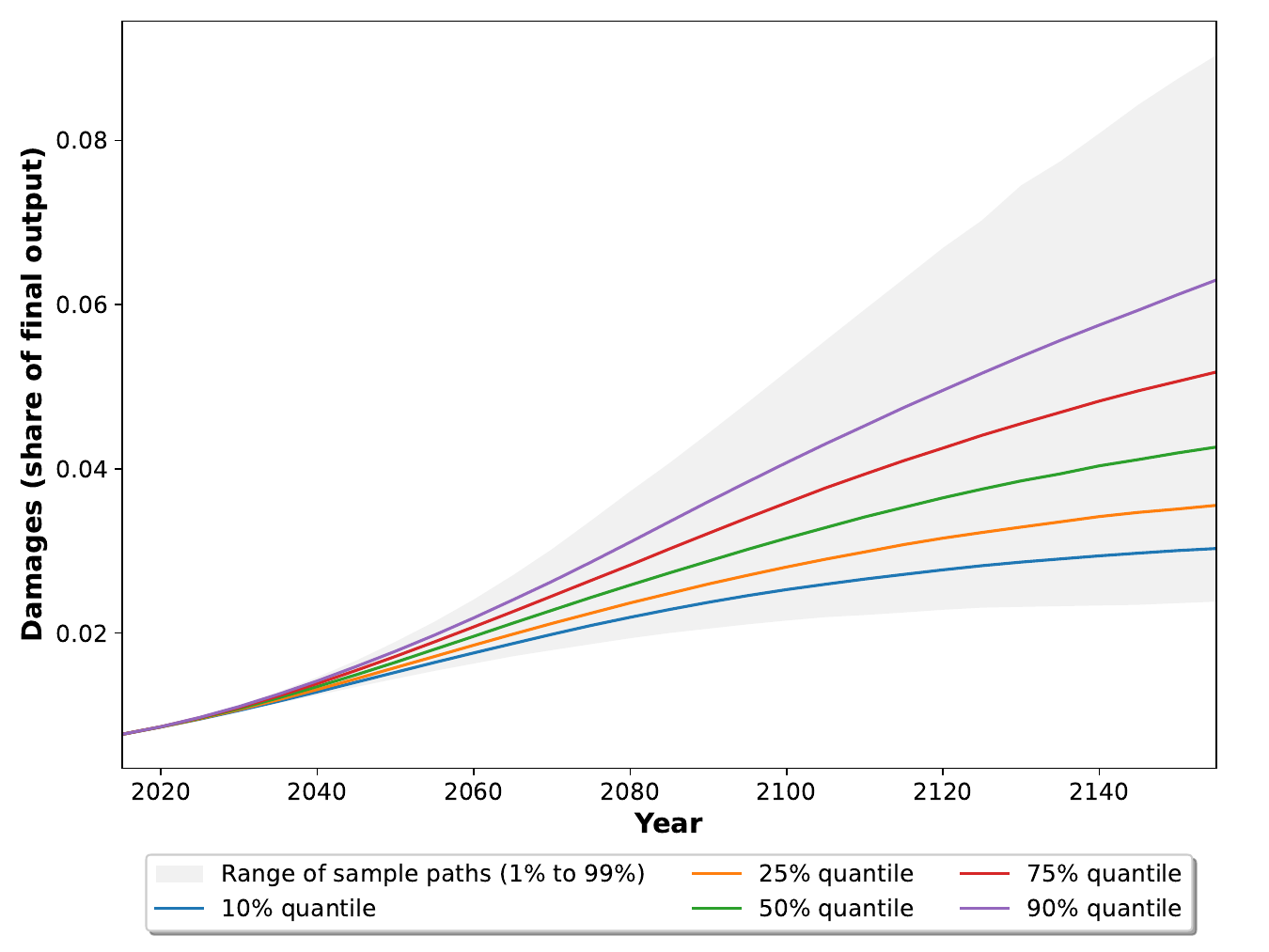} 
        \caption{Damages}
    \end{subfigure}
    \hfill
    \begin{subfigure}{0.45\textwidth}
        \includegraphics[width=0.95\textwidth]{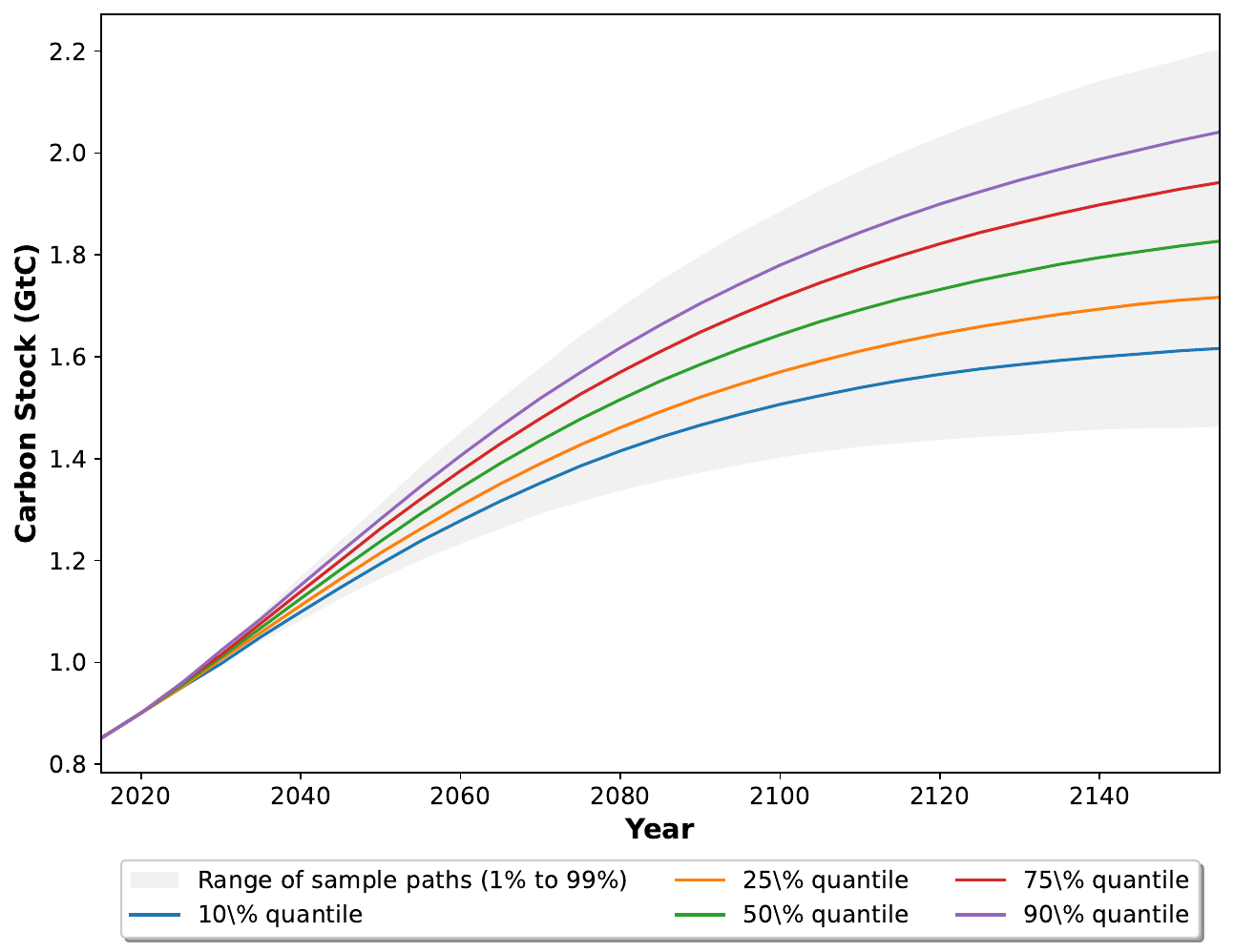} 
        \caption{Stock of Carbon}
    \end{subfigure}
    \caption{\textbf{Projected Climate and Economic Outcomes under the BAU Scenario.}
    The figure displays key projections from the SOLG IAM over a 150-year horizon.
    \textit{Top left:} Endogenous emissions (mean and full distribution), with the mean trajectory closely following the RCP4.5 scenario.
    \textit{Top right:} Global warming, projected to reach a mean of approximately 3°C, with a substantial uncertainty range of 2.5°C to 3.6°C.
    \textit{Bottom left:} Economic damages as a percentage of GDP, averaging 4\% but with worst-case outcomes exceeding 15\%.
    \textit{Bottom right:} The corresponding stock of carbon in the atmosphere.}
    \label{fig:results_BAU}
\end{figure}

We now turn to the model's solution. Figure~\ref{fig:results_BAU} presents the main results under the BAU scenario.
%
%
The top-left panel shows that mean endogenous emissions closely follow the RCP4.5 trajectory, an intermediate socioeconomic scenario from the Intergovernmental Panel on Climate Change (IPCC) for climate change.\footnote{The Representative Concentration Pathways are standardized scenarios used in climate modeling to project future greenhouse gas concentrations and radiative forcing, ranging from RCP2.6 (stringent mitigation) to RCP8.5 (high emissions). See \url{https://tntcat.iiasa.ac.at/RcpDb} for the related data sets.} The top-right panel illustrates the resulting evolution of global temperature. While the mean Global warming is projected to reach approximately 3°C in 150 years, the associated uncertainty is substantial, with the distribution spanning from 2.5°C to over 3.6°C. The economic consequences are depicted in the bottom-left panel, where damages are projected to reach a mean of 4\% of GDP over the same period. In the worst-case scenarios, however, these damages could exceed 15\% of GDP. Finally, the bottom-right panel shows the corresponding stock of atmospheric carbon.
As the BAU scenario involves no policy optimization, there is no SWF to maximize. Consequently, steps 2 and 3 of our methodology are not applicable in this case.

\subsection{Welfare-improving Linear Taxes on Cumulative Emissions}
\label{sec:cum_emissions}

Following the no-policy (BAU) baseline analysis in the previous section, we now illustrate the complete 3-step methodology from Section~\ref{sec:methodology} with a simple policy instrument. We assume the planner can implement a carbon tax that is a linear function of cumulative emissions, $E_t$:
\begin{equation}
    \tau_t(E_t) = \vartheta_0 + \vartheta_E E_t.
    \label{eq:tax_fun_lin_S}
\end{equation}
This tax function introduces two parameters, $\vartheta_0$ and $\vartheta_E$, which are treated as pseudo-states and added to the model's state space. Their permissible ranges $[a,b]\in\mathbb{R}^2$ are specified in Table~\ref{tab:param_bounds_main} in Appendix~\ref{sec:domain_taxes} for the model designated as ``Linear in E''.\footnote{Appendix~\ref{appendix:Tax_params_sampling} provides further details on the derivation of these parameter bounds.}
In addition, for illustrative purposes, we assume a specific transfer scheme: the youngest generation receives the largest share of total tax revenue, and the share for each older generation decreases by 10\% relative to the next younger one. The shares are normalized to sum to one. Formally, our transfer scheme, $\mathbb{T}$, is defined as:
\begin{equation}
    \mathbb{T}_{t,j} = 0.9^{(j-1)} \cdot 0.1394 \cdot \tau_t e_t, \quad j \in \{2,...,12 \}.
\end{equation}
Running the DEQN algorithm (step 1 of our methodology) with the loss function from Equation~\eqref{eq:loss_optimal_DEQN} yields highly accurate global solutions, with mean out-of-sample errors for the relative Euler equations~\eqref{eq:EE_error_optimal_ERROR} and value functions~\eqref{eq:value_error_optimal_ERROR} in the order of $\mathcal{O}(10^{-4})$ and a maximum error in the order of $\mathcal{O}(10^{-3})$, as detailed in Table~\ref{tab:EE_errors_lin_S_only_maxwelf}.
\begin{table}[ht!]
\centering
\begin{adjustbox}{width=\textwidth}
\begin{tabular}{@{}lccccccccccc@{}}
\toprule
& Gen 1 & Gen 2 & Gen 3 & Gen 4 & Gen 5 & Gen 6 & Gen 7 & Gen 8 & Gen 9 & Gen 10 & Gen 11 \\
\midrule
Rel EE Mean          & 0.0007 & 0.0004 & 0.0003 & 0.0005 & 0.0005 & 0.0005 & 0.0005 & 0.0012 & 0.0013 & 0.0006 & 0.0006 \\
Rel EE 99.9     & 0.0031 & 0.0021 & 0.0013 & 0.0027 & 0.0017 & 0.0019 & 0.0018 & 0.0029 & 0.0023 & 0.0019 & 0.0017 \\
Rel Value Function Mean       & 0.0004 & 0.0004 & 0.0003 & 0.0003 & 0.0003 & 0.0003 & 0.0004 & 0.0004 & 0.0006 & 0.0006 & 0.0008 \\
Rel Value Function 99.9  & 0.0024 & 0.0023 & 0.0021 & 0.0023 & 0.0026 & 0.0025 & 0.0023 & 0.0019 & 0.0019 & 0.0024 & 0.0023 \\
\bottomrule
\end{tabular}
\end{adjustbox}
\caption{
    \textbf{Accuracy Metrics for the DEQN Model Solution under the welfare improving scenario with a linear tax on cumulative emissions.}
    The table presents the mean and maximum (defined as the 99.9th percentile) for two error metrics: the Relative Euler Equation Error (Rel EE) and the Relative Value Function Error across the various overlapping generations (Gen $i$). Statistics are computed over 10,000 simulated sample paths of the model's equilibrium conditions.
}
\label{tab:EE_errors_lin_S_only_maxwelf}
\end{table}
%

In step 2 of our method, the neural-network surrogate for the SOLG policies, $\mathcal{N}(\cdot)$, is used to compute the GP surrogate for the SWF. Following the general form of Equation~\eqref{eq:surrogate_E}, we build a surrogate that maps the tax parameters, $\vartheta = \{\vartheta_0,\vartheta_E \}$, to the SWF. Specifically, we approximate our quantity of interest, $\widehat{\mathrm{QoI}}(\vartheta) = \mathbb{E}[W(\vartheta)]$, by constructing a GP surrogate for which we assume uniform welfare weights ($\gamma_t = 0.025 \; \forall t$). The resulting predictive mean of the GP, $\mu_*(\vartheta)$, serves as our fast-to-evaluate surrogate for the SWF (cf. Section~\ref{sec:GPR_intro}):
\begin{equation}
   \mu_*(\vartheta) \;=\; \widehat{\mathrm{QoI}}(\vartheta).
\end{equation}
To assess its accuracy, we compute the leave-one-out cross-validation error, which is an excellent $3.2 \cdot 10^{-5}$.
\begin{figure}[th!]
    \centering
    \includegraphics[width=0.5\linewidth]{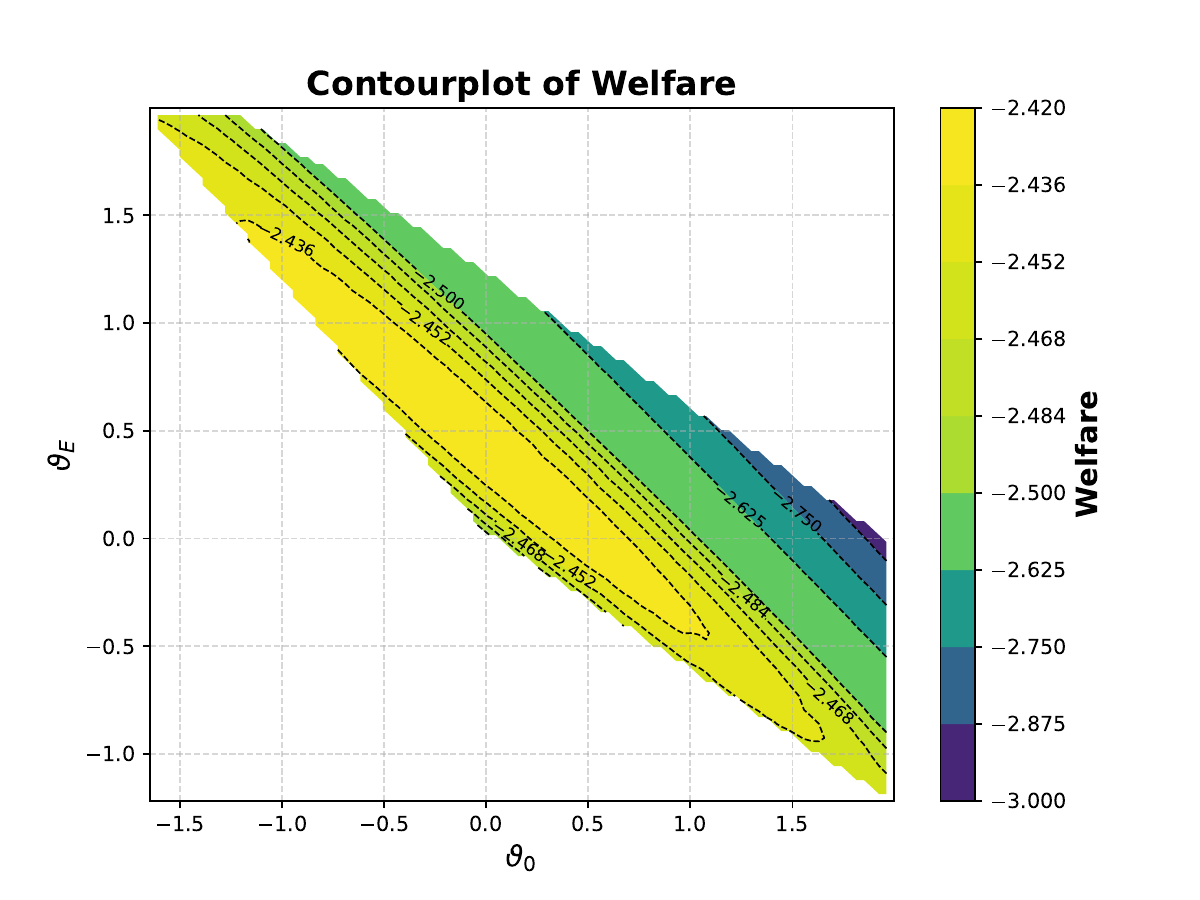}
    \caption{Contour plot of the welfare surrogate.}
    \label{fig:welfare_surrogate_S_main}
\end{figure}

Given this accurate surrogate, the third step of our method is to find the tax parameters by maximizing it, and thus, the SWF. To ensure the optimizer does not search outside the surrogate (that is, does not extrapolate), we enforce a set of linear constraints on the tax rate. The resulting constrained optimization problem reads as:
\begin{equation}
\label{eq:optim_cum_emm_maxwelf}
\begin{aligned}
  \vartheta^{\ast}
  \;=\;
  & \arg\max_{\vartheta\in[a,b]}\,
  \widehat{\mathrm{QoI}}(\vartheta) \\
& \quad \text{s.t.} \\
& 0 \leq \vartheta_0 + \vartheta_E  E_0 \leq \bar{\tau}, \\
& 0 \leq \vartheta_0 + \vartheta_E  E_{29} \leq \bar{\tau}.
\end{aligned}
\end{equation}
The constraints limit the tax rate in the first and last periods, using cumulative emissions values of $E_0 = 0.851$ and $E_{29} = 1.6$, and a maximum tax rate of $\bar{\tau} = 1.5$ (see Appendix~\ref{appendix:Tax_params_sampling} for details).
The resulting surrogate landscape is visualized in Figure~\ref{fig:welfare_surrogate_S_main}. 

\begin{figure}[h!]
    \centering
    \begin{subfigure}{0.45\textwidth}
        \includegraphics[width=0.95\textwidth]{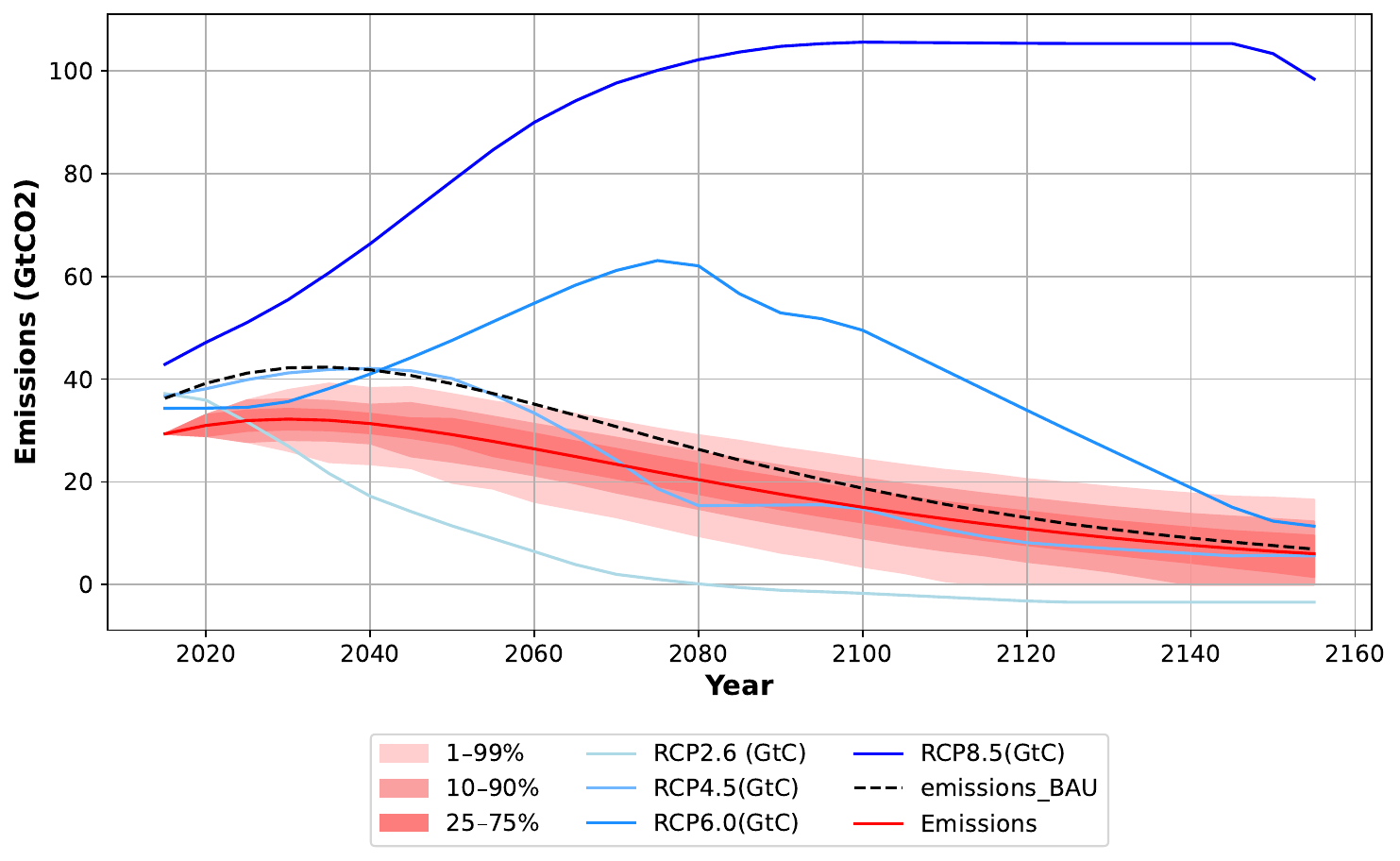} 
        \caption{Emissions}
    \end{subfigure}
        \hfill
    \begin{subfigure}{0.45\textwidth}
        \includegraphics[width=0.95\textwidth]{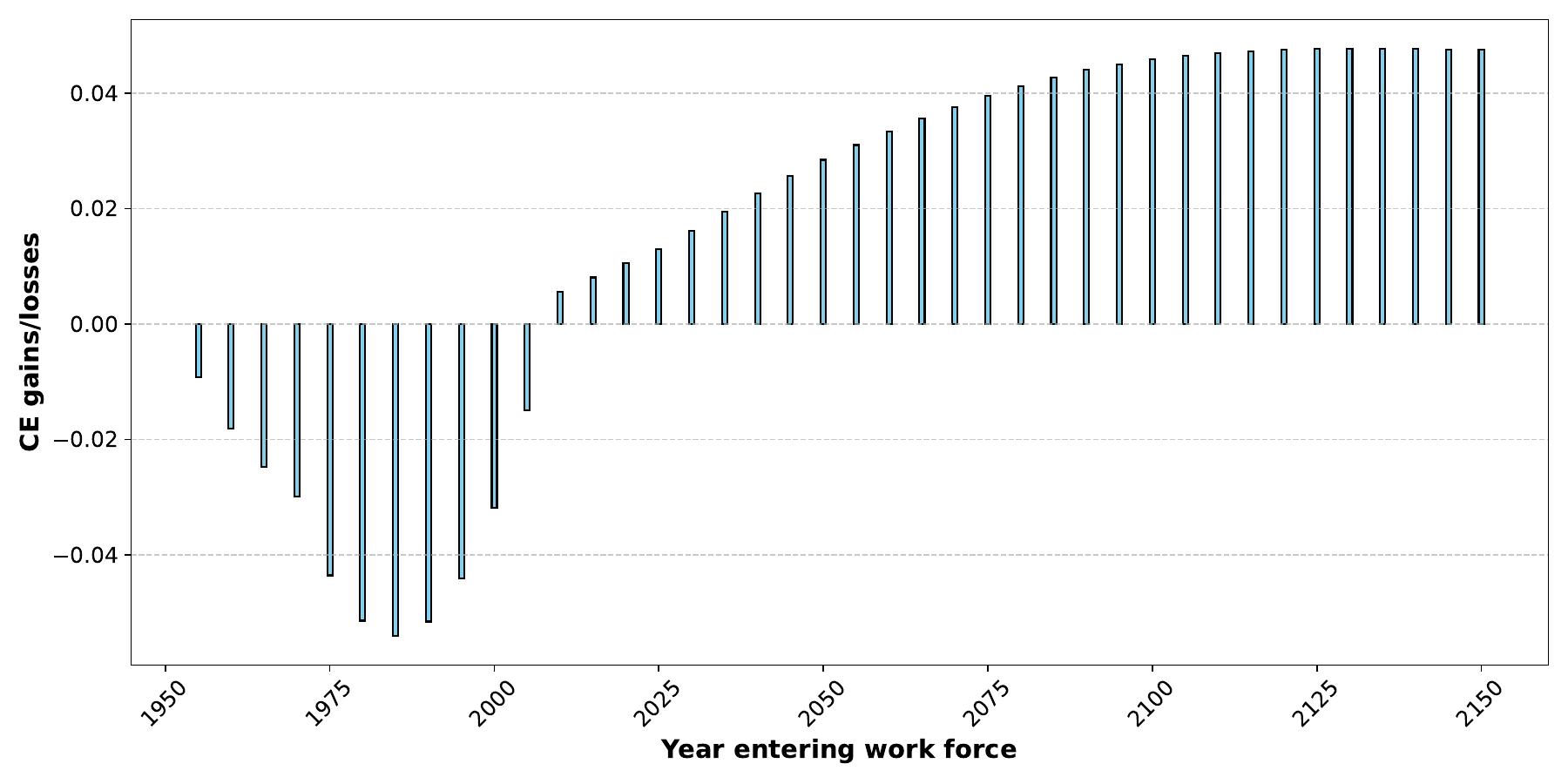} 
        \caption{Welfare Gains}
        \end{subfigure}
    \begin{subfigure}{0.45\textwidth}
        \includegraphics[width=0.95\textwidth]{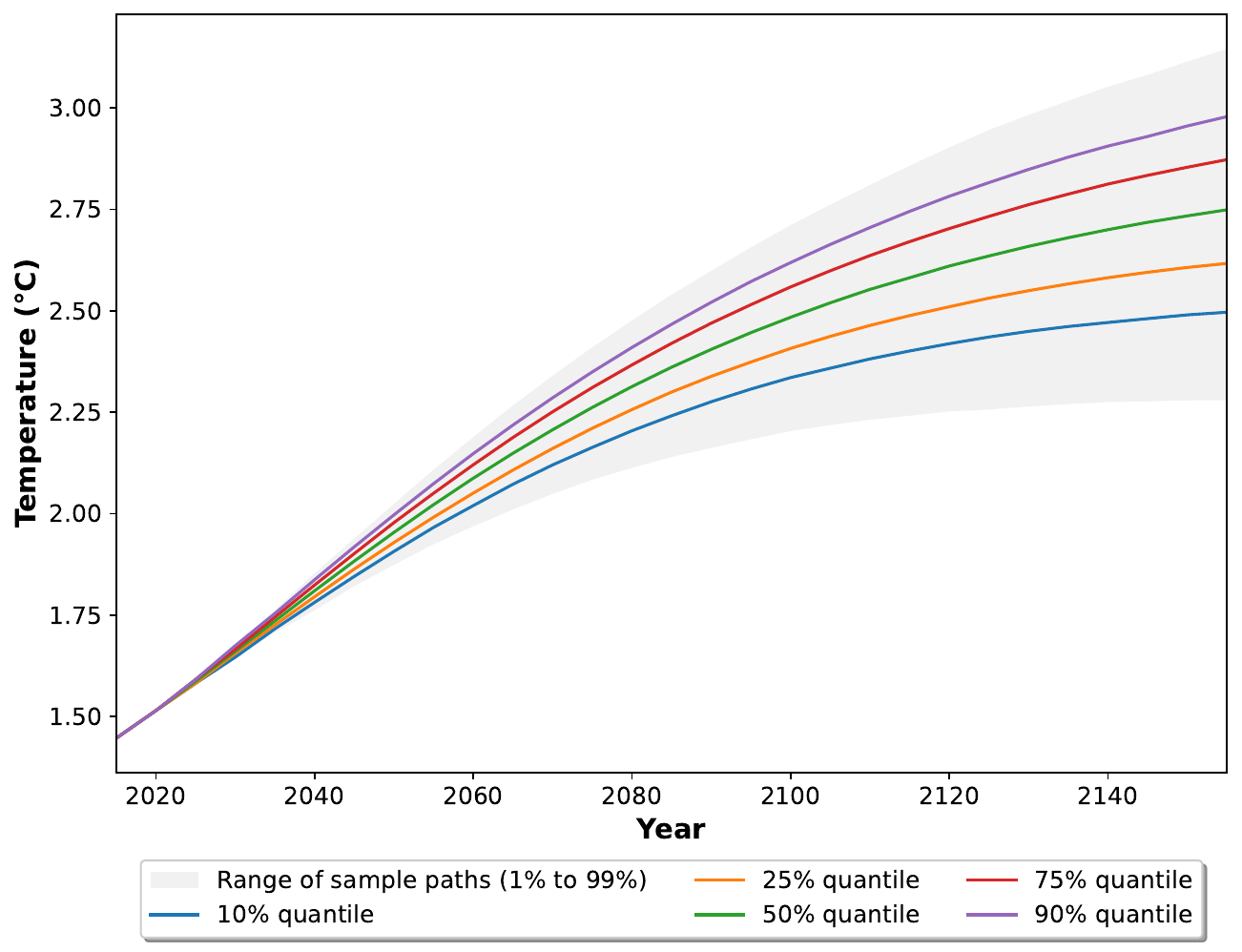} 
        \caption{Temperature}  
    \end{subfigure}
    \hfill
    \begin{subfigure}{0.45\textwidth}
        \includegraphics[width=0.95\textwidth]{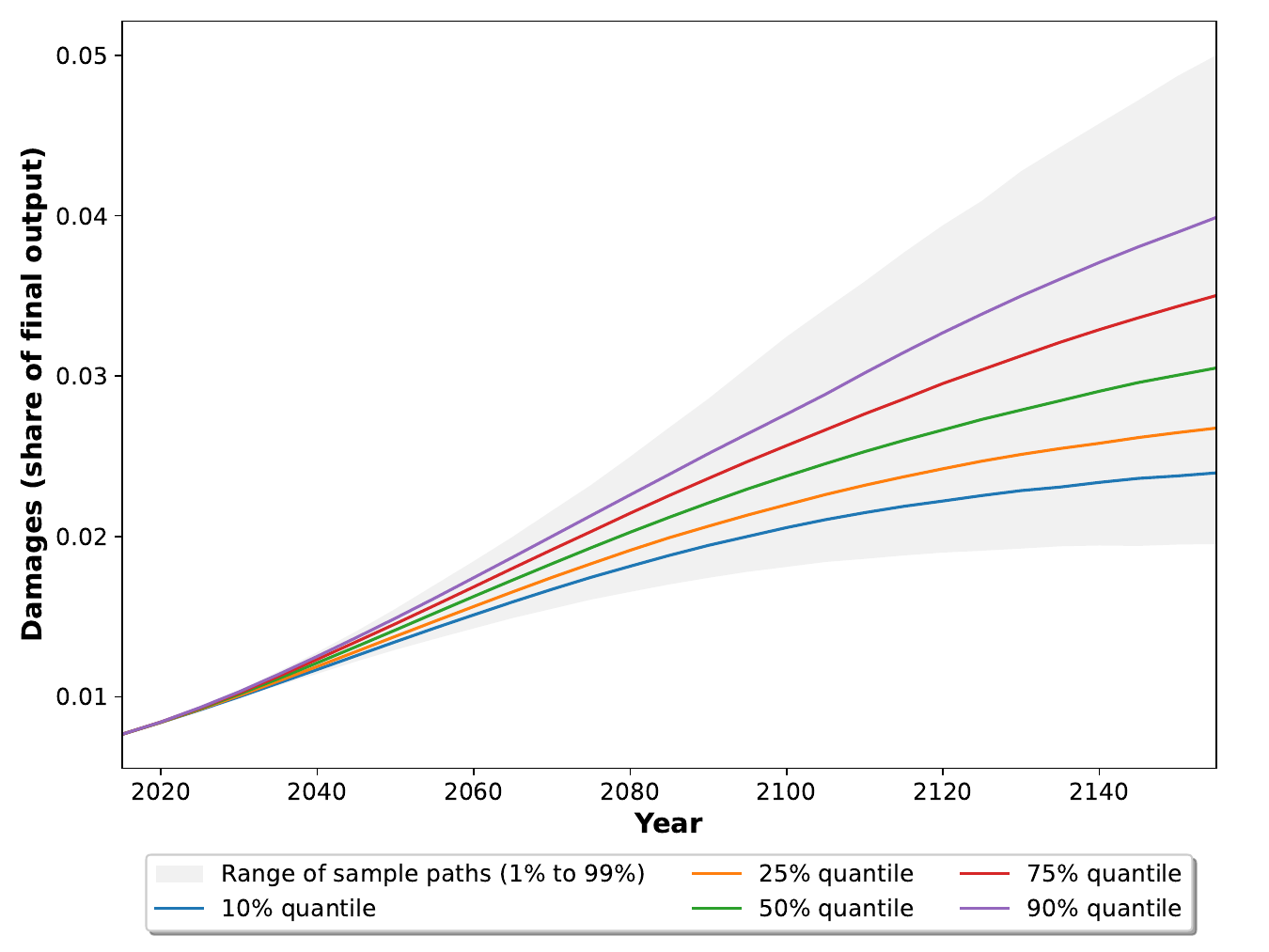} 
        \caption{Damages}  
    \end{subfigure}
     \begin{subfigure}{0.45\textwidth}
        \includegraphics[width=0.95\textwidth]{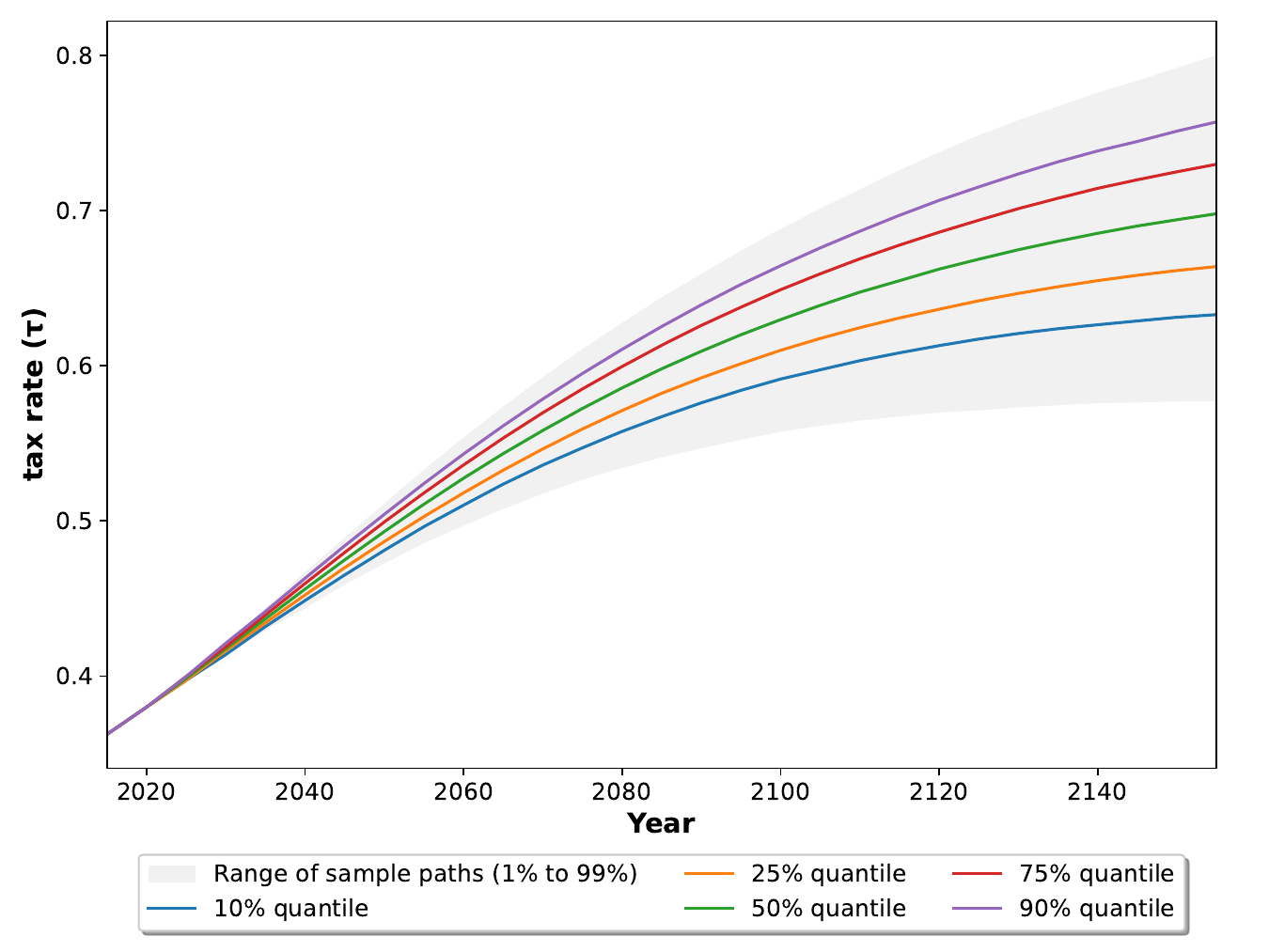} 
        \caption{Carbon taxes}
    \end{subfigure}
    \hfill
    \begin{subfigure}{0.45\textwidth}
        \includegraphics[width=0.95\textwidth]{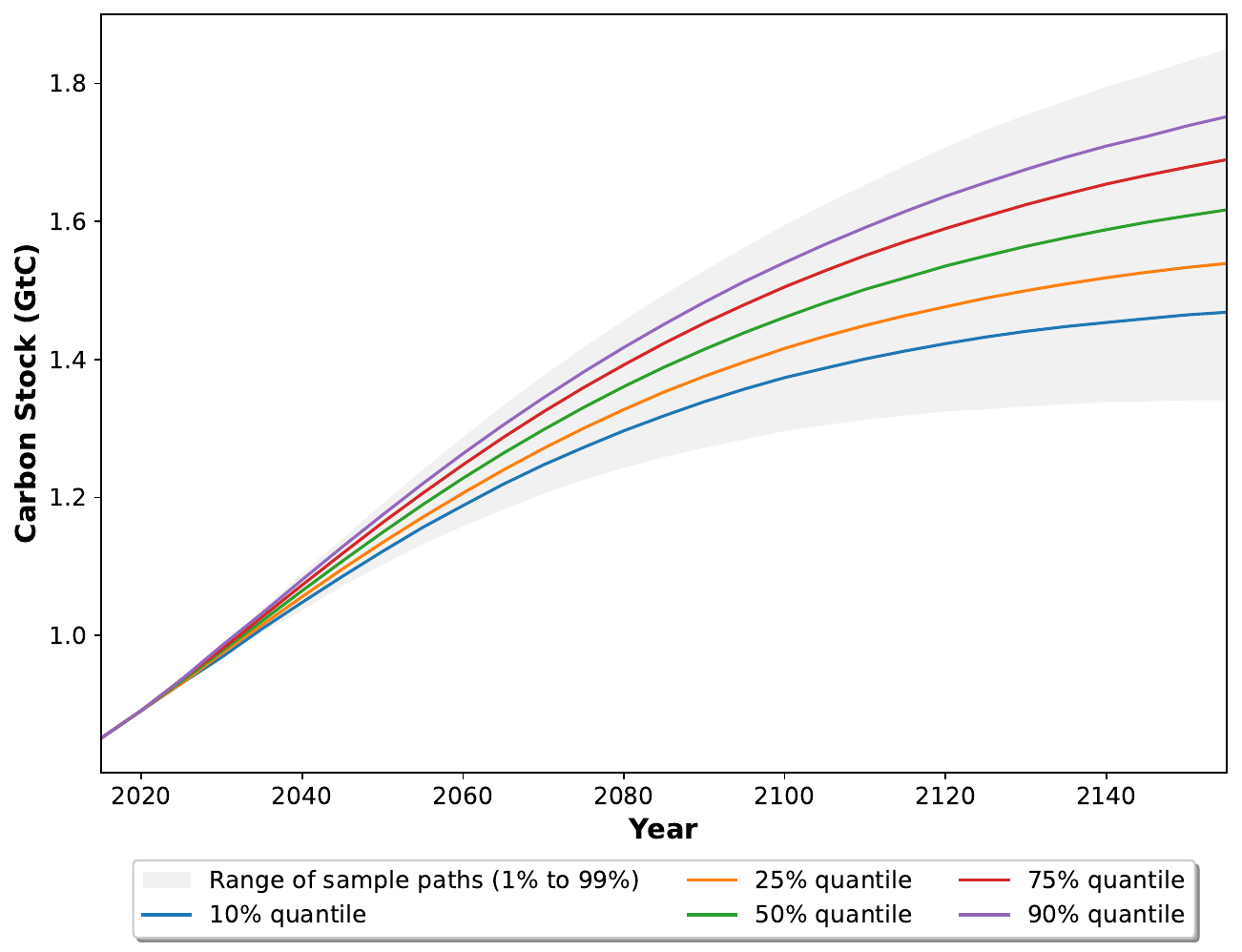} 
        \caption{Stock of Carbon}
    \end{subfigure}
    \caption{\textbf{Projected Climate and Economic Outcomes under the Optimal Tax on Cumulative Emissions.}
    The figure displays key projections from the SOLG IAM over a 150-year horizon, assuming the optimal tax policy and the exogenous transfer scheme $\mathbb{T}$.
    \textit{Top left:} Endogenous emissions (mean and full distribution) under the optimal tax policy.
    \textit{Top right:} Per-generation welfare gains and losses, in consumption equivalents, relative to the no-policy (BAU) scenario.
    \textit{Middle left:} Global warming, projected to stabilize at a mean of approximately 2.7°C (range: 2.2°C to 3.1°C).
    \textit{Middle right:} Economic damages as a percentage of GDP, averaging approximately 3\%.
    \textit{Bottom left:} The evolution of the optimal tax rate on cumulative emissions over the 150-year horizon.
    \textit{Bottom right:} The corresponding evolution of the atmospheric carbon stock.}
    \label{fig:results_lin_tax_10trans}
\end{figure}

Solving this problem yields the optimal parameters for our tax function (Equation~\eqref{eq:tax_fun_lin_S}): $\vartheta_0 = -0.01$ and $\vartheta_E = 0.43$. The main results of this welfare-improving tax policy are presented in Figure~\ref{fig:results_lin_tax_10trans}. The top-left panel of the figure shows that the tax policy significantly reduces emissions compared to BAU. For the first 50 years, average emissions are approximately 25\% lower than under the BAU scenario; even the 99th percentile of emissions remains below the BAU average. While emissions converge back towards the BAU case after 2100, the policy greatly reduces the risk of extreme emission scenarios. This mitigation effort leads to better climate outcomes, as shown in the middle-left panel. The average Global warming stabilizes at approximately 2.7°C, with considerably less uncertainty, ranging between 2.2°C and 3.1°C after 150 years, than in the BAU case.

This reduction in warming translates directly to lower climate damages, as shown in the middle-right panel. Compared to the BAU scenario, damages remain substantially lower, averaging about 3\% of GDP and spanning a range of 1.9\% to 5\% in 150 years. The bottom-right panel shows the corresponding stock of atmospheric carbon, which is maintained at a considerably lower level than in the BAU case. These physical improvements in climate outcomes lead to significant welfare effects across generations, as presented in the top-right panel (cf. Equation~\eqref{eq:cons_equiv_variation}). While initial generations (e.g., those entering the workforce in 1985) experience welfare losses of up to 5\%, the youngest generations alive at the start of the policy already gain. Welfare gains for future generations increase and stabilize at around 4.8\% for those born near 2100. Overall social welfare increases by 1.6\% in consumption equivalents compared to the BAU scenario. The policy responsible for these changes is shown in the bottom-left panel. The tax rate starts at approximately 35\%, increases over time, and reaches an average of 70\% after 150 years, with a range of 55\% to 80\%.

\subsection{Pareto-improving Linear Taxes on Cumulative Emissions and Optimal Transfers}
\label{sec:cum_emissions_optimal_transfer}

Building on the illustrative example from Section~\ref{sec:cum_emissions}, we extend the set 
of policy instruments available to the planner by enabling the selection of specific transfer shares for each age group. We apply the procedure from Section~\ref{sec:methodology} to identify Pareto-improving tax and transfer policies. 

The planner implements a tax that is linear in cumulative emissions, as defined in 
Equation~\eqref{eq:tax_fun_lin_S}. The sampling of the tax parameters is specified in 
Table~\ref{tab:param_bounds_main} for the model designated as ``linear in E + Transfers.''
In addition to the tax, the planner establishes a transfer scheme that allocates a constant share
of the resulting tax revenue to each cohort alive in a given period:
\begin{equation}
\label{eq:transfers_planner}
    \mathbb{T}^{\text{planner}} = (\vartheta_1, \vartheta_2, \dots, \vartheta_{A})^\top \in \mathbb{R}^A 
    \quad \text{s.t.} \; \sum_{i=1}^{A} \vartheta_i = 1,
\end{equation}
where $A=12$. The resulting transfer amount for generation $j$ in period $t$ is therefore 
$\mathbb{T}_{t,j} = \vartheta_j \tau_t e_t$, as per the resource constraint in 
Equation~\eqref{eq:resource_constraint_planner}.
These transfer shares, $\vartheta_j$, are treated as pseudo-state variables and added to the 
model's state space. The shares are sampled from a Dirichlet distribution, as detailed in 
Appendix~\ref{appendix:transfer_share_sampling}. This framework enables a systematic search 
for Pareto-improving carbon policies.

Applying the DEQN algorithm (Step~1 of our methodology) to this model using the loss function~\eqref{eq:loss_optimal_DEQN} yields highly accurate global solutions.
The mean out-of-sample errors for the Relative Euler Equation errors (Equation~\eqref{eq:EE_error_optimal_ERROR}) and the Relative Value Function errors (Equation~\eqref{eq:value_error_optimal_ERROR}) are of the order $\mathcal{O}(10^{-4})$, while the maximum errors are of the order $\mathcal{O}(10^{-3})$, as detailed in 
Table~\ref{tab:EE_errors_lin_S_transfers_model}.
%
\begin{table}[htbp]
\centering
\begin{adjustbox}{width=\textwidth}
\begin{tabular}{@{}lccccccccccc@{}}
\toprule
& Gen 1 & Gen 2 & Gen 3 & Gen 4 & Gen 5 & Gen 6 & Gen 7 & Gen 8 & Gen 9 & Gen 10 & Gen 11 \\
\midrule
Rel EE Mean          & 0.0006 & 0.0002 & 0.0006 & 0.0005 & 0.0001 & 0.0009 & 0.0006 & 0.0003 & 0.0004 & 0.0004 & 0.0003 \\
Rel EE 99.9     & 0.0035 & 0.0015 & 0.0037 & 0.0043 & 0.0008 & 0.0032 & 0.0014 & 0.0015 & 0.0014 & 0.0013 & 0.0011 \\
Rel Value Function Mean       & 0.0003 & 0.0003 & 0.0003 & 0.0003 & 0.0003 & 0.0005 & 0.0003 & 0.0002 & 0.0002 & 0.0004 & 0.0005 \\
Rel Value Function 99.9  & 0.0019 & 0.0018 & 0.0017 & 0.0020 & 0.0019 & 0.0023 & 0.0016 & 0.0013 & 0.0011 & 0.0013 & 0.0018 \\
\bottomrule
\end{tabular}
\end{adjustbox}
\caption{\textbf{Accuracy Metrics for the DEQN Model Solution in a Pareto-Improving Scenario.} 
    The scenario features a linear tax on cumulative emissions and optimal transfers.
    The table presents the mean and maximum (defined as the 99.9th percentile) for two metrics: 
    the Relative Euler Equation Error (Rel EE) and the Relative Value Function Error, evaluated
    for each overlapping generation (Gen~$i$). The statistics are computed over 10,000 simulated 
    sample paths of the model's equilibrium conditions.}
\label{tab:EE_errors_lin_S_transfers_model}
\end{table}

In Step 2, the neural-network surrogate $\mathcal{N}(\cdot)$ provides policy evaluations that are used to train 
GP surrogates for the SWF. We follow the general
setup of Equation~\eqref{eq:opt_pareto_y} and map each cohort’s
expected lifetime utility onto the policy-parameter vector
\begin{equation}    
  \vartheta
  = \{\vartheta_0,\vartheta_E,\vartheta_1,\ldots,\vartheta_{12}\}
  \in\mathbb{R}^{14},
\end{equation}
which contains the two tax coefficients and the twelve transfer shares. Concretely, we fit forty independent GPs, one for each cohort $t=-10,\ldots,29$, according to Equation~\eqref{eq:pareto_map}:
\begin{equation}
  \mu_{*,t}(\vartheta)
  \;=\;
  \widehat{\mathrm{QoI}}_{t}(\vartheta)
  \;=\;
  \mathbb{E}\!\bigl[\tilde{U}_{t}(\vartheta)\bigr],
  \qquad t=-10,\ldots,29.
\end{equation}
Each predictive mean $\mu_{*,t}(\vartheta)$ therefore serves as a
computationally inexpensive proxy for the Monte-Carlo estimate of cohort $t$’s expected utility under the policy parameters~$\vartheta$.

In Step 3, we solve the constrained optimization problem
\begin{equation}
\label{eq:opt_pareto_y2}
\begin{aligned}
  \vartheta^{\ast}
  &=\arg\max_{\vartheta\in[a,b]}
      \sum_{t=-10}^{29}\gamma_t\,\widehat{\mathrm{QoI}}_t(\vartheta)
    =\arg\max_{\vartheta\in[a,b]}
      \sum_{t=-10}^{29}\gamma_t\,\mathbb{E}_0\!\bigl[\tilde{U}_t(\vartheta)\bigr] \\[4pt]
  \text{s.t.}\qquad
  &\mathbb{E}_0\!\bigl[\tilde{U}_t(\vartheta)\bigr]\;\ge\;\mathbb{E}_0[U_t] 
     \quad\forall\,t, \\[2pt]
  &0 \;\le\; \vartheta_0 + \vartheta_E\,E_0  \;\le\; \bar{\tau}, \\[2pt]
  &0 \;\le\; \vartheta_0 + \vartheta_E\,E_{29}\;\le\; \bar{\tau}, \\[2pt]
  &\sum_{i=1}^{12}\vartheta_i = 1,
\end{aligned}
\end{equation}
where the first line enforces a Pareto improvement, the next two
inequalities bound taxes (as in Equation~\eqref{eq:optim_cum_emm_maxwelf} but with
$\bar{\tau}=0.8$), and the final equality guarantees that the transfer shares sum to one (cf.\ Appendix~\ref{appendix:sampling}).  
We set the welfare weights to $\gamma_t = 0.025$ for all $t$.
The optimizer yields the tax coefficients reported in
Table~\ref{tab:optimal_tax_params}; the corresponding optimal transfer shares are listed in Table~\ref{tab:optimal_transfer_lin_S}.
%
\begin{table}[ht!]
  \centering
  \begin{tabular}{lcc}
    \toprule
    Coefficient & Symbol & Value \\
    \midrule
    Intercept   & $\vartheta_0$ & $-0.186$ \\
    Slope       & $\vartheta_E$ & $0.225$  \\
    \bottomrule
  \end{tabular}
\caption{\textbf{Pareto-optimal coefficients for the linear tax on cumulative emissions.}
The table lists the estimated intercept $\vartheta_0$ and slope $\vartheta_E$ that jointly maximize social welfare subject to the Pareto constraints.}
  \label{tab:optimal_tax_params}
\end{table}
\begin{table}[ht!]
    \centering
    \begin{tabular}{|c|c|c|c|c|c|c|c|c|c|c|c|c|}
        \hline
        \textbf{t1} & \textbf{t2} & \textbf{t3} & \textbf{t4} & \textbf{t5} & \textbf{t6} & \textbf{t7} & \textbf{t8} & \textbf{t9} & \textbf{t10} & \textbf{t11} & \textbf{t12} \\
        \hline
        0.128 & 0.051 & 0.058 & 0.089 & 0.149 & 0.09 & 0.066 & 0.143 & 0.076 & 0.048 & 0.039 & 0.061 \\
        \hline
    \end{tabular}
\caption{\textbf{Pareto-optimal transfer scheme under a linear tax on cumulative emissions.}
The table reports the optimal transfer share allocated to each concurrent cohort $i$; shares sum to one.}
    \label{tab:optimal_transfer_lin_S}
\end{table}

\begin{figure}[th!]
    \centering
    \begin{subfigure}{0.45\textwidth}
        \includegraphics[width=0.95\textwidth]{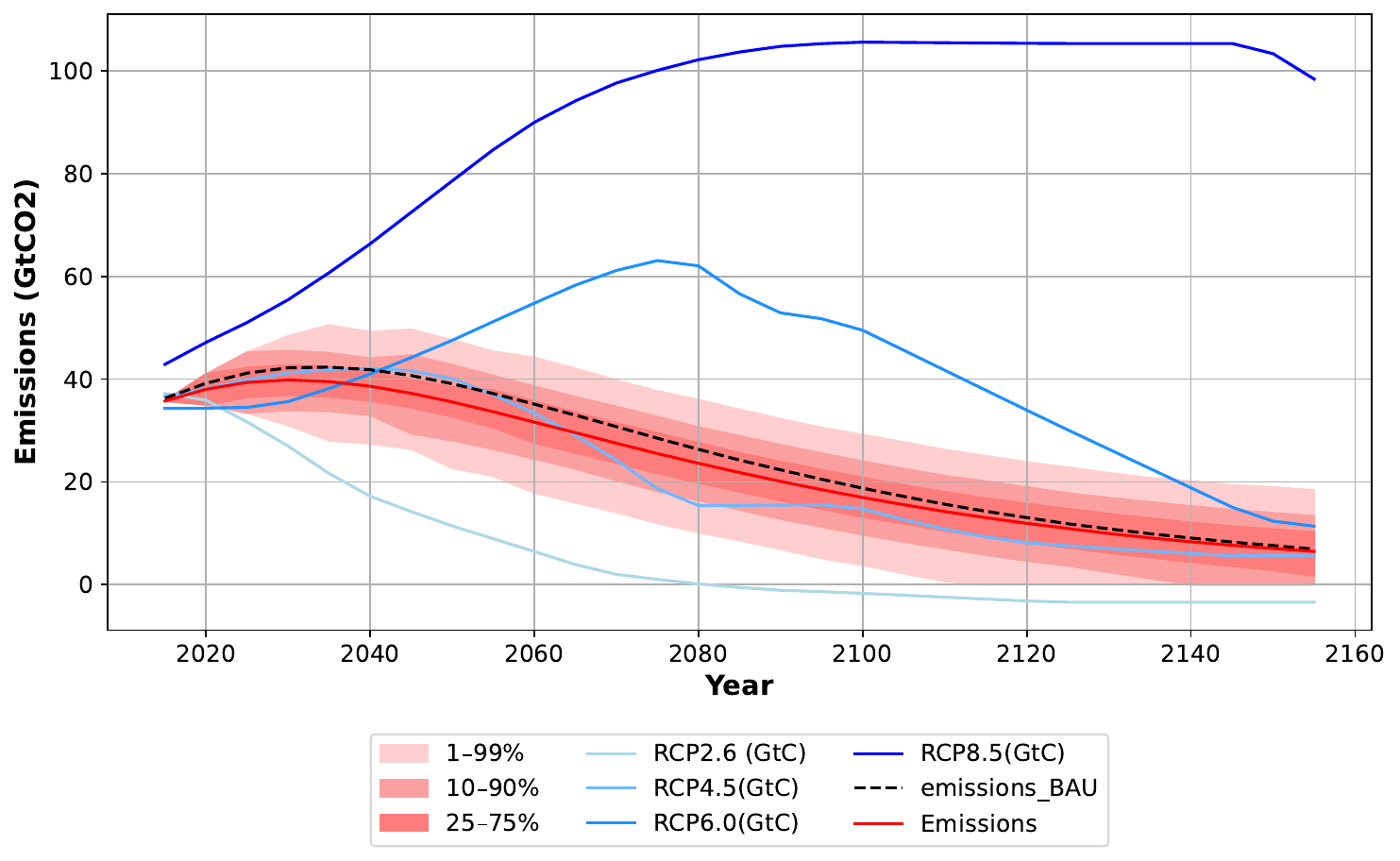} 
        \caption{Emissions}  
    \end{subfigure}
    \hfill\begin{subfigure}{0.45\textwidth}
        \includegraphics[width=0.95\textwidth]{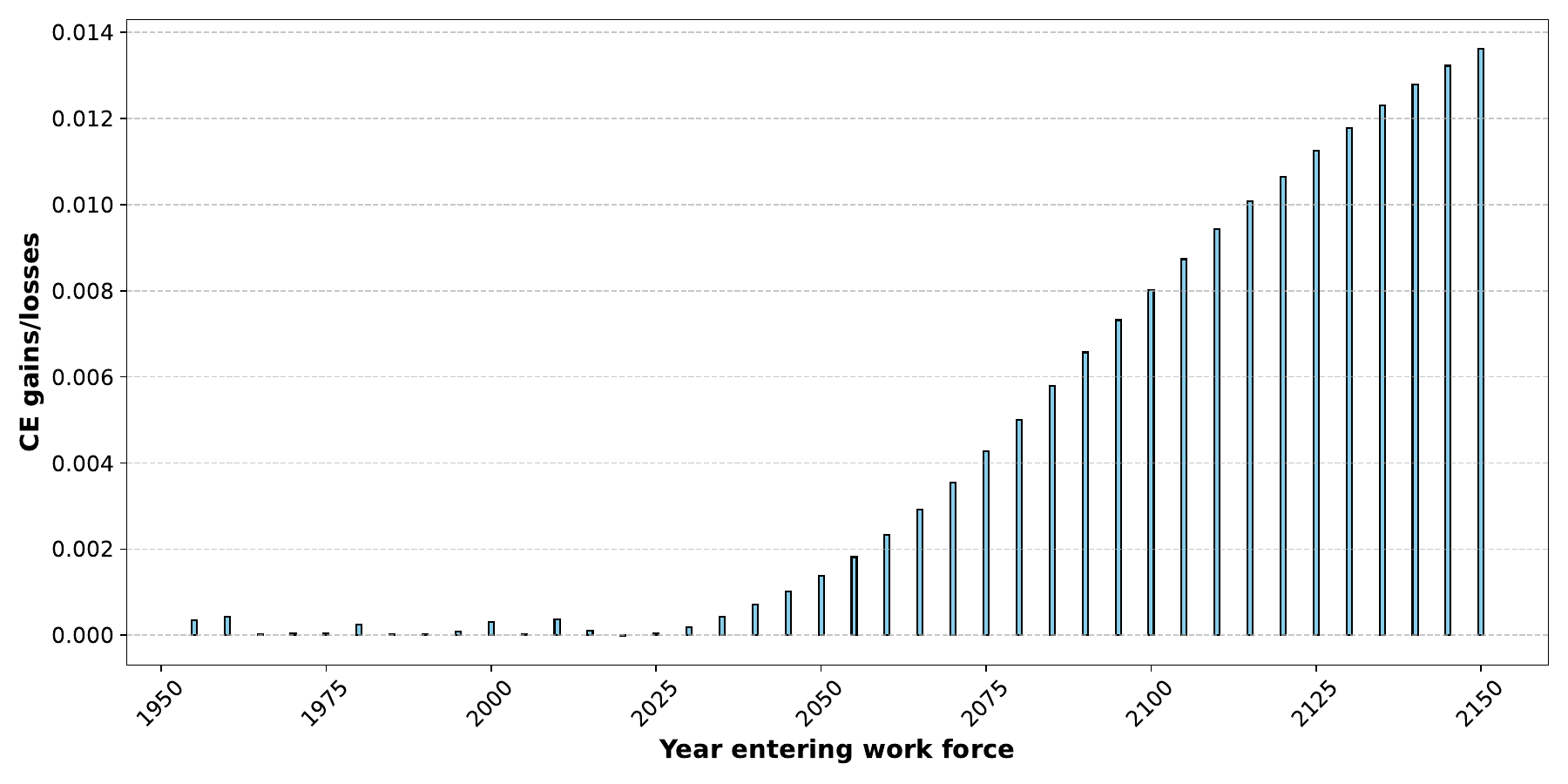} 
        \caption{Welfare Gains}
    \end{subfigure}
    \begin{subfigure}{0.45\textwidth}
        \includegraphics[width=0.95\textwidth]{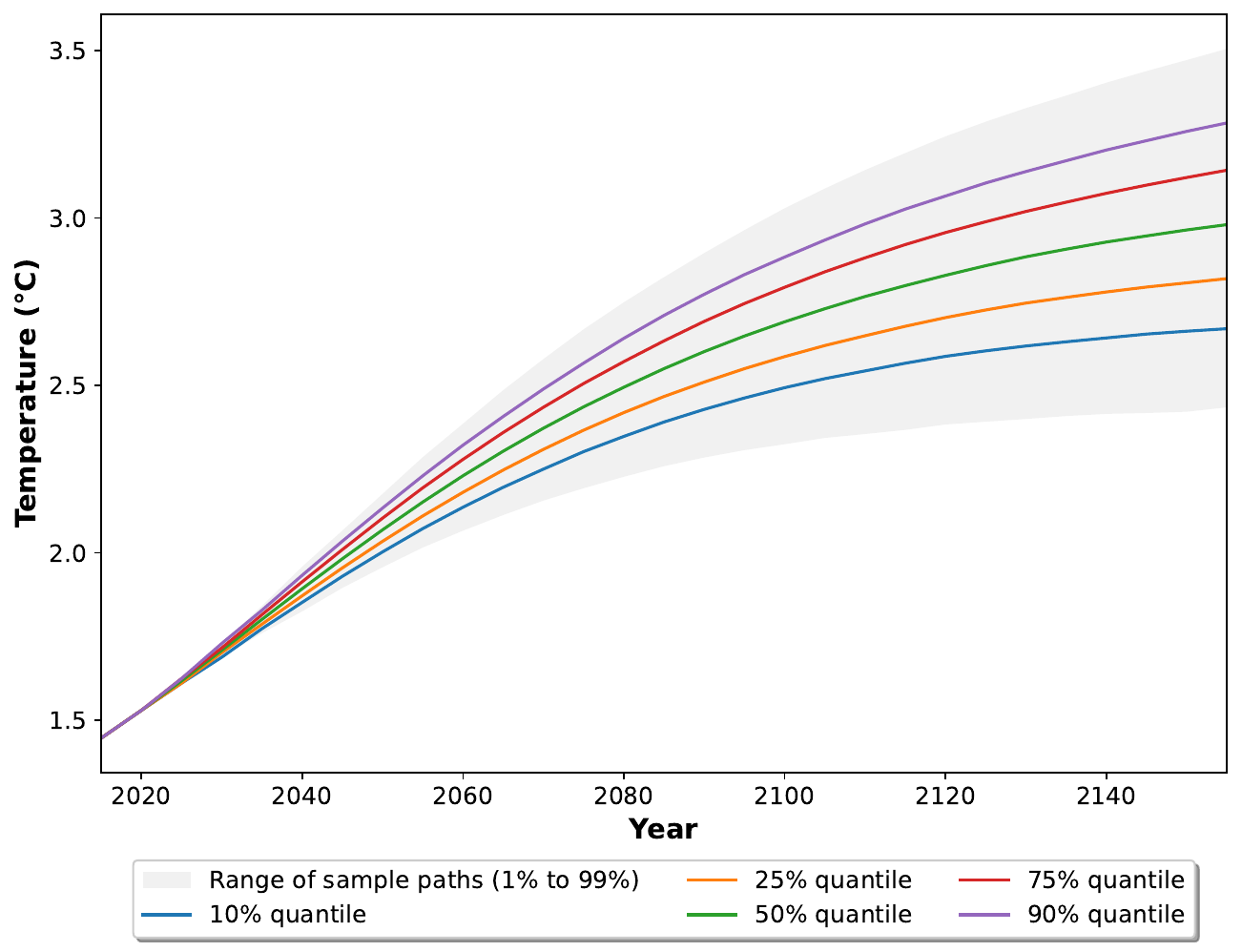}
        \caption{Temperature}  
    \end{subfigure}
    \hfill
    \begin{subfigure}{0.45\textwidth}
        \includegraphics[width=0.95\textwidth]{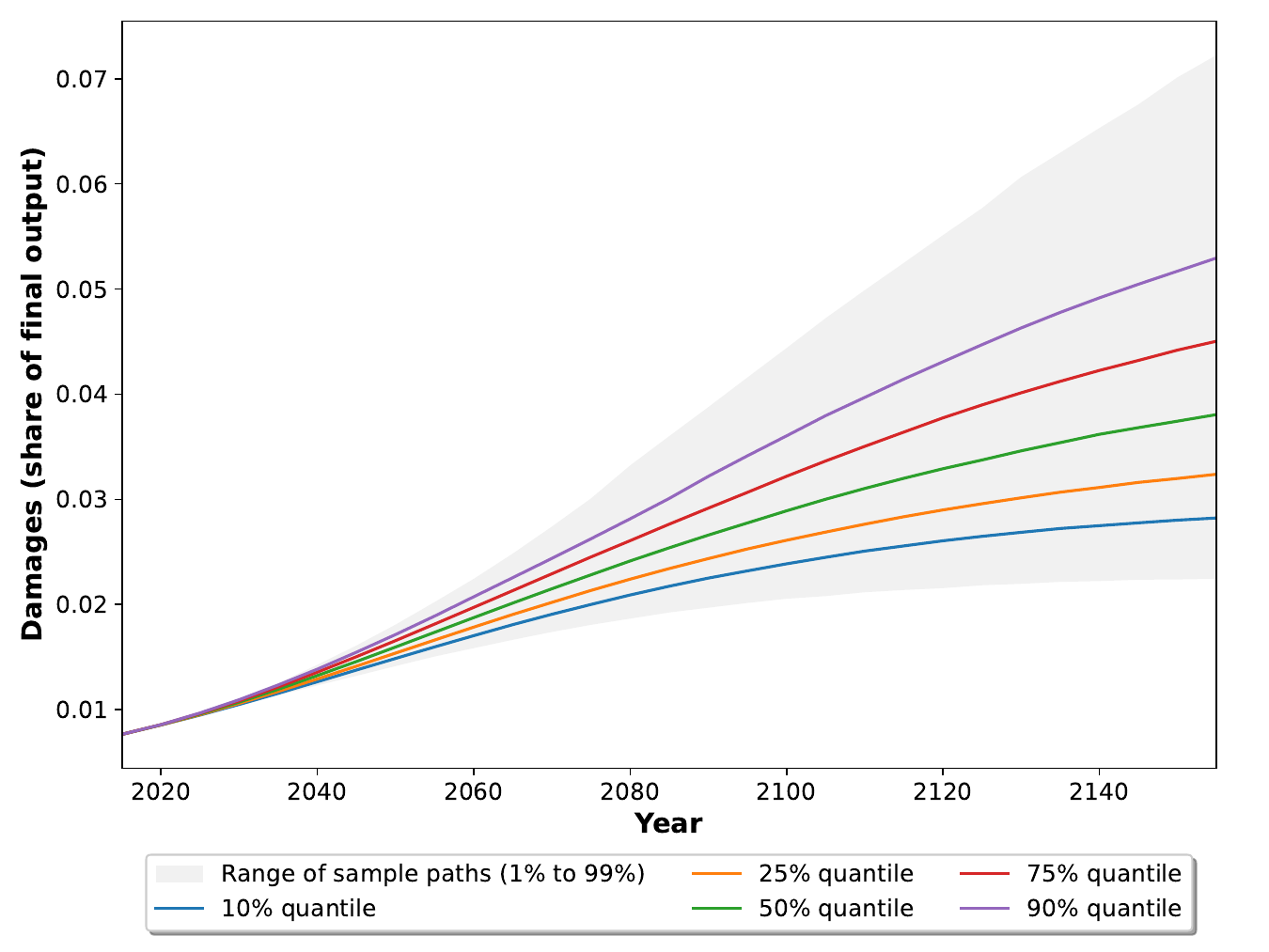} 
        \caption{Damages}
        \end{subfigure}
    \begin{subfigure}{0.45\textwidth}
        \includegraphics[width=0.95\textwidth]{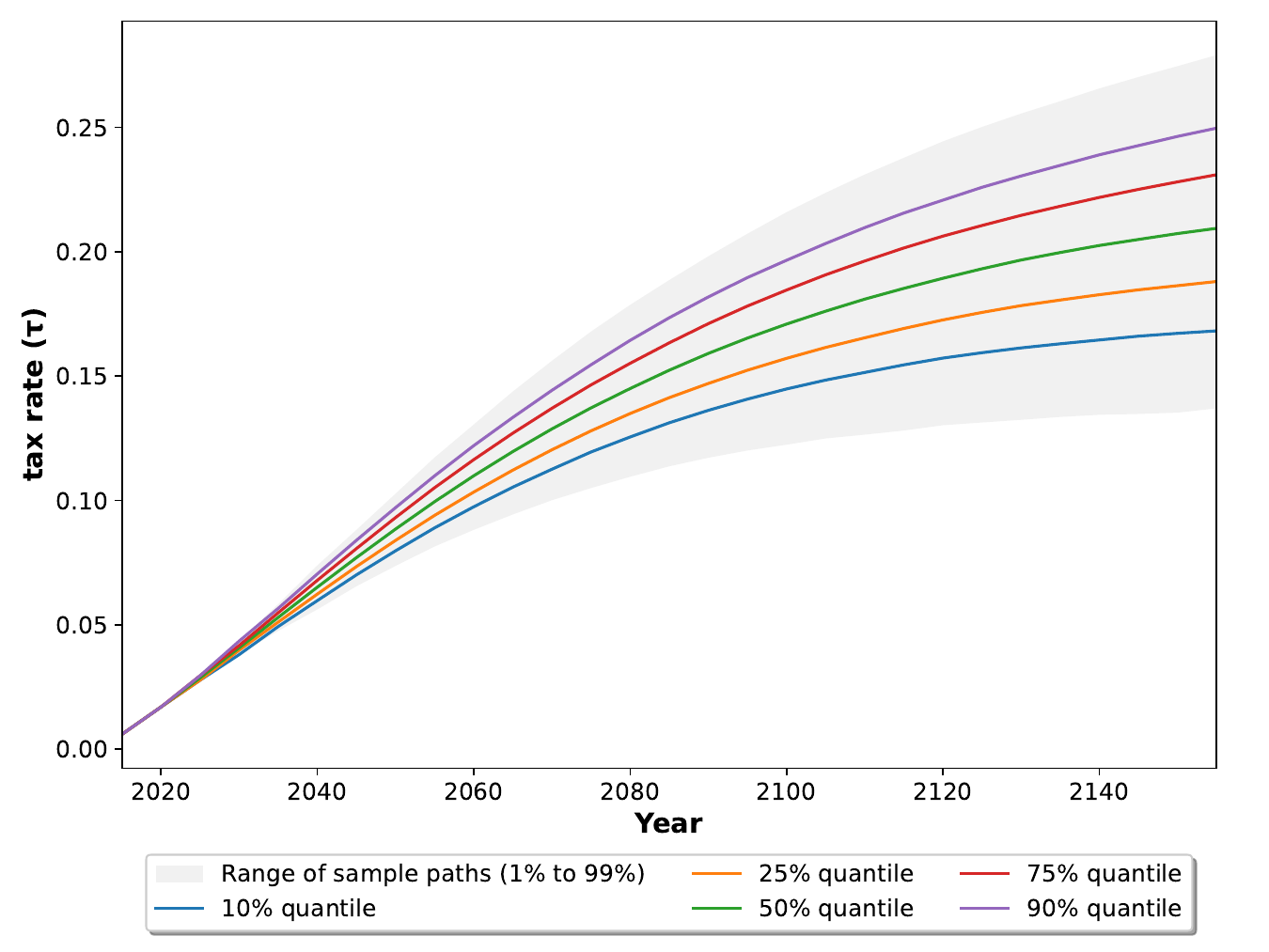} 
        \caption{Tax rate}
        \end{subfigure}
        \hfill
    \begin{subfigure}{0.45\textwidth}
        \includegraphics[width=0.95\textwidth]{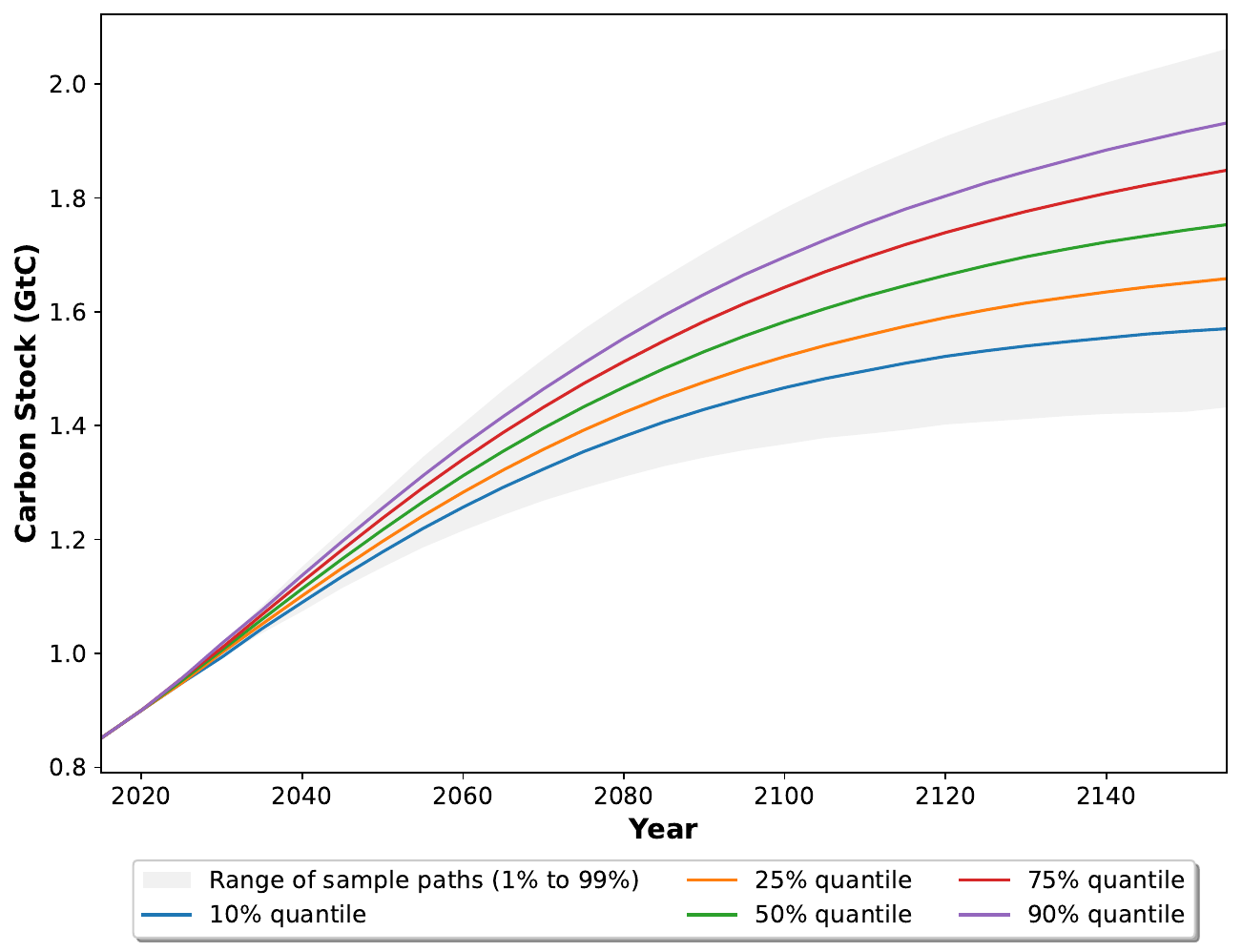} 
        \caption{Stock of Carbon}
    \end{subfigure}    
    \caption{\textbf{Projected Climate and Economic Outcomes under the Optimal Tax on Cumulative Emissions with Endogenous Transfers.} The figure displays key projections from the SOLG IAM over a 150-year horizon for the Pareto-improving tax and transfer policy.
\textit{Top left:} Endogenous emissions (mean and full distribution) under the joint optimal tax and transfer scheme.
\textit{Top right:} Per-generation welfare gains and losses, in consumption equivalents, relative to the business-as-usual scenario.
\textit{Middle left:} Global warming, projected to stabilise at a mean just below 3 °C (range: 2.3 °C – 3.5 °C).
\textit{Middle right:} Economic damages as a percentage of GDP, averaging below 4 \% with a 99\textsuperscript{th} percentile of about 7 \%.
\textit{Bottom left:} Evolution of the optimal tax rate on cumulative emissions over the 150-year horizon.
\textit{Bottom right:} Corresponding evolution of the atmospheric carbon stock.}
    \label{fig:lin_S_transfer_pareto}
\end{figure}

After solving the SOLG IAM and determining the optimal tax and transfer scheme, we analyze the resulting climate and economic outcomes. Figure~\ref{fig:lin_S_transfer_pareto} presents the main results of this Pareto-improving policy, which combines a linear tax on cumulative emissions with targeted inter-generational transfers.

The top-left panel shows that emissions are, on average, lower than under the BAU scenario, although not by a large margin. The policy's primary impact is on mitigating upside risk; the 99th percentile of emissions drops below 40 GtCO$_2$ by 2070, approximately 20 years earlier than in the BAU case. This modest mitigation effort leads to slightly improved climate outcomes, as shown in the middle-left panel. The average Global warming stabilizes at a level marginally below 3°C, and importantly, the 99th percentile remains below 3.5°C. The bottom-right panel shows the corresponding stock of atmospheric carbon, which is maintained at a level consistent with these climate outcomes.

This reduction in warming translates to lower climate damages, as illustrated in the middle-right panel. While average damages are only slightly lower than in the BAU scenario, the policy effectively truncates the worst-case outcomes. The 99th percentile of damages reaches approximately 7\% of GDP (compared to 9\% under BAU), with the worst-case scenarios remaining below 12\% (compared to 15\% under BAU). Furthermore, the 90th percentile is reduced from 6\% to 5\%.

These physical outcomes are achieved via a carefully structured tax and transfer scheme designed to ensure Pareto improvement, with welfare effects shown in the top-right panel. The transfers are allocated to shield the initial generations, including those born in the next 10 years, from welfare losses, keeping them approximately at their BAU utility levels. Subsequent generations experience progressively larger welfare gains, which increase to nearly 1.4\% in consumption equivalents for the last generation considered (cf. Equation~\eqref{eq:cons_equiv_variation}). Overall, this policy increases social welfare by 0.42\% in consumption equivalents compared to the BAU scenario. The carbon tax responsible for these changes is depicted in the bottom-left panel.

\subsection{Pareto-improving Linear Taxes on Cumulative Emissions, Carbon Intensity, and Climate Tipping, and Optimal Transfers}
\label{sec:emissions_carbon_intensity_damages_optimal_transfer}

In this section, we analyze a markedly richer policy instrument set than in Section~\ref{sec:cum_emissions_optimal_transfer}: the planner now controls four levers. 
The carbon tax is specified as a linear function of cumulative emissions, carbon intensity, and proximity to a climate-tipping point,
\begin{equation}
    \label{eq:tax_fun_lin_S_kappa_tipping}
    \tau_t(E_t,\kappa_t,TP_t)
    \;=\;
    \vartheta_0
    + \vartheta_E \,\frac{E_t}{E_0}
    + \vartheta_\kappa \,\frac{\kappa_t}{\kappa_0}
    + \vartheta_{TP}\,\bigl(1-\mathbb{D}_{TP}\bigr),
\end{equation}
where the normalised distance to tipping is defined as
\begin{equation}
    \mathbb{D}_{TP}
\;=\;
\frac{\max\!\bigl(TP_t-T_t^{AT},\,0\bigr)}
     {TP_{\max}-T_0^{AT}}
\in[0,1],
\end{equation}
and attains unity once a simulation path crosses the tipping threshold. 
We set $TP_{\max}=3.5$ and scale $E_t$ and $\kappa_t$ by their initial values $E_0$ and $\kappa_0$.

Equation~\eqref{eq:tax_fun_lin_S_kappa_tipping} introduces two additional coefficients, $\vartheta_\kappa$ and $\vartheta_{TP}$, which, together with the other tax and transfer parameters, are treated as pseudo-states and appended to the model’s state vector, raising its dimensionality to $d_\vartheta = 16$. 
The sampling bounds for all parameters are reported in Table~\ref{tab:param_bounds_main} under ``Full linear + Transfers,'' and the transfer-allocation rule is given in Equation~\eqref{eq:transfers_planner}.

Applying the same DEQN procedure to the ``Full linear + Transfers’’ specification delivers accuracy that is virtually identical to the simpler model: mean out-of-sample Relative Euler Equation and Value-Function errors are $\mathcal{O}(10^{-4})$, and the 99.9th-percentile errors are $\mathcal{O}(10^{-3})$ (Table~\ref{tab:EE_errors_linear_transfers_model}).
\begin{table}[htbp]
\centering
\begin{adjustbox}{width=\textwidth}

\begin{tabular}{@{}lccccccccccc@{}}
\toprule
& Gen 1 & Gen 2 & Gen 3 & Gen 4 & Gen 5 & Gen 6 & Gen 7 & Gen 8 & Gen 9 & Gen 10 & Gen 11 \\
\midrule
Rel EE Mean          & 0.0004 & 0.0005 & 0.0005 & 0.0003 & 0.0002 & 0.0003 & 0.0004 & 0.0004 & 0.0006 & 0.0006 & 0.0007 \\
Rel EE 99.9     & 0.0035 & 0.0024 & 0.0032 & 0.0015 & 0.0013 & 0.0030 & 0.0019 & 0.0024 & 0.0022 & 0.0020 & 0.0036 \\
Rel Value Function Mean       & 0.0004 & 0.0003 & 0.0002 & 0.0003 & 0.0003 & 0.0004 & 0.0003 & 0.0003 & 0.0004 & 0.0005 & 0.0006 \\
Rel Value Function 99.9  & 0.0016 & 0.0015 & 0.0017 & 0.0019 & 0.0014 & 0.0015 & 0.0013 & 0.0014 & 0.0018 & 0.0020 & 0.0022 \\
\bottomrule
\end{tabular}

\end{adjustbox}
\caption{\textbf{Accuracy Metrics for the DEQN Solution in the ``Full linear + Transfers'' Scenario.}
    The planner’s instruments are a linear tax on cumulative emissions, carbon intensity, and tipping, together with optimal transfers.
    The table reports the mean and the 99.9th-percentile (labelled ``Max'') of the Relative Euler Equation Error (Rel EE) and the Relative Value-Function Error for each overlapping generation (Gen $i$), computed from 10,000 simulated equilibrium paths.}
\label{tab:EE_errors_linear_transfers_model}
\end{table}

The policy network $\mathcal{N}(\cdot)$ supplies Monte-Carlo estimates of each cohort’s expected lifetime utility, which we interpolate with GPs.  
The 16-dimensional policy vector now reads as follows:
\begin{equation}
  \vartheta
  \;=\;
  \{\vartheta_0,\vartheta_E,\vartheta_\kappa,\vartheta_{TP},
    \vartheta_1,\ldots,\vartheta_{12}\}\in\mathbb{R}^{16},
\end{equation}
containing four tax coefficients and the twelve transfer shares.
As in Equation~\eqref{eq:opt_pareto_y2}, we fit forty independent GPs, one per cohort~$t=-10,\ldots,29$:
\begin{equation}
  \mu_{*,t}(\vartheta)
  \;=\;
  \widehat{\mathrm{QoI}}_{t}(\vartheta)
  \;=\;
  \mathbb{E}\!\bigl[\tilde{U}_{t}(\vartheta)\bigr],
  \qquad t=-10,\ldots,29.
\end{equation}
Each predictive mean $\mu_{*,t}(\vartheta)$ serves again as a
computationally inexpensive proxy for the corresponding Monte-Carlo utility estimate.

In Step 3, we now solve the Pareto–constrained planner’s problem:
\begin{equation}
\label{eq:opt_pareto_y_linear_transfers}
\begin{aligned}
  \vartheta^{\ast}
  &=\arg\max_{\vartheta\in[a,b]}
      \sum_{t=-10}^{29}\gamma_t\,\widehat{\mathrm{QoI}}_t(\vartheta)
    =\arg\max_{\vartheta\in[a,b]}
      \sum_{t=-10}^{29}\gamma_t\,\mathbb{E}_0\!\bigl[\tilde{U}_t(\vartheta)\bigr] \\[4pt]
  \text{s.t.}\qquad
  &\mathbb{E}_0\!\bigl[\tilde{U}_t(\vartheta)\bigr]\;\ge\;\mathbb{E}_0[U_t] 
     \quad\forall\,t, \\[2pt]
  &0 \leq \vartheta_0 + \vartheta_E  \tfrac{E_0}{E_0} + \vartheta_\kappa \tfrac{\kappa_0}{\kappa_0}
        + \vartheta_{TP}(1-\mathbb{D}_{TP,0}) \le \bar{\tau}, \\[2pt]
  &0 \leq \vartheta_0 + \vartheta_E  \tfrac{E_{29}}{E_0} + \vartheta_\kappa \tfrac{\kappa_{29}}{\kappa_0}
        + \vartheta_{TP}(1-\mathbb{D}_{TP,29}) \le \bar{\tau}, \\[2pt]
  &\sum_{i=1}^{12}\vartheta_i = 1,
\end{aligned}
\end{equation}
with $E_0=0.851$, $E_{29}=1.6$, $\kappa_0=0.35032$, $\kappa_{29}=0.05$, 
$\{\mathbb{D}_{TP,0},\mathbb{D}_{TP,29}\}=\{0.2435,1.0\}$, and $\bar{\tau}=0.8$
(cf.\ Table~\ref{tab:param_bounds_main} in Appendix~\ref{appendix:Tax_params_sampling}).  
Uniform welfare weights $\gamma_t = 0.025$ are retained.

The optimizer delivers the tax coefficients reported in
Table~\ref{tab:optimal_tax_params_linear}; the optimal transfer shares are
listed in Table~\ref{tab:optimal_transfer_linear}.

\begin{table}[ht!]
  \centering
  \begin{tabular}{lcc}
    \toprule
    Coefficient & Symbol & Value \\
    \midrule
    Intercept               & $\vartheta_0$     & $-0.237$ \\
    Cumulative emissions     & $\vartheta_E$     & $0.203$  \\
    Carbon intensity         & $\vartheta_\kappa$& $0.037$  \\
    Distance to tipping      & $\vartheta_{TP}$  & $0.012$  \\
    \bottomrule
  \end{tabular}
\caption{\textbf{Pareto-optimal coefficients for the full linear tax on cumulative emissions, carbon intensity, and tipping.}
The table reports the estimated intercept $\vartheta_0$ and the three slopes 
$\{\vartheta_E,\vartheta_\kappa,\vartheta_{TP}\}$ that jointly maximize social welfare 
while satisfying the Pareto constraints.}
  \label{tab:optimal_tax_params_linear}
\end{table}

\begin{table}[h!]
    \centering
    \begin{tabular}{|c|c|c|c|c|c|c|c|c|c|c|c|c|}
        \hline
        \textbf{t1} & \textbf{t2} & \textbf{t3} & \textbf{t4} & \textbf{t5} & \textbf{t6} & \textbf{t7} & \textbf{t8} & \textbf{t9} & \textbf{t10} & \textbf{t11} & \textbf{t12} \\
        \hline
        0.068 & 0.127 & 0.064 & 0.089 & 0.086 & 0.156 & 0.091 & 0.072 & 0.065 & 0.092 & 0.0 & 0.088 \\
        \hline
    \end{tabular}
    \caption{\textbf{Pareto optimal transfer scheme with a linear tax on cumulative Emissions, carbon intensity, and tipping.} The table presents the optimal transfer shares for each concurrent cohort $i$ with a linear tax on cumulative emissions.}
    \label{tab:optimal_transfer_linear}
\end{table}

\begin{figure}[th!]
    \centering
    \begin{subfigure}{0.45\textwidth}
        \includegraphics[width=0.95\textwidth]{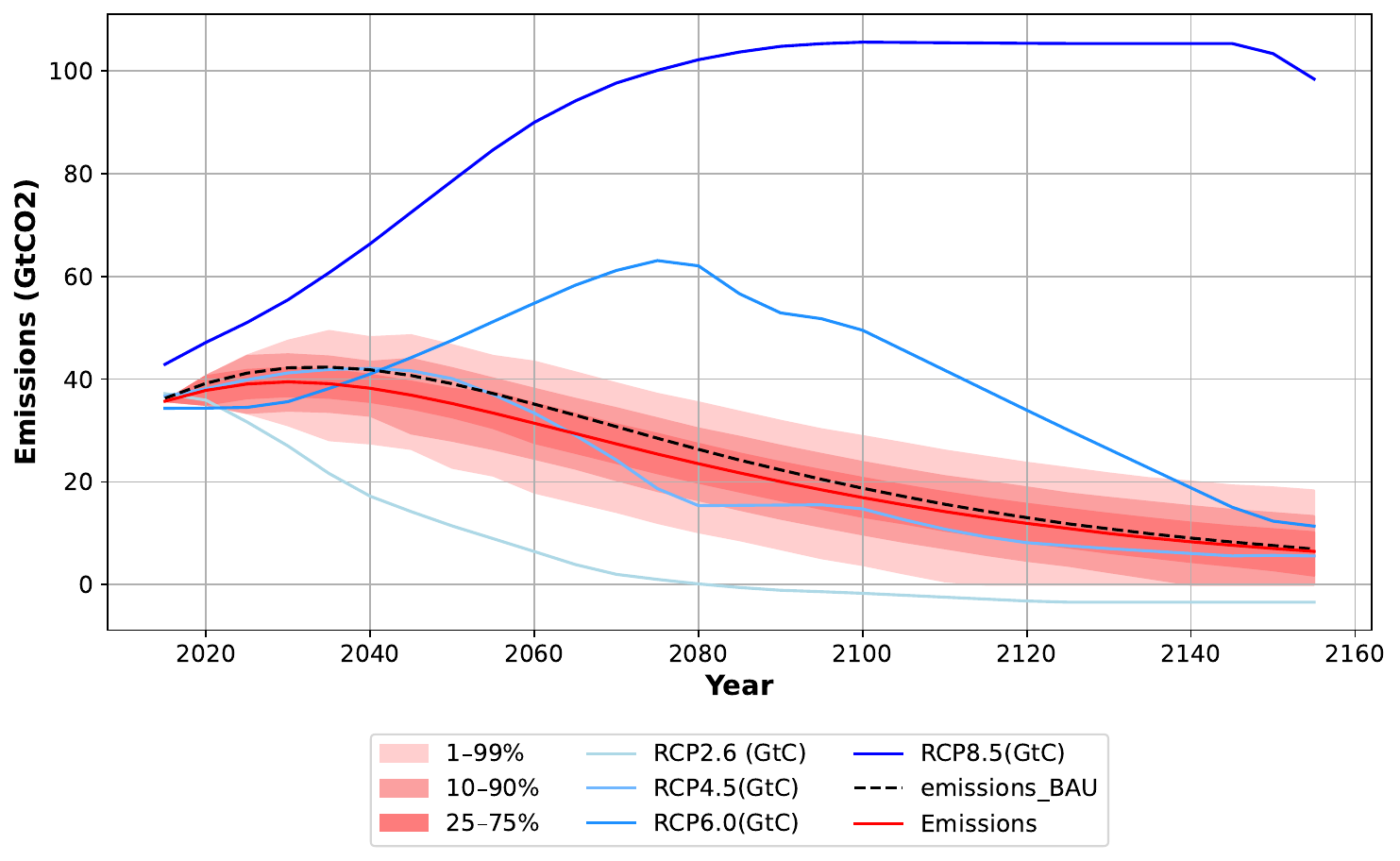} 
        \caption{Emissions}  
    \end{subfigure}
    \hfill
    \begin{subfigure}{0.45\textwidth}
        \includegraphics[width=0.95\textwidth]{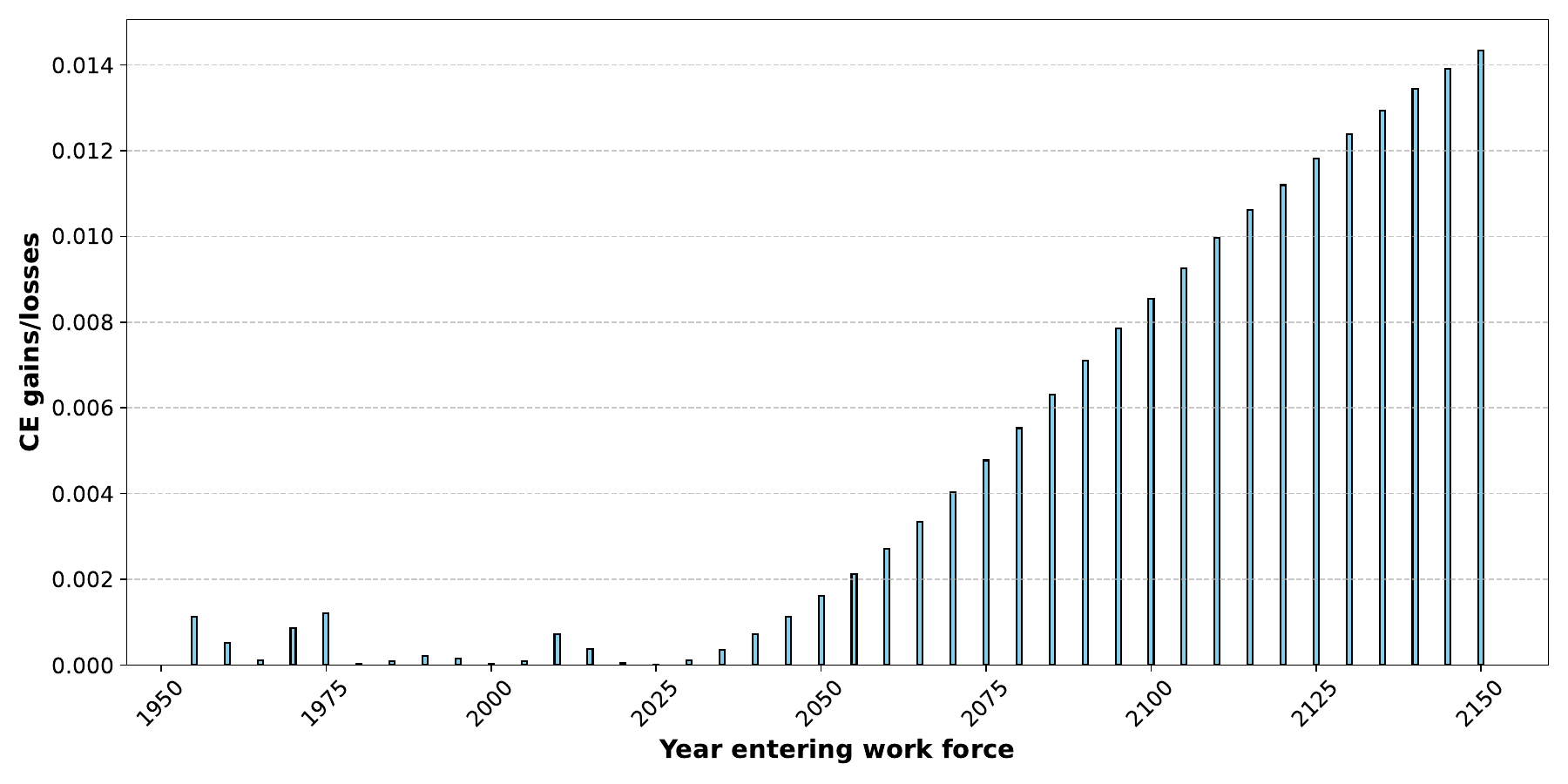} 
        \caption{Welfare Gains}
    \end{subfigure}
    \begin{subfigure}{0.45\textwidth}
        \includegraphics[width=0.95\textwidth]{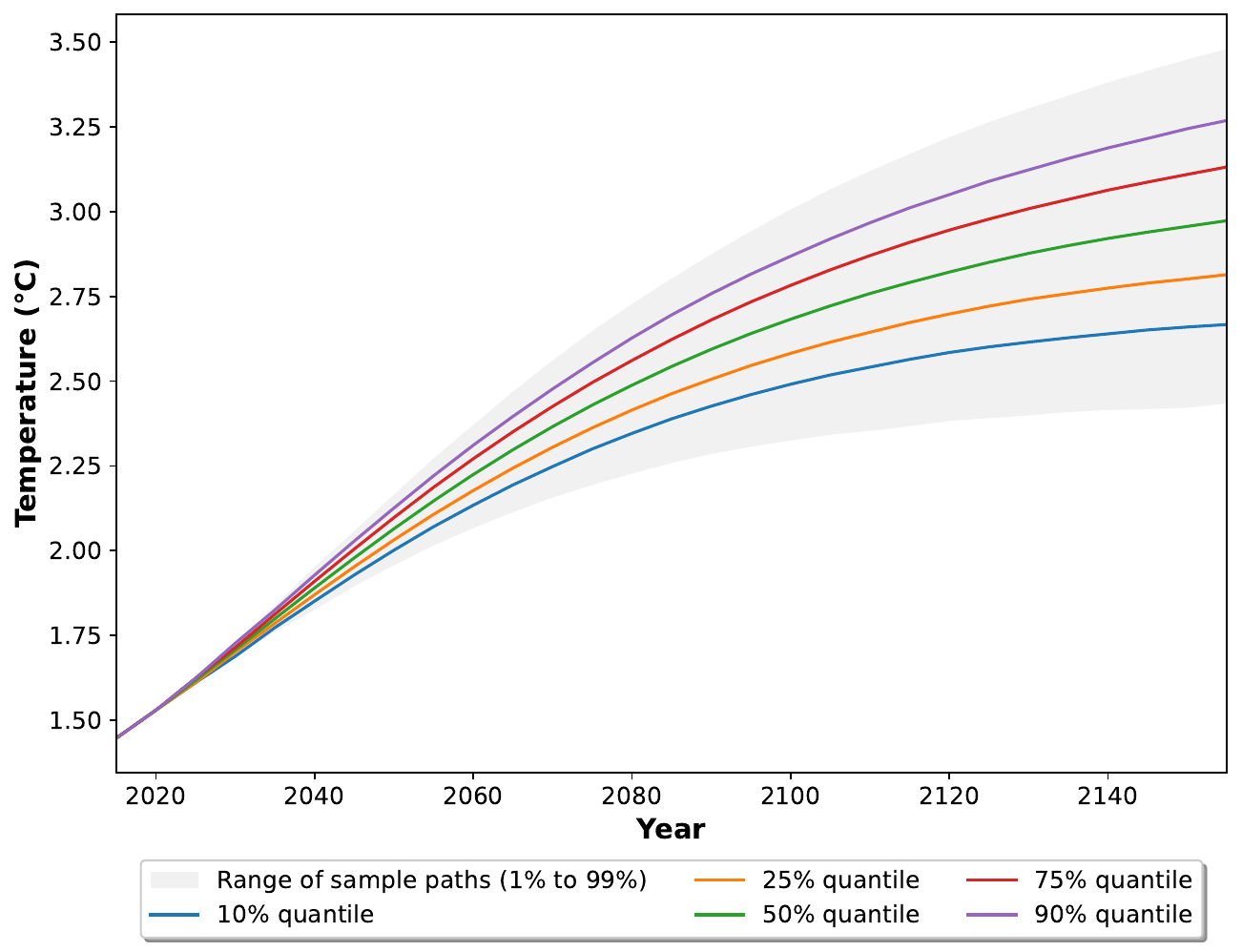} 
        \caption{Temperature}  
    \end{subfigure}
    \hfill
    \begin{subfigure}{0.45\textwidth}
        \includegraphics[width=0.95\textwidth]{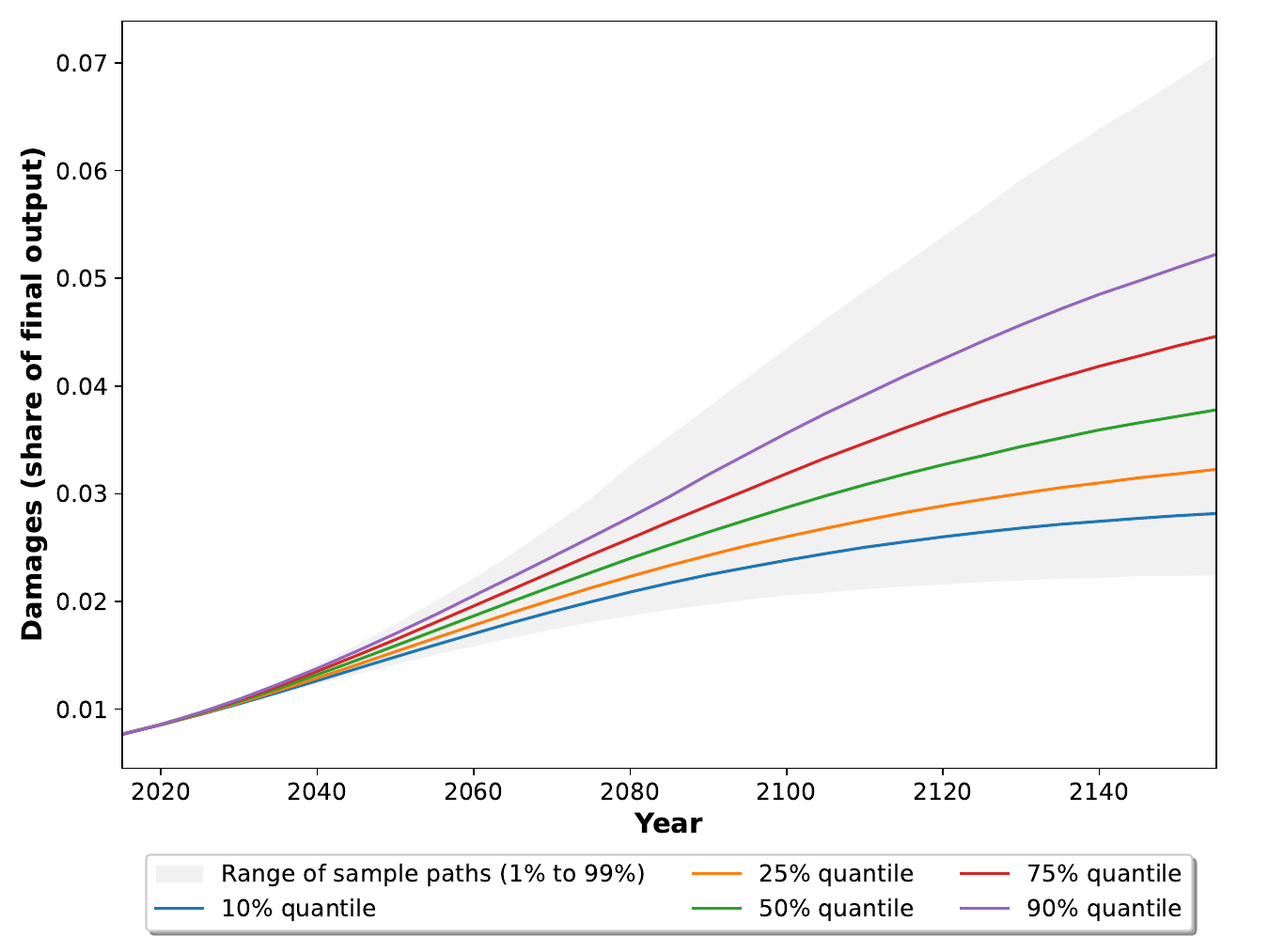} 
        \caption{Damages}  
    \end{subfigure}
        \begin{subfigure}{0.45\textwidth}
        \includegraphics[width=0.95\textwidth]{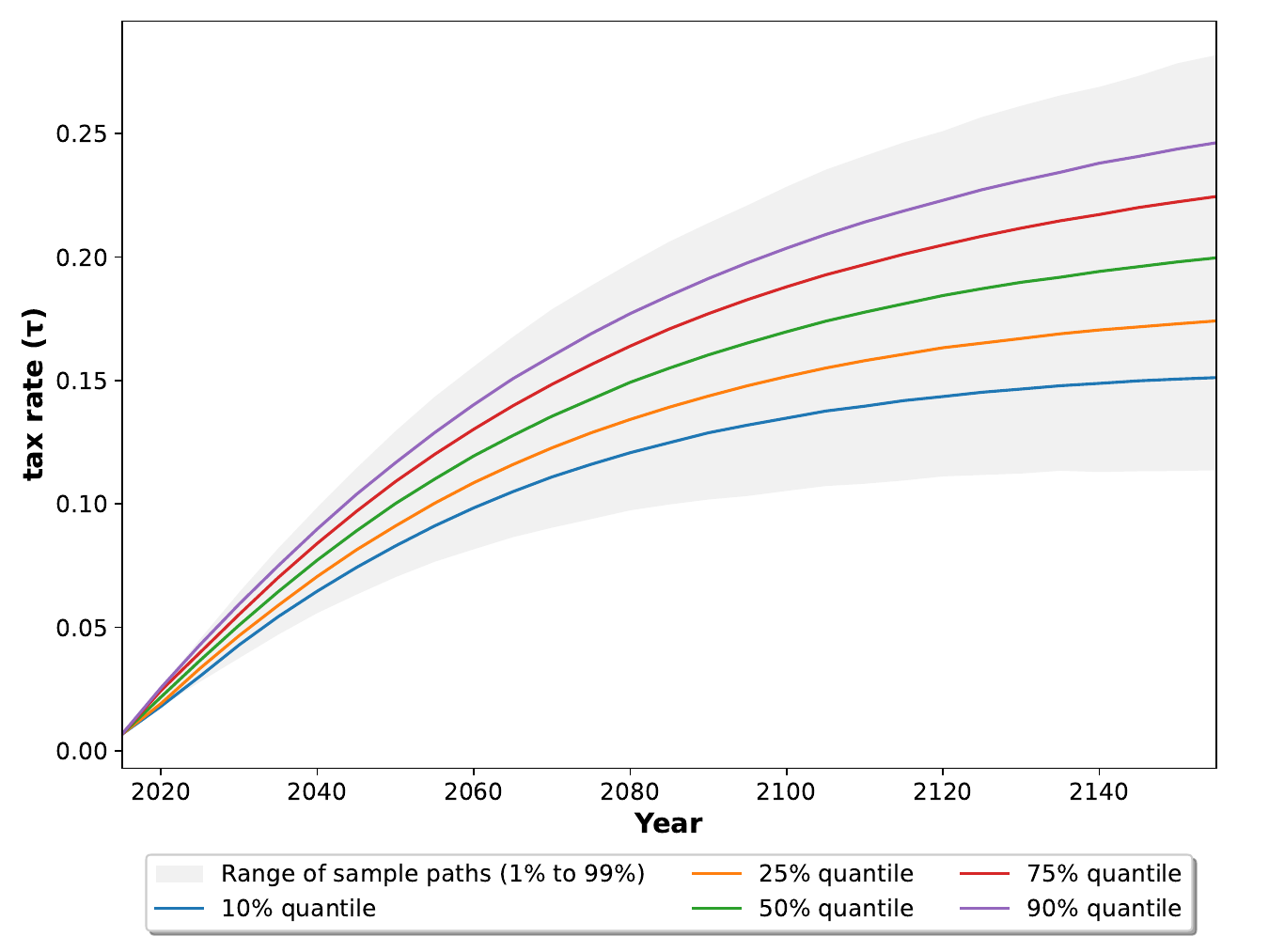} 
        \caption{Tax rate}  
    \end{subfigure}
        \hfill
    \begin{subfigure}{0.45\textwidth}
        \includegraphics[width=0.95\textwidth]{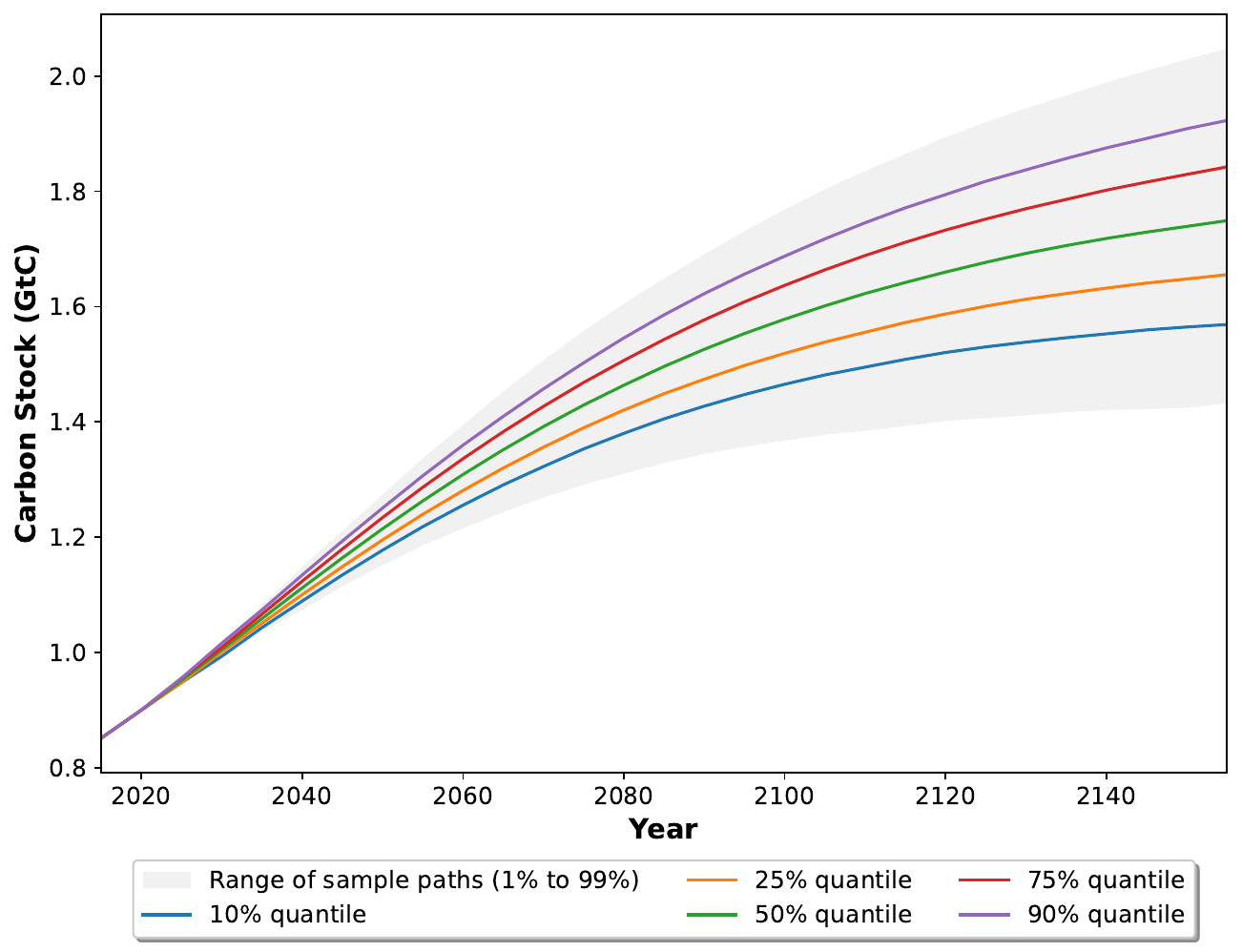} 
        \caption{Stock of Carbon}
    \end{subfigure}
\caption{\textbf{Projected Climate and Economic Outcomes under the Optimal Linear Tax on Cumulative Emissions, Carbon Intensity, and Tipping with Endogenous Transfers.} 
The figure reports key projections from the SOLG IAM over a 150-year horizon for the Pareto-improving policy that combines four tax instruments with an optimal inter-generational transfer scheme. 
\textit{Top left:} Endogenous emissions (mean and full distribution) under the joint optimal tax and transfer scheme. 
\textit{Top right:} Per-generation welfare gains and losses, in consumption equivalents, relative to the business-as-usual scenario. 
\textit{Middle left:} Global warming, projected to stabilise at a mean just below 3 °C (range: 2.3 °C – 3.5 °C). 
\textit{Middle right:} Economic damages as a percentage of GDP, averaging below 4 \% with a 99\textsuperscript{th} percentile of about 7 \%. 
\textit{Bottom left:} Evolution of the optimal linear tax rate on cumulative emissions, carbon intensity, and tipping over the 150-year horizon. 
\textit{Bottom right:} Corresponding evolution of the atmospheric carbon stock.}
\label{fig:lin_all_transfer_pareto}

\end{figure}

After obtaining the Pareto-optimal coefficients for the full linear tax on cumulative emissions, carbon intensity, and tipping, along with the associated transfer scheme, we simulate the economy–climate system and analyze the resulting outcomes.  Figure~\ref{fig:lin_all_transfer_pareto} summarizes the main results.

The top-left panel shows that mean emissions lie only modestly below the BAU path; the policy’s primary effect is to compress the upper tail.  The \(99^{\text{th}}\) percentile falls beneath \(40\;\text{GtCO}_{2}\) by 2070, roughly two decades earlier than under BAU. This mitigation translates into slightly improved climate outcomes (\textit{middle-left}): the global-mean temperature warming stabilizes just under \(3^{\circ}\text{C}\) on average, and the \(99^{\text{th}}\) percentile remains below \(3.5^{\circ}\text{C}\).

Lower warming curtails economic damages (middle-right). Average losses remain only marginally below BAU, yet extreme outcomes are noticeably truncated: the \(99^{\text{th}}\) percentile reaches about \(7\%\) of GDP, and the worst observed realisations stay near \(11\%\) (versus \(9\%\) and \(15\%\) in BAU).  The \(90^{\text{th}}\) percentile declines from roughly \(6\%\) to \(5\%\).

Welfare effects appear in the top-right panel. Transfers are calibrated so that cohorts already alive and those born within the next decade retain utilities close to their BAU levels; several even realize slight gains. Thereafter, welfare rises monotonically, exceeding \(1.4\%\) (in consumption-equivalent terms; cf. Equation~\eqref{eq:cons_equiv_variation}) for the youngest generation considered. Aggregating across cohorts yields a social-welfare gain of \(0.45\%\), only about \(7\%\) larger than the \(0.42\%\) gain achieved by the simpler cumulative-emissions tax. Hence, augmenting the tax base to include carbon intensity and tipping points provides only limited additional ``bang for the buck'' in this calibration, as documented in Table~\ref{tab:tax_instruments}.
%
\begin{table}[h!]
\centering
\begin{tabular}{lcc}
\toprule
 & \textbf{2 instruments} & \textbf{4 instruments} \\
\midrule
Welfare gain & 0.42\% & 0.45\% \\
\bottomrule
\end{tabular} 
\caption{\textbf{Aggregate welfare gains under Pareto-improving tax-and-transfer schemes with increasing policy complexity.}
The table reports the overall consumption-equivalent welfare improvement for two specifications:  
(i) a policy with two tax instruments, intercept and slope in cumulative emissions (Section~\ref{sec:cum_emissions_optimal_transfer});  
(ii) a policy with four instruments, taxes linear in cumulative emissions, carbon intensity, and tipping (Section~\ref{sec:emissions_carbon_intensity_damages_optimal_transfer}).}
\label{tab:tax_instruments}
\end{table}

The bottom-left and bottom-right panels plot the optimal tax rate and the implied atmospheric carbon stock, respectively, both of which are consistent with the moderate but effective mitigation profile described above.

\section{Conclusion}
\label{sec:conclusion}

Designing effective and politically feasible climate policy requires navigating complex trade-offs in a world characterized by uncertainty and heterogeneous agents. The computational burden of solving for optimal policies in such settings has traditionally limited quantitative analysis to simplified models. This paper confronts this challenge by developing and applying a novel, scalable, three-step computational framework capable of deriving constrained optimal tax and transfer rules in a stochastic overlapping generations model with climate risk. By integrating Deep Equilibrium Nets for global solutions and Gaussian Process surrogates for efficient welfare optimization, our methodology makes the systematic search for optimal policies computationally tractable.

Our application of this framework to a stochastic overlapping generations model with uncertain carbon intensity and climate tipping points yields several key insights. First, we confirm that an unconstrained, welfare-maximizing carbon tax, in our case, a simple linear function of cumulative emissions, can substantially reduce climate risk, stabilizing the average temperature increase at 2.7$^{\circ}$C compared to 3$^{\circ}$C under business-as-usual. However, this policy comes at a significant distributional cost, imposing welfare losses of up to 5\% on initial generations, thereby highlighting the political implementation challenge.

The primary contribution of our analysis is demonstrating a constructive path to overcoming this hurdle. By simultaneously optimizing the tax rule and an accompanying inter-generational transfer scheme, our framework successfully identifies a Pareto-improving policy. This policy guarantees that no generation is made worse off relative to the business-as-usual scenario, shielding initial cohorts from losses through carefully targeted transfers of tax revenue. While the resulting climate mitigation is more modest, this politically viable policy still achieves a notable social welfare gain of 0.42\% in consumption-equivalent terms and effectively truncates the worst-case climate and economic damage scenarios. This result underscores that the design of the transfer scheme is as critical as the design of the tax itself for creating implementable climate policy.

Furthermore, our analysis provides valuable insights into the optimal complexity of tax rules. When we expand the tax base from a simple function of cumulative emissions to a richer specification that also includes carbon intensity and proximity to tipping points, we find that the additional welfare gains are surprisingly small. The overall social welfare gain increases from 0.42\% to only 0.45\% in consumption-equivalent terms. This suggests that a simple, well-designed tax rule, when paired with an optimal transfer system, can capture the vast majority of achievable welfare gains. For policymakers, this implies that the focus may be more productively placed on the political economy of implementing simpler, robust rules and their associated transfers, rather than on designing ever-more-complex tax formulas that may offer little marginal benefit.

While our model is deliberately stylized to isolate key inter-generational mechanisms, the framework we propose is both general and extensible. Future research could incorporate richer features, such as intra-generational heterogeneity, a more detailed multi-sector energy transition, endogenous technological change, or the design of state-dependent transfer schemes to allow for a more ambitious initial tax path. Furthermore, a next step could also be to explore nonlinear policy rules, such as quadratic forms, to determine if they can unlock significant welfare gains where additional linear instruments showed diminishing returns. The methodology itself provides a template for tackling a wide range of previously intractable Ramsey problems in modern macroeconomics, from optimal fiscal and monetary policy in heterogeneous-agent models to the design of social insurance systems. By making the computation of constrained optimal policies in complex environments feasible, this work opens new avenues for quantitative policy analysis equipped to address the pressing economic challenges of the twenty-first century.

\newpage
\appendix
\section{Additional Model Details}

This appendix first supplies additional modelling details and then explains how the SOLG IAM is mapped onto the DEQN solution framework.  
Appendix~\ref{sec:appendix_recursive} sets out the recursive formulation of the household problem,  
Appendix~\ref{sec:Climate_module_appendix} elaborates on the climate emulator used in the SOLG,  
and Appendix~\ref{sec:DEQN_for_SOLG_IAM} describes the transformations that render the model DEQN-amenable.

\subsection{Recursive Formulation Of The Household Problem}
\label{sec:appendix_recursive}

In order to assess lifetime utility (cf. Section~\ref{sec:households}), we reformulate the household problem into a recursive problem:
\begin{equation}
    V_{t,j} = \max_{c_{t,j},a_{t+1,j}} \left\{ \left( \frac{C_{t,j}^{1-\sigma}}{1-\sigma} \right) + \beta \mathbb{E}_t V_{t+1,j+1} \right\} \quad \forall j \in 1,...,A-1
    \label{eq:Value_function}
\end{equation}
\begin{equation}
    V_{t,A} = \left( \frac{C_{t,A}^{1-\sigma}}{1-\sigma} \right)
    \label{eq:Value_function_oldest}
\end{equation}
subject to the budget constraint \eqref{eq:BC_main} initial and terminal conditions $a_{t,1} = 0$, $a_{t+1,A+1} = 0$.

\subsection{The Climate Module}
\label{sec:Climate_module_appendix}
Global emissions, $e_t$, in period $t$ are defined as
\begin{equation}
    e_t = (1 - \mu_t) \kappa_t K_t^{\alpha} L_t^{1-\alpha}.
\end{equation}
The stock of atmospheric carbon, $E_t$, and the atmospheric temperature, $T_t^{AT}$, then evolve according to
\begin{align}
    E_{t}      & = E_{t-1} + e_t / \text{C2CO2}, \\
    T_t^{AT} & = \sigma_{CCR} E_t,
\end{align}
where $\text{C2CO2} \approx 3.666$ is the conversion factor from carbon to CO$_2$ concentration and $\sigma_{CCR}$ is the transient climate response to cumulative carbon emissions (cf. Table~\ref{tab:Parameters}).

\subsection{Deep Equilibrium Nets for SOLG IAMs}
\label{sec:DEQN_for_SOLG_IAM}
In this appendix, we detail the procedure for mapping a stochastic, nonlinear, and non-stationary SOLG IAM presented 
in Section~\ref{sec:model} onto the DEQN framework. 
Recall that the state of the economy (for the optimal policy case) at time \( t \) is given by
\begin{equation}
  x_t = \bigl[
      t,\;
      TP_t,\;
      TP_{\text{reached},t},\;
      \kappa_t,\;
      \{a_{t,j}\}_{j=1}^{A},\;
      E_t,\;
      \vartheta
    \bigr]^{\top} \in \mathbb{R}^{A+5+d_\vartheta},
\label{eq:state_optimal}
\end{equation}
In the BAU case, the state vector simplifies to
\begin{equation}
  x_t^{BAU}= \bigl[
      t,\;
      TP_t,\;
      TP_{\text{reached},t},\;
      \kappa_t,\;
      \{a_{t,j}\}_{j=1}^{A},\;
      E_t
    \bigr]^{\top} \in \mathbb{R}^{A+5},
\end{equation}
as no tax-and-transfer schemes are present (\(\vartheta = \emptyset\)) and no mitigation occurs. This reduces the state space dimension by excluding the tax-and-transfer parameters and reflects the absence of policy interventions in the model.

To address the non-stationary nature of the IAM, as in~\cite{traeger4StatedDICEQuantitatively2014,Folini_2021}, we include time as a state variable and map the unbounded physical time \( t \in [0,\infty) \) to a computational time scale via the monotonic transformation
\begin{equation}
    t_{\text{comp}} = 1 - \exp(-\zeta t),
    \label{eq:time_trafo}
\end{equation}
where \( t_{\text{comp}} \in (0,1] \) and \( \zeta > 0 \) is a scaling parameter. The physical time is recovered using the inverse transformation
\begin{equation}
    t = -\frac{\ln(1 - t_{\text{comp}})}{\zeta}.
\end{equation}
In our numerical applications (cf. Section~\ref{sec:results}), we set $\zeta=0.015$ 
This transformation ensures computational tractability while preserving the model’s temporal dynamics.

To stabilize the training, we normalize consumption, asset holdings, and capital in effective labour units:
\begin{equation}
c_t = \frac{C_t}{L_t}, \hat{a}_{t,j} = \frac{a_{t,j}}{L_t}, k_t = \frac{K_t}{L_t}, \hat{l}_j = \frac{l_j}{L_t}, v_{t,j} = \frac{V_{t,j}}{L_t^{1/(1-\sigma)}}.
\end{equation}
Emissions are given by
\begin{equation}    
e_t = (1 - \mu_t) \kappa_t L_t k_t^{\alpha}. 
\end{equation}
The period utility functions (and therefore also the value functions) are scaled with a constant $B=20$:
\begin{equation}    
U(C_t) = U(B c_t).
\end{equation}
\vspace{-1.25em}
\paragraph{First-order Equations, and Loss Function for the BAU case.}
Given those normalizations, the first-order equations in the BAU case read as follows:
\begin{align}
    0 & = \frac{\left(\beta \mathbb{E}_t \left[ (1+r_{t+1}) c_{t+1,j+1}^{-\sigma}  \right]\right)^{-1/\sigma}}{c_{t,j}} - 1, \quad \forall j \in 1,...,A-1, \label{eq:EE_error_BAU}\\
0 & = c_{t,j}+\hat{a}_{t+1, j+1} - (1+r_t)\hat{a}_{t,j} -w_{t} \hat{l}_{j} , \quad \forall j \in 1,...,A,\label{eq:bc_error_BAU}\\
0 & = u_{t,j} + \beta  \mathbb{E}_t \left[ v_{t+1,j+1} \right] - v_{t,j} , \quad \forall j \in 1,...,A-1,\label{eq:value_error_BAU}
\end{align}
with the initial conditions $\hat{a}_{t,1} = 0$, and the terminal conditions $\hat{a}_{t+1,A+1} = 0$. 
In addition, we use the following definitions:
\begin{align}
k_t & = \sum_{j = 1}^{A} \hat{a}_{t,j},  \\
1  & = \sum_{j = 1}^{A} \hat{l}_j ,\\
r_t & =  \Omega_t T^{\mathrm{AT}}_t \alpha k_t^{\alpha-1}  - \delta, \\
w_t & =  \Omega_t T^{\mathrm{AT}}_t (1-\alpha) k_t^{\alpha} , \\
c_{t,j} & =  (1+r_t)\hat{a}_{t,j} + w_{t} \hat{l}_{j} -\hat{a}_{t+1, j+1} ,\\
u_{t,j}  & = c_{t,j}^{1-\sigma}{1-\sigma},\\ 
\sum_{j = 1}^{A} c_{t,j} + \sum_{j = 1}^{A-1} \hat{a}_{t+1,j}  & = \Omega_t T_{t}^{\mathrm{AT}} \left( k_t^{\alpha} \right) + (1- \delta) k_t.
\end{align}
%

The (normalized) policy function $P(x_t)$ we intend to approximate with the aid of deep neural networks is given by:
\begin{equation}
  \mathcal{N}(x_t^{BAU})
  \;=\;
  \bigl[
      \{\hat{a}_{t+1,j+1}, v_{t,j}\}_{j=1}^{A-1}
    \bigr]^{\top}
  \in \mathbb{R}^{2(A-1)}.
  \label{eq:NN_approx_appendix}
\end{equation}

Having presented the preceding definitions, we can now specify the individual terms that constitute the loss function used by the DEQN algorithm:
\begin{align}
    l_1, ..., l_{11}:= &  \frac{\left(\beta \mathbb{E}_t \left[ (1+r_{t+1}) c_{t+1,j+1}^{-\sigma} \right]\right)^{-1/\sigma}}{c_{t,j}} - 1  \quad \forall j \in \{1,...,11\}, \label{eq:EE_error_BAU_ERROR}\\
l_{12}:= & c_{t,1}+\hat{a}_{t+1, 2} -w_{t} \hat{l}_{1} ,\label{eq:bc_error_BAU_2}\\
l_{13}, ..., l_{22}:= & c_{t,j}+\hat{a}_{t+1, j+1} - (1+r_t)\hat{a}_{t,j} -w_{t} \hat{l}_{j} \quad \forall j \in \{2,...,11\} ,\label{eq:bc_error_BAU_3}\\
l_{23}:= & c_{t,12} - (1+r_t)\hat{a}_{t,12} -w_{t} \hat{l}_{12} ,\label{eq:bc_error_BAU_4}\\
l_{24}, ..., l_{34}:= & \frac{ u_{t,j} + \beta  \mathbb{E}_t \left[ v_{t+1,j+1} \right] }{v_{t,j}}- 1 \quad \forall j \in \{1,...,11\}. \label{eq:value_error_BAU_ERROR}
\end{align}
Taken together, these expressions result in the total loss function, that is,
\begin{equation}
    \ell_{\nu}^{BAU} := \frac{1}{N_{\text{path length}}} \sum_{{x}^{BAU}_t \in \text{sim. path}} \sum_{m=1}^{N_{\text{eq}} = 34} \left( \ell_m({x}^{BAU}_t, \mathcal{N}({x}^{BAU}_t) \right)^2.
    \label{eq:total_BAU_LOSS}
\end{equation}

The final ingredient required by the DEQN algorithm is the one–step state transition for the BAU economy, which allows the loss
function~\eqref{eq:total_BAU_LOSS} to be evaluated along simulated
trajectories, together with the initial state at \(t=0\)
(cf.\ Section~\ref{sec:DEQN}).  In the BAU SOLG, the forward mapping for the DEQN simulation is given by
\begin{equation}
  x_{t+1}^{\text{BAU}}
  =[
      t+1,\,
      TP_{t+1},\,
      TP_{\text{reached},t+1},\,
      \kappa_{t+1},\,
      \{a_{t+1,j}\}_{j=1}^{A},\,
      E_{t+1}
    ]^T.
    \label{eq:Bau_LOM}
\end{equation}
All simulations start in the year 2015 (\(t=0\)), so the initial state vector is
\begin{equation}
  \label{eq:x0_display}
  x_0^{BAU}
  =[
      0,\,
      3.0,\,
      0,\,
      0.35032,\,
      \{a_{0,j}\}_{j=1}^{A},\,
      0.851
    ]^T,
\end{equation}
corresponding to
\begin{equation}
      [t,\;TP_0,\;TP_{\text{reached},0},\;\kappa_0,\;
    \{a_{0,j}\},\;E_0]^T .
\end{equation}
The endogenous state variables are rolled forward according to
\begin{align}
  \hat{a}_{t+1,j} &\Rightarrow \hat{a}_{t,j+1}, \\
  k_{t+1}         &\Rightarrow k_t,             \\
  E_{t+1}         &\Rightarrow E_t,             \\
  T_{t+1}^{\mathrm{AT}} &\Rightarrow T_{t}^{\mathrm{AT}},
\end{align}
where the ``$\Rightarrow$'' indicates that a choice variable at step $t$ becomes an endogenous state at $t+1$.
The initial asset distribution is obtained from a deterministic simulation without climate change, and provided in Table~\ref{tab:init_assets}.\footnote{The initial asset distribution for our simulations is the non-stochastic steady state derived from a simplified, deterministic version of our BAU model in which temperature and damages are held constant at their $t=0$ levels (see Equation~\eqref{eq:x0_display}).}
\begin{table}[h!]
  \centering
  \begin{tabular}{lcccccccccccc}
    \toprule
    \(a_{0,1}\) & \(a_{0,2}\) & \(a_{0,3}\) & \(a_{0,4}\) & \(a_{0,5}\) & \(a_{0,6}\) &
    \(a_{0,7}\) & \(a_{0,8}\) & \(a_{0,9}\) & \(a_{0,10}\) & \(a_{0,11}\) & \(a_{0,12}\) \\
    \midrule
    0 & 0.002 & 0.009 & 0.021 & 0.037 & 0.056 & 0.076 & 0.095 & 0.111 & 0.093 & 0.070 & 0.039 \\
    \bottomrule
  \end{tabular}
  \caption{Initial asset distribution obtained from a deterministic simulation without climate change.}
  \label{tab:init_assets}
\end{table}

\vspace{-1.25em}
\paragraph{First-order Equations, and Loss Function for the SOLG with Policy.}

Next, we derive the input necessary for the DEQN for the SOLG model with taxes and transfers. The first-order equations in this case read as follows:
\begin{align}
0 & = \frac{\left(\beta \mathbb{E}_t \left[ (1+r_{t+1}) c_{t+1,j+1}^{-\sigma}  \right]\right)^{-1/\sigma}}{c_{t,j}} - 1 , \quad \forall j \in 1,...,A-1 ,\label{eq:ee_error_tax}\\
0 & = c_{t,j}+\hat{a}_{t+1, j+1} - (1+r_t)\hat{a}_{t,j} -w_{t} \hat{l}_{j} - \mathbb{T}_{t,j}, \quad \forall j \in 1,...,A ,\label{eq:BC_error_tax} \\
0 & = u_{t,j} + \beta  \mathbb{E}_t \left[ v_{t+1,j+1} \right] - v_{t,j} , \quad \forall j \in 1,...,A-1,\label{eq:value_error_tax} 
\end{align}
with initial conditions $\hat{a}_{t,1} = 0$, and terminal conditions $\hat{a}_{t+1,A+1} = 0$. 
In addition, we use the following definitions:
\begin{align}
k_t & = \sum_{j = 1}^{A} \hat{a}_{t,j}, \\
1 & = \sum_{j = 1}^{A} \hat{l}_j ,  \\
r_t & =  \left(  \Omega_t T^{AT}_t) (1-\theta_1 \mu_t^{\theta_2}) - \tau_t \kappa_t (1-\mu_t) \right) \alpha  k_t^{\alpha-1}  - \delta ,\\
w_t & =  \left(  \Omega_t T^{\mathrm{AT}}_t (1-\theta_1 \mu_t^{\theta_2}) - \tau_t \kappa_t (1-\mu_t)  \right) (1-\alpha)  k_t^{\alpha} ,  \\
\sum_{j = 1}^{A} c_{t,j} + \sum_{j = 1}^{A-1} \hat{a}_{t+1,j}  & = \Omega_t T_{t}^{\mathrm{AT}} (1-\theta_1\mu_t^{\theta_2}) k_t^{\alpha}+ (1- \delta) k_t.
\end{align}
As before, the DEQN approximates the normalized policies stated in Equation~\eqref{eq:NN_approx_appendix}. 
The individual components entering the loss function for training the DEQN read as follows:
\begin{align}
    l_1, ..., l_{11}:= &  \frac{(\beta \mathbb{E}_t \left[ (1+r_{t+1}) c_{t+1,j+1}^{-\sigma} \right])^{-1/\sigma}}{c_{t,j}} - 1  \quad \forall j \in \{1,...,11\} ,\label{eq:EE_error_optimal_ERROR}\\
l_{12}:= & c_{t,1}+\hat{a}_{t+1, 2} -w_{t} \hat{l}_{1} - \mathbb{T}_{t,1},\label{eq:bc_error_BAU_5}\\
l_{13}, ..., l_{22}:= & c_{t,j}+\hat{a}_{t+1, j+1} - (1+r_t)\hat{a}_{t,j} -w_{t} \hat{l}_{j} - \mathbb{T}_{t,j} \quad \forall j \in \{2,...,11\} ,\label{eq:bc_error_BAU_6}\\
l_{23}:= & c_{t,12} - (1+r_t)\hat{a}_{t,12} -w_{t} \hat{l}_{12} - \mathbb{T}_{t,12} ,\label{eq:bc_error_BAU_7}\\
l_{24} ... ,l_{34}:= & \frac{ u_{t,j} + \beta  \mathbb{E}_t \left[ v_{t+1,j+1} \right] }{v_{t,j}}- 1 \quad \forall j \in \{1,...,11\} ,\label{eq:value_error_optimal_ERROR}
\end{align}
and result in the total loss function given by
\begin{equation}
    \ell_{\nu} := \frac{1}{N_{\text{path length}}} \sum_{{x}_t \in \text{sim. path}} \sum_{m=1}^{N_{\text{eq}} = 34} \left( \ell_m({x}_t, \mathcal{N}({x}_t)) \right)^2,
    \label{eq:loss_optimal_DEQN}
\end{equation}
where the state $x_t$ is now given by expression~\eqref{eq:state_optimal}, which encompasses the parameters for the tax and transfer schemes.
Recall that, in the present SOLG, the state vector is partitioned as \(x_t=[x_{\text{state},t},\vartheta]^{\top}\).
The law of motion for \(x_{\text{state},t+1}\) is therefore identical to the BAU specification in Equation~\eqref{eq:Bau_LOM}.
The corresponding initial state \(x_{\text{state},0}\) is given by Equation~\eqref{eq:x0_display} together with Table~\ref{tab:init_assets}.

Updating the pseudo–state vector \(\vartheta\) within the DEQN framework proceeds as follows.  
We launch \(N = 512\) parallel simulations, all starting from the same economic–climate state \(x_{\text{state},0}\) but with distinct draws \(\{\vartheta_i\}_{i=1}^{N}\) from the respective distribution (cf. Section~\ref{sec:results}). Each simulation is then propagated for \(N_{\texttt{path length}} = 70\) periods under the current tax and transfer scheme.  
After these trajectories are completed and the network weights have been updated, a new batch of parameter draws is generated, and the process is repeated, \(\vartheta\) remains fixed along any single trajectory but is resampled before every training episode (cf.\ Section~\ref{sec:DEQN}).

\vspace{-1.25em}
\paragraph{Hyperparameters.}

Table~\ref{tab:hyperparams_DEQN} summarises the hyperparameters used in all numerical experiments reported in Section~\ref{sec:results}.  
The entry “Parallel simulations” indicates that \(512\) independent trajectories, each of length \(70\), are generated per training episode, yielding \(512 \times 70 = 35,840\) state transitions before the policies are updated and a new batch of data is produced.
\begin{table}[h]
  \centering
  \begin{tabular}{@{}ll@{}}
    \toprule
    \textbf{Hyperparameter} & \textbf{Value} \\ \midrule
    Optimiser & Adam \\
    Initial learning rate & \(10^{-4}\) \\
    Minibatch size & 64 \\
    Hidden layers & 2 \\
    Neurons per hidden layer & 512 \\
    Activation function & GELU \\
    Path length ($N_{\texttt{path length}}$ ) & 70 \\
    Parallel simulations & 512 \\ \bottomrule
  \end{tabular}
  \caption{Hyper-parameters of the DEQN training routine.}
  \label{tab:hyperparams_DEQN}
\end{table}

\vspace{-1.25em}
\paragraph{DEQN runtimes.}

Despite the complexity of the high-dimensional, non-stationary SOLG, DEQN delivers global solutions in remarkably short wall-clock times.  
On readily available consumer hardware, such as a laptop equipped with an Apple M1 chip, the full training procedure completes in approximately four hours.  
Running the same code on a high-end accelerator, such as an NVIDIA GH200 Grace Hopper GPU, cuts the runtime to the order of minutes, well below one hour, demonstrating that even modest resources suffice for timely solutions, while specialized hardware can accelerate the process dramatically.


\section{Domain of the Pseudo-States}
\label{appendix:sampling}

Because the SOLG IAM is non-stationary, it is non-trivial to specify a priori a computational domain $[a,b]\subset\mathbb{R}^{d_\vartheta}$ for the pseudo-state vector $\vartheta$ over which we build our DEQN and GP surrogate models. The remainder of this appendix details our procedure: Appendix~\ref{appendix:Tax_params_sampling} derives model-consistent bounds for the tax-function parameters, whereas Appendix~\ref{appendix:transfer_share_sampling} establishes the corresponding limits for the optimal transfers.

\subsection{Domain of the Tax Parameters}
\label{appendix:Tax_params_sampling}

We impose economically meaningful bounds on the tax function, guided by the permissible range of the mitigation effort, $0 \leq \mu_t \leq 1$. First, we enforce a non-negativity constraint on the tax rate ($\tau_t \geq 0$), as negative taxes would effectively subsidize pollution. Second, since mitigation cannot exceed 100\% ($\mu_t \leq 1$), any tax rate beyond the level required to achieve full mitigation is economically redundant (see Equation~\eqref{eq:FOC_mu}). Together, these two restrictions define the following permissible domain for the tax rate $\tau_t$, which holds for all models:
\begin{equation}
    \label{eq:tax_main_constraint}
    0 \leq \tau_t(\vartheta) \leq \frac{\Omega_t(T^{AT}_t) \theta_1 \theta_2}{\kappa_t} \quad \forall \; t.
\end{equation}

To avoid unnecessary computational costs in the DEQN algorithm (cf. Section~\ref{sec:DEQN_for_SOLG_IAM}), we constrain the parameter space to exclude regions that would produce economically implausible tax policies. For example, since our framework rules out negative taxes (which would subsidize pollution), parameter combinations that yield negative tax rates are computationally wasteful to explore.

Therefore, we employ a strategy that imposes joint constraints on the tax parameters, $\vartheta$, based on the implied tax rates at the beginning and end of the relevant simulation horizon. Specifically, we enforce bounds on the tax rate, $\tau_t(\vartheta)$, at the initial period ($t=0$) and at the final period relevant for welfare evaluations ($t=29$). This is achieved by evaluating the tax function at heuristic values for the state variables in those two periods, with the state values themselves being informed by simulations of the BAU scenario.

The constraints require the implied tax rate at both $t=0$ and $t=29$ to lie within a plausible range $[0, \bar{\tau}]$. This gives rise to the following bounds:

\vspace{-1.25em}
\paragraph{Lower Bound.}
The constraint $\tau_t(\vartheta) \geq 0$ for $t \in \{0, 29\}$ ensures that we discard from the outset any parameter set that implies negative taxes at the start or end of the horizon. The rationale is that a plausible tax path should not begin or end with subsidies for pollution.

\vspace{-1.25em}
\paragraph{Upper Bound.}
Similarly, the constraint $\tau_t(\vartheta) \leq \bar{\tau}$ for $t \in \{0, 29\}$ discards parameter sets that would result in excessively high taxes at these points. The rationale is that an optimal tax path is unlikely to start or end at an extreme level.

\vspace{-1.25em}
\paragraph{Validation.}
This strategy effectively prunes the parameter space without sacrificing relevant solutions. We validate this approach post-hoc, as the optimal tax policies identified in our results (Section~\ref{sec:results}) are never constrained by these ex-ante bounds.

\vspace{-1.25em}
\paragraph{Tax-parameter sampling procedure.}
We generate candidate parameter vectors for the tax rule in three steps.  

\emph{(i) Uniform draw.}  
Each coefficient is drawn independently from a uniform distribution over the bounds reported in Table~\ref{tab:param_bounds_main}, which the modeler has to choose ex-ante.  
The intervals are deliberately wide so that, during the subsequent optimisation, interior solutions are never pushed against these limits; any binding restriction should arise only from the economic constraints (e.g.\ the requirement that $\tau_0>0$), not from the sampling range itself.  

\emph{(ii) Feasibility filter.}  
A draw is retained only if the implied tax rates in the first and last simulated periods lie in economically meaningful intervals,
\begin{equation}    
  0 \;\le\; \tau_0(\vartheta) \;\le\; \bar{\tau},
  \qquad
  0 \;<\; \tau_{29}(\vartheta) \;\le\; \bar{\tau},
\end{equation}
where the common upper bound $\bar{\tau}$ is set high enough that it never becomes binding in the optimisation.  
To calibrate the bound for the final period, $\tau_{29}$, we evaluate the tax function at representative state values informed by the BAU scenario. For instance, in the model with three tax instruments, these heuristic state values are $E_{29}=1.6$, $\kappa_{29}=0.05$, and $\mathbb{D}_{TP,29}=1$. This check allows us to discard parameter draws that would imply economically implausible taxes, such as those that are redundant (i.e., full abatement is already achieved) or would lead to negative emissions.

\emph{(iii) Objective-specific caps.}  
The value of $\bar{\tau}$ depends on the planner’s objective.  
When the search is restricted to Pareto-improving policies, the admissible tax range is typically narrower than in exercises that allow some generations to incur welfare losses.  
The exact bounds used for each model studied in Section~\ref{sec:results} are listed in Appendix~\ref{sec:domain_taxes}, Table~\ref{tab:param_bounds_main}.

\subsubsection{Heuristic Range of the Tax Parameters}
\label{sec:domain_taxes}

Table~\ref{tab:param_bounds_main} provides the permissible ranges for the pseudo-states that parameterize the various tax functions assumed in Section~\ref{sec:results}.

\begin{table}[ht!]
\centering
\begin{tabular}{lllll}
\toprule
Model & Planner Target & Parameter & $\underline{\vartheta_i}$ & $\bar{\vartheta_i}$ \\
\midrule
Linear in E & Max. Welfare & $\vartheta_0$     & -2.0   & 2.0   \\
            & Max. Welfare & $\vartheta_E$     & -2.0   & 2.0  \\
           & Max. Welfare & $\tau$         & 0.0  & 1.5  \\
            \midrule
Linear in E + Transfers & Pareto & $\vartheta_0$ & -0.5 & 0.1 \\
                        & Pareto & $\vartheta_E$ & 0.0    & 0.5 \\
                        & Pareto & $\tau$         & 0.0  & 0.8  \\
\midrule
Full linear + Transfers & Pareto & $\vartheta_0$     & -1.0   & 0.5 \\
                        & Pareto & $\vartheta_E$     & -0.5 & 1.0   \\
                        & Pareto & $\vartheta_\kappa$ & -0.6 & 0.6 \\
                        & Pareto & $\vartheta_{TP}$   & -1.0   & 1.0   \\
                        & Pareto & $\tau$         & 0.0  & 0.8  \\
\bottomrule
\end{tabular}
\caption{Parameter bounds across models and objectives. }
\label{tab:param_bounds_main}
\end{table}

\subsection{Range of Transfer Parameters}
\label{appendix:transfer_share_sampling}

The vector of transfer shares, $\mathbb{T} = (\mathbb{T}_1, \dots, \mathbb{T}_{12})^\top$, is sampled from the unit simplex using a Dirichlet distribution, $\mathbb{T} \sim \text{Dir}(\boldsymbol{\alpha})$. This formulation naturally enforces the constraints that all shares are non-negative ($\mathbb{T}_j \geq 0$) and sum to one, that is, $\sum_{j=1}^{12} \mathbb{T}_j = 1$. The vector $\boldsymbol{\alpha} \in \mathbb{R}^{12}_{++}$ contains the concentration parameters that shape the distribution.

A standard choice without prior knowledge would be a uniform distribution over the simplex (i.e., setting $\alpha_j = 1$ for all $j$). However, this approach is inefficient for our purposes, as it generates highly skewed marginal distributions for each share ($\mathbb{T}_j \sim \text{Beta}(1, 11)$), where large transfers to any single generation are rarely sampled. This poor exploration of the policy space's boundaries can destabilize the DEQN training and hinders the discovery of potentially optimal, highly targeted transfer schemes.

To ensure both stable training and efficient exploration, we adopt a multi-stage sampling strategy. Initially, we use concentration parameters that encourage more uniform draws, preventing any single share from dominating early in the training. For the final analysis, we switch to a more targeted sampling distribution, using the calibrated concentration parameters $\boldsymbol{\alpha}$ shown in Table~\ref{tab:dirichlet_sampling_range}. These values were informed by preliminary optimization runs to focus the search on regions of the policy space likely to yield Pareto improvements.
\begin{table}[h!]
\centering
\begin{tabular}{lcccccccccccc}
\toprule
\textbf{Gen} & \textbf{1} & \textbf{2} & \textbf{3} & \textbf{4} & \textbf{5} & \textbf{6} & \textbf{7} & \textbf{8} & \textbf{9} & \textbf{10} & \textbf{11} & \textbf{12} \\
\midrule
$\boldsymbol{\alpha}$ & 1.33 & 1.25 & 1.22 & 0.61 & 2.7 & 1.83 & 0.61 & 1.79 & 1.16 & 0.65 & 0.67 & 1.08 \\
\bottomrule
\end{tabular}
\caption{Dirichlet concentration parameters ($\boldsymbol{\alpha}$) for transfer share sampling to find Pareto-improving policies.}
\label{tab:dirichlet_sampling_range}
\end{table}

\section{Details on Approximating Social Welfare Functions}
\label{sec:Appendix_Soc_welfare}

This appendix describes the construction of the high-accuracy GP surrogates used to identify the SWF-optimal tax and transfer parameters.

Appendix~\ref{sec:BAL_text} gives a concise introduction to Bayesian active learning, a reinforcement-learning technique that efficiently selects training points.  Appendices~\ref{sec:HP_selection_GPs} and~\ref{sec:HP_Pareto_GP} then document the kernel choices and hyper-parameter settings that generate the numerical results reported in Section~\ref{sec:results}.

\subsection{Bayesian Active Learning for Training Point Selection}
\label{sec:BAL_text}

When training data is expensive to generate, as is the case for us (cf. Section~\ref{sec:SWF_objective_sim}), an efficient sampling strategy is required to build an accurate GP surrogate with a minimal number of training points. While uniform sampling (e.g., via grids or quasi-random sequences) covers the parameter space, it may inefficiently place points in regions where the function is already well-approximated. Bayesian Active Learning (BAL) offers a principled solution to this problem by sequentially selecting the most informative points for evaluation.

The goal of BAL is to iteratively augment the training dataset, $\mathcal{D}_n = \{(\vartheta_i, y_i)\}_{i=1}^{n}$, in a way that most rapidly improves the surrogate model and aids in finding the global maximum of the objective function. 
The BAL process follows an iterative loop:

\begin{enumerate}
    \item \textbf{Train the Surrogate:} Given the current training dataset $\mathcal{D}_n$, we train a GP surrogate. This yields a predictive mean function $\mu_n(\vartheta)$ (our best estimate of the objective function) and a predictive variance function $\sigma_n^2(\vartheta)$ (our uncertainty about that estimate).

    \item \textbf{Guide the Search:} We use an \textbf{acquisition function}, $\alpha(\vartheta)$, to decide which point to evaluate next. This function is cheap to compute and uses the surrogate's mean and variance to score every potential candidate point $\vartheta$ in the parameter space $[a,b] \subset \mathbb{R}^{d_\vartheta}$.

    \item \textbf{Select the Next Point:} The next point to be evaluated with the expensive simulation, $\vartheta_{n+1}$, is the one that maximizes the acquisition function:
    \begin{equation}
        \vartheta_{n+1} = \arg\max_{\vartheta \in [a,b]} \alpha(\vartheta).
        \label{eq:argmax_alpha}
    \end{equation}

    \item \textbf{Evaluate and Augment:} We perform one expensive evaluation of the true function to obtain $y_{n+1} = f(\vartheta_{n{+1}})$. This new pair is added to the training set to create an augmented dataset, $\mathcal{D}_{n+1} = \mathcal{D}_n \cup \{(\vartheta_{n+1}, y_{n+1})\}$.
\end{enumerate}

These steps are repeated until a stopping criterion is met, such as reaching a predefined budget of function evaluations.

\paragraph{Our Practical Implementation: The Upper Confidence Bound Acquisition Function.}

The acquisition function drives the learning process by balancing two opposing objectives. Exploitation steers evaluations toward parameter vectors where the surrogate already predicts a high objective value, sharpening our estimate of the optimum.  Exploration, in contrast, targets regions with large predictive uncertainty, thereby reducing overall error and preventing premature convergence to a local optimum.  A well-designed acquisition rule blends these forces in a single score, favouring points that are either promising in expected performance or highly informative because of their uncertainty.

A simple and intuitive example that formalizes this trade-off and which we use in our numerical experiments (cf. Section~\ref{sec:results}) is the Upper Confidence Bound (UCB) acquisition function:
\begin{equation}
    \alpha_{\text{UCB}}(\vartheta) = \alpha_{UCB}\mu_n(\vartheta) + \kappa_{UCB} \sigma_n(\vartheta).
    \label{eq:ucb}
\end{equation}
Here, the first term, $\mu_n(\vartheta)$, encourages exploitation, while the second term, $\sigma_n(\vartheta)$ (the predictive standard deviation), encourages exploration. The user-set parameters $ \alpha_{UCB}$ and $\kappa \geq 0$ control this trade-off. By selecting the point with the highest ``optimistic'' estimate of the objective, the UCB algorithm efficiently steers the parameter space to find the global maximum.
In our practical applications (cf. Section~\ref{sec:results}), the BAL procedure is implemented as follows. We begin with an initial training set of 450 samples drawn from the parameter space $[a,b] \subset \mathbb{R}^{d_\vartheta}$, on which we train an initial GP. At each iteration of the active learning loop, we set $ \alpha_{UCB} =1,  \kappa_{UCB}=100$, and evaluate the acquisition function from Equation~\eqref{eq:ucb} for a candidate set of 1,000 new points, drawn uniformly from the same domain. The candidate point with the highest acquisition score is then added to the training set, and the GP is re-trained with the augmented data. This iterative process is repeated until the surrogate's leave-one-out cross-validation error falls below a threshold of $10^{-4}$. Further details on the BAL implementation specific to each SOLG IAM are provided in Appendix~\ref{sec:HP_selection_GPs}.


\subsection{Hyperparameters: Welfare-improving Policies}
\label{sec:HP_selection_GPs}

For the welfare-improving surrogate required in Section~\ref{sec:cum_emissions}, we employ a zero-mean Gaussian process with a Matérn-$5/2$ kernel and automatic-relevance detection, so that each policy parameter receives its own length–scale.  
The initial training design comprises \(450\) parameter vectors drawn uniformly from the admissible hyper-cube (cf.\ Appendix~\ref{sec:domain_taxes}).  
BAL with \(\alpha_{\text{UCB}}=1\) and \(\kappa_{\text{UCB}}=100\) then adds another \(50\) information-rich points, bringing the final training set to \(500\).

Kernel hyper-parameters are estimated by maximising the log marginal likelihood with \texttt{scipy.optimize.minimize} (SLSQP) using \(500\) random restarts to avoid poor local optima.  
The leave-one-out cross-validation error is typically \(\mathcal{O}(10^{-4})\); for the baseline model discussed in Section~\ref{sec:cum_emissions}, it falls to \(3\times10^{-5}\).  
In a sweep over ten feed-forward neural networks (best architecture: two hidden layers with~128 GeLU units, Adam optimizer, learning rate~\(10^{-4}\)), the GP achieved comparable out-of-sample accuracy with roughly ten times fewer model evaluations and still yielded slightly lower prediction errors.

\subsection{Hyperparameters: Pareto-improving Policies}
\label{sec:HP_Pareto_GP}

In the models with Pareto-improving tax and transfer schemes, 
we approximate the 40 GPs per QoI to form the SWF surrogate also 
with a Matérn-$5/2$ kernel and automatic-relevance detection, and the same BAL setting as described before in Appendix~\ref{sec:HP_selection_GPs}. 
Since those models (cf. Sections~\ref{sec:cum_emissions_optimal_transfer} and~\ref{sec:emissions_carbon_intensity_damages_optimal_transfer} are slightly more complex than the welfare-improving case, larger training sets are required. Those vary between 500 (cf. Section~\ref{sec:cum_emissions_optimal_transfer}) and 800 (cf. Section~\ref{sec:emissions_carbon_intensity_damages_optimal_transfer}) until the desired leave-one-out error has been reached.
As before, the kernel hyper-parameters are estimated by maximising the log marginal likelihood with \texttt{scipy.optimize.minimize} (SLSQP) using \(500\) random restarts to avoid poor local optima.

\newpage
\bibliography{bib_econ.bib}

\begin{thebibliography}{48}
\providecommand{\natexlab}[1]{#1}
\providecommand{\url}[1]{\texttt{#1}}
\expandafter\ifx\csname urlstyle\endcsname\relax
  \providecommand{\doi}[1]{doi: #1}\else
  \providecommand{\doi}{doi: \begingroup \urlstyle{rm}\Url}\fi

\bibitem[Azinovic et~al.(2022)Azinovic, Gaegauf, and
  Scheidegger]{azinovicDEEPEQUILIBRIUMNETS2022}
Marlon Azinovic, Luca Gaegauf, and Simon Scheidegger.
\newblock {{DEEP EQUILIBRIUM NETS}}.
\newblock \emph{International Economic Review}, 63\penalty0 (4):\penalty0
  1471--1525, November 2022.
\newblock ISSN 0020-6598, 1468-2354.
\newblock \doi{10.1111/iere.12575}.

\bibitem[Barnett et~al.(2020)Barnett, Brock, and Hansen]{Barnett2020}
Michael Barnett, William Brock, and Lars~Peter Hansen.
\newblock {Pricing uncertainty induced by climate change}.
\newblock \emph{Review of Financial Studies}, 33\penalty0 (3):\penalty0
  1024--1066, 2020.
\newblock ISSN 14657368.
\newblock \doi{10.1093/rfs/hhz144}.

\bibitem[Bellman(1961)]{bellman1961adaptive}
R.~Bellman.
\newblock \emph{Adaptive Control Processes: A Guided Tour}.
\newblock 'Rand Corporation. Research studies. Princeton University Press,
  1961.

\bibitem[Benzell et~al.(2015)Benzell, Goryunov, Kazakova, Kotlikoff, LaGarda,
  Nesterova, and Zubarev]{benzell2015simulating}
Seth~G Benzell, Eugene Goryunov, Maria Kazakova, Laurence~J Kotlikoff,
  Guillermo LaGarda, Kristina Nesterova, and Andrey Zubarev.
\newblock Simulating russia's and other large economies' challenging and
  interconnected transitions.
\newblock Technical report, National Bureau of Economic Research, 2015.

\bibitem[Bhandari et~al.(2013)Bhandari, Evans, Golosov, and
  Sargent]{bhandari2013}
Anmol Bhandari, David Evans, Mikhail Golosov, and Thomas~J Sargent.
\newblock Taxes, debts, and redistributions with aggregate shocks.
\newblock Technical report, National Bureau of Economic Research, 2013.

\bibitem[Brumm et~al.(2017)Brumm, Kubler, and Scheidegger]{BKS1}
Johannes Brumm, Felix Kubler, and Simon Scheidegger.
\newblock \emph{Computing Equilibria in Dynamic Stochastic Macro-Models with
  Heterogeneous Agents}, volume~2 of \emph{Econometric Society Monographs},
  pages 185--230.
\newblock Cambridge University Press, 2017.

\bibitem[Burton(1993)]{Burton1993}
Peter~S. Burton.
\newblock {Intertemporal preferences and intergenerational equity
  considerations in optimal resource harvesting}.
\newblock \emph{Journal of Environmental Economics and Management}, 24\penalty0
  (2):\penalty0 119--132, 1993.
\newblock ISSN 0095-0696.
\newblock \doi{10.1006/jeem.1993.1008}.
\newblock URL
  \url{http://www.sciencedirect.com/science/article/B6WJ6-45P13FR-1D/2/3df3a733c83612ef9a6056ec25a0c47c}.

\bibitem[Cai and Lontzek(2018)]{doi:10.1086/701890}
Yongyang Cai and Thomas Lontzek.
\newblock The social cost of carbon with economic and climate risks.
\newblock \emph{Journal of Political Economy}, 2018.
\newblock \doi{10.1086/701890}.
\newblock URL \url{https://doi.org/10.1086/701890}.

\bibitem[Chen et~al.(2025)Chen, Didisheim, and Scheidegger]{chen2025Deep}
Hui Chen, Antoine Didisheim, and Simon Scheidegger.
\newblock Deep {{Surrogates}} for {{Finance}}: {{With}} an {{Application}} to
  {{Option Pricing}}.
\newblock \emph{Journal of Financial Economics}, 2025.
\newblock \doi{10.2139/ssrn.3782722}.
\newblock Forthcoming.

\bibitem[Dietz(2024)]{DIETZ20241}
Simon Dietz.
\newblock Chapter 1 - introduction to integrated assessment modeling of climate
  change.
\newblock volume~1 of \emph{Handbook of the Economics of Climate Change}, pages
  1--51. North-Holland, 2024.
\newblock \doi{https://doi.org/10.1016/bs.hesecc.2024.10.002}.
\newblock URL
  \url{https://www.sciencedirect.com/science/article/pii/S3050729424000023}.

\bibitem[Dietz and Asheim(2012)]{DIETZ2012321}
Simon Dietz and Geir~B. Asheim.
\newblock Climate policy under sustainable discounted utilitarianism.
\newblock \emph{Journal of Environmental Economics and Management}, 63\penalty0
  (3):\penalty0 321--335, 2012.
\newblock ISSN 0095-0696.
\newblock \doi{https://doi.org/10.1016/j.jeem.2012.01.003}.
\newblock URL
  \url{https://www.sciencedirect.com/science/article/pii/S0095069612000058}.

\bibitem[Dietz and Venmans(2019)]{dietz2019cumulative}
Simon Dietz and Frank Venmans.
\newblock Cumulative carbon emissions and economic policy: in search of general
  principles.
\newblock \emph{Journal of Environmental Economics and Management},
  96:\penalty0 108--129, 2019.

\bibitem[Dietz et~al.(2021)Dietz, van~der Ploeg, Rezai, and
  Venmans]{doi:10.1086/713977}
Simon Dietz, Frederick van~der Ploeg, Armon Rezai, and Frank Venmans.
\newblock Are economists getting climate dynamics right and does it matter?
\newblock \emph{Journal of the Association of Environmental and Resource
  Economists}, 8\penalty0 (5):\penalty0 895--921, 2021.
\newblock \doi{10.1086/713977}.
\newblock URL \url{https://doi.org/10.1086/713977}.

\bibitem[Douenne et~al.(2024)Douenne, Hummel, and
  Pedroni]{douenne_hummel_pedroni_2024}
Thomas Douenne, Albert~Jan Hummel, and Marcelo Pedroni.
\newblock Optimal fiscal policy in a climate-economy model with heterogeneous
  households.
\newblock CEPR Discussion Paper 19151, 2024.
\newblock URL \url{https://cepr.org/publications/dp19151}.

\bibitem[Dyrda and Pedroni(2023)]{dyrda2023optimal}
Sebastian Dyrda and Marcelo Pedroni.
\newblock Optimal fiscal policy in a model with uninsurable idiosyncratic
  income risk.
\newblock \emph{The Review of Economic Studies}, 90\penalty0 (2):\penalty0
  744--780, 2023.

\bibitem[Eftekhari et~al.(2024)Eftekhari, Folini, Friedl, K{\"u}bler,
  Scheidegger, and Schenk]{eftekhari2024building}
Aryan Eftekhari, Doris Folini, Aleksandra Friedl, Felix K{\"u}bler, Simon
  Scheidegger, and Olaf Schenk.
\newblock Building interpretable climate emulators for economics.
\newblock \emph{arXiv preprint arXiv:2411.10768}, 2024.

\bibitem[Elsken et~al.(2019)Elsken, Metzen, and Hutter]{elsken2019neural}
Thomas Elsken, Jan~Hendrik Metzen, and Frank Hutter.
\newblock Neural architecture search: A survey.
\newblock \emph{Journal of Machine Learning Research}, 20\penalty0
  (55):\penalty0 1--21, 2019.

\bibitem[Feng et~al.(2024)Feng, Han, and Zhu]{feng2024optimal}
Zhigang Feng, Jiequn Han, and Shenghao Zhu.
\newblock Optimal taxation with incomplete markets--an exploration via
  reinforcement learning.
\newblock \emph{Available at SSRN 4758552}, 2024.

\bibitem[Fern{\'a}ndez-Villaverde et~al.(2024)Fern{\'a}ndez-Villaverde,
  Nu{\~n}o, and Perla]{fernandez2024taming}
Jes{\'u}s Fern{\'a}ndez-Villaverde, Galo Nu{\~n}o, and Jesse Perla.
\newblock Taming the curse of dimensionality: quantitative economics with deep
  learning.
\newblock Technical report, National Bureau of Economic Research, 2024.

\bibitem[Fernández-Villaverde et~al.(2025)Fernández-Villaverde, Gillingham,
  and
  Scheidegger]{annurev:/content/journals/10.1146/annurev-economics-091124-045357}
Jesús Fernández-Villaverde, Kenneth~T. Gillingham, and Simon Scheidegger.
\newblock Climate change through the lens of macroeconomic modeling.
\newblock \emph{Annual Review of Economics}, 2025.
\newblock ISSN 1941-1383.
\newblock \doi{https://doi.org/10.1146/annurev-economics-091124-045357}.
\newblock URL
  \url{https://www.annualreviews.org/content/journals/10.1146/annurev-economics-091124-045357}.

\bibitem[Folini et~al.(2024)Folini, Friedl, Kübler, and
  Scheidegger]{Folini_2021}
Doris Folini, Aleksandra Friedl, Felix Kübler, and Simon Scheidegger.
\newblock {The Climate in Climate Economics}.
\newblock \emph{The Review of Economic Studies, forthcoming}, 2024.

\bibitem[Friedl et~al.(2023)Friedl, K{\"u}bler, Scheidegger, and
  Usui]{friedlDeep2023}
Aleksandra Friedl, Felix K{\"u}bler, Simon Scheidegger, and Takafumi Usui.
\newblock Deep uncertainty quantification: With an application to integrated
  assessment models.
\newblock 2023.

\bibitem[Golosov et~al.(2014)Golosov, Hassler, Krusell, and
  Tsyvinski]{golosov2014optimal}
Mikhail Golosov, John Hassler, Per Krusell, and Aleh Tsyvinski.
\newblock Optimal taxes on fossil fuel in general equilibrium.
\newblock \emph{Econometrica}, 82\penalty0 (1):\penalty0 41--88, 2014.

\bibitem[Gottardi and Kubler(2011)]{gottardi2011social}
Piero Gottardi and Felix Kubler.
\newblock Social security and risk sharing.
\newblock \emph{Journal of Economic Theory}, 146\penalty0 (3):\penalty0
  1078--1106, 2011.

\bibitem[Harenberg et~al.(2019)Harenberg, Marelli, Sudret, and
  Winschel]{Harenberg17}
Daniel Harenberg, Stefano Marelli, Bruno Sudret, and Viktor Winschel.
\newblock Uncertainty quantification and global sensitivity analysis for
  economic models.
\newblock \emph{Quantitative Economics}, 10\penalty0 (1):\penalty0 1--41, 2019.
\newblock \doi{https://doi.org/10.3982/QE866}.
\newblock URL \url{https://onlinelibrary.wiley.com/doi/abs/10.3982/QE866}.

\bibitem[Heijdra et~al.(2006)Heijdra, Kooiman, and Ligthart]{Heijdra2006}
Ben~J. Heijdra, Jan~Peter Kooiman, and Jenny~E. Ligthart.
\newblock {Environmental quality, the macroeconomy, and intergenerational
  distribution}.
\newblock \emph{Resource and Energy Economics}, 28\penalty0 (1):\penalty0
  74--104, 2006.
\newblock ISSN 09287655.
\newblock \doi{10.1016/j.reseneeco.2005.05.001}.

\bibitem[Itskhoki and Moll(2019)]{itskhoki2019Optimal}
Oleg Itskhoki and Benjamin Moll.
\newblock Optimal {{Development Policies With Financial Frictions}}.
\newblock \emph{Econometrica}, 87\penalty0 (1):\penalty0 139--173, 2019.
\newblock ISSN 0012-9682.
\newblock \doi{10.3982/ECTA13761}.

\bibitem[Karp and Rezai(2014)]{Karp2014}
Larry Karp and Armon Rezai.
\newblock {The political economy of environmental policy with overlapping
  generations}.
\newblock \emph{International Economy Review}, 55\penalty0 (3):\penalty0
  711--733, 2014.

\bibitem[Karp et~al.(2024)Karp, Peri, and Rezai]{Karp_Peri_Rezai_2024}
Larry Karp, Alesandro Peri, and Armon Rezai.
\newblock Selfish incentives for climate policy: Empower the young!
\newblock \emph{Journal of the Association of Environmental and Resource
  Economists}, 11\penalty0 (5):\penalty0 1165--1200, 2024.
\newblock \doi{10.1086/728740}.
\newblock URL \url{https://doi.org/10.1086/728740}.

\bibitem[Kase et~al.(2022)Kase, Melosi, and Rottner]{kase2022estimating}
Hanno Kase, Leonardo Melosi, and Matthias Rottner.
\newblock \emph{Estimating nonlinear heterogeneous agents models with neural
  networks}.
\newblock Centre for Economic Policy Research, 2022.

\bibitem[Kingma and Ba(2014)]{kingma2014adam}
Diederik~P Kingma and Jimmy Ba.
\newblock Adam: A method for stochastic optimization.
\newblock \emph{arXiv preprint arXiv:1412.6980}, 2014.

\bibitem[Kotlikoff et~al.(2021{\natexlab{a}})Kotlikoff, Kubler, Polbin, Sachs,
  and Scheidegger]{kotlikoff2021making}
Laurence Kotlikoff, Felix Kubler, Andrey Polbin, Jeffrey Sachs, and Simon
  Scheidegger.
\newblock Making carbon taxation a generational win win.
\newblock \emph{International Economic Review}, 62\penalty0 (1):\penalty0
  3--46, 2021{\natexlab{a}}.

\bibitem[Kotlikoff et~al.(2021{\natexlab{b}})Kotlikoff, Kubler, Polbin, and
  Scheidegger]{kotlikoff2020pareto}
Laurence Kotlikoff, Felix Kubler, Andrey Polbin, and Simon Scheidegger.
\newblock {Pareto-Improving Carbon-Risk Taxation}.
\newblock \emph{Economic Policy}, 02 2021{\natexlab{b}}.
\newblock ISSN 0266-4658.
\newblock \doi{10.1093/epolic/eiab008}.
\newblock URL \url{https://doi.org/10.1093/epolic/eiab008}.
\newblock eiab008.

\bibitem[Kotlikoff et~al.(2024)Kotlikoff, Kubler, Polbin, and
  Scheidegger]{kotlikoffCanTodaysTomorrows2024}
Laurence Kotlikoff, Felix Kubler, Andrey Polbin, and Simon Scheidegger.
\newblock Can today's and tomorrow's world uniformly gain from carbon taxation?
\newblock \emph{European Economic Review}, 168:\penalty0 104819, September
  2024.
\newblock ISSN 00142921.
\newblock \doi{10.1016/j.euroecorev.2024.104819}.

\bibitem[Krueger and Kubler(2006)]{krueger2006pareto}
Dirk Krueger and Felix Kubler.
\newblock Pareto-improving social security reform when financial markets are
  incomplete!?
\newblock \emph{American Economic Review}, 96\penalty0 (3):\penalty0 737--755,
  2006.

\bibitem[Lenton et~al.(2008)Lenton, Held, Kriegler, Hall, Lucht, Rahmstorf, and
  Schellnhuber]{lenton2008tipping}
Timothy~M Lenton, Hermann Held, Elmar Kriegler, Jim~W Hall, Wolfgang Lucht,
  Stefan Rahmstorf, and Hans~Joachim Schellnhuber.
\newblock Tipping elements in the earth's climate system.
\newblock \emph{Proceedings of the national Academy of Sciences}, 105\penalty0
  (6):\penalty0 1786--1793, 2008.

\bibitem[Matthews et~al.(2009)Matthews, Gillett, Stott, and
  Zickfeld]{matthews2009proportionality}
H~Damon Matthews, Nathan~P Gillett, Peter~A Stott, and Kirsten Zickfeld.
\newblock The proportionality of global warming to cumulative carbon emissions.
\newblock \emph{Nature}, 459\penalty0 (7248):\penalty0 829--832, 2009.

\bibitem[Murphy(2022)]{pml2Book}
Kevin~P. Murphy.
\newblock \emph{Probabilistic Machine Learning: Advanced Topics}.
\newblock MIT Press, 2022.
\newblock URL \url{probml.ai}.

\bibitem[Nordhaus(2017)]{nordhausRevisitingSocialCost2017}
William~D. Nordhaus.
\newblock Revisiting the social cost of carbon.
\newblock \emph{Proceedings of the National Academy of Sciences}, 114\penalty0
  (7):\penalty0 1518 LP -- 1523, February 2017.
\newblock \doi{10.1073/pnas.1609244114}.

\bibitem[Nu{\~n}o and Thomas(2020)]{nuno2020Optimal}
Galo Nu{\~n}o and Carlos Thomas.
\newblock Optimal {{Monetary Policy}} with {{Heterogeneous Agents}}, 2020.

\bibitem[Nu{\~n}o et~al.(2024)Nu{\~n}o, Renner, and
  Scheidegger]{nuno2024monetary}
Galo Nu{\~n}o, Philipp Renner, and Simon Scheidegger.
\newblock Monetary policy with persistent supply shocks.
\newblock 2024.

\bibitem[Rasmussen and Williams(2005)]{Rasmussen:2005:GPM:1162254}
Carl~Edward Rasmussen and Christopher K.~I. Williams.
\newblock \emph{Gaussian Processes for Machine Learning (Adaptive Computation
  and Machine Learning)}.
\newblock The MIT Press, 2005.
\newblock ISBN 026218253X.

\bibitem[Renner and Scheidegger(2018)]{rennerscheidegger_2018}
Philipp Renner and Simon Scheidegger.
\newblock Machine learning for dynamic incentive problems.
\newblock Working paper. Available at SSRN:
  http://dx.doi.org/10.2139/ssrn.3282487, 2018.

\bibitem[Scheidegger et~al.(2018)Scheidegger, Mikushin, Kubler, and
  Schenk]{IPDPS2018}
S.~Scheidegger, D.~Mikushin, F.~Kubler, and O.~Schenk.
\newblock Rethinking large-scale economic modeling for efficiency:
  Optimizations for gpu and xeon phi clusters.
\newblock In \emph{2018 IEEE International Parallel and Distributed Processing
  Symposium (IPDPS)}, pages 610--619, May 2018.
\newblock \doi{10.1109/IPDPS.2018.00070}.

\bibitem[Scheidegger and Bilionis(2019)]{SCHEIDEGGER201968}
Simon Scheidegger and Ilias Bilionis.
\newblock Machine learning for high-dimensional dynamic stochastic economies.
\newblock \emph{Journal of Computational Science}, 33:\penalty0 68 -- 82, 2019.
\newblock ISSN 1877-7503.
\newblock \doi{https://doi.org/10.1016/j.jocs.2019.03.004}.
\newblock URL
  \url{http://www.sciencedirect.com/science/article/pii/S1877750318306161}.

\bibitem[Traeger(2014)]{traeger4StatedDICEQuantitatively2014}
Christian~P. Traeger.
\newblock A 4-{{Stated DICE}}: {{Quantitatively Addressing Uncertainty
  Effects}} in {{Climate Change}}.
\newblock \emph{Environmental and Resource Economics}, 59\penalty0
  (1):\penalty0 1--37, September 2014.
\newblock ISSN 0924-6460.
\newblock \doi{10.1007/s10640-014-9776-x}.

\bibitem[W{\"a}chter and Biegler(2006)]{wachter2006implementation}
Andreas W{\"a}chter and Lorenz~T Biegler.
\newblock On the implementation of an interior-point filter line-search
  algorithm for large-scale nonlinear programming.
\newblock \emph{Mathematical programming}, 106:\penalty0 25--57, 2006.

\bibitem[Weitzman(2012)]{weitzman2012ghg}
Martin~L Weitzman.
\newblock Ghg targets as insurance against catastrophic climate damages.
\newblock \emph{Journal of Public Economic Theory}, 14\penalty0 (2):\penalty0
  221--244, 2012.

\end{thebibliography}

\end{document}